\documentstyle[preprint,prc,eqsecnum,aps,floats,epsf,tighten,symbols,srev]{revtex}
\symbolprintfalse  

\begin{document}
%


%

%
%
%
%
%
%
%
%
%
\let\dsize=\displaystyle
\let\tsize=\textstyle
\let\ssize=\scriptstyle
\let\sssize\scriptscriptstyle
%
%
%
%
%
\def\spacesymbol{\leavevmode\hbox{\tt\char'040}}
\def\tildesymbol{\mathchar"0218 }
\def\coeff#1#2{{\textstyle{#1\over #2}}}
\def\tover#1#2{{\textstyle {#1\over #2}}}
\def\shalf{\tover12}
\def\e{\,{\rm e}}
\def\Tr{\mathop{\rm Tr}\nolimits}
\def\tr{\mathop{\rm tr}\nolimits}
\def\lsim{\lower0.6ex\vbox{\hbox{$\ \buildrel{\textstyle <}
         \over{\sim}\ $}}}
\def\gsim{\lower0.6ex\vbox{\hbox{$\ \buildrel{\textstyle >}
         \over{\sim}\ $}}}
%
\def\Order#1{{\cal O}(#1)}
\def\pairof#1{#1^+ #1^-}
\def\ee{\pairof{e}}
\def\CM{{\rm CM}}
\def\bar#1{\overline{#1}}

\def\bo{{\raise.15ex\hbox{\large$\Box$}}}   
\def\dell{\bigtriangledown}   

\def\th{$^{\rm th}$}
\def\st{$^{\rm st}$}
\def\nd{$^{\rm nd}$}
\def\rd{$^{\rm rd}$}

\def\unittie{\nobreak\,}
\def\fm{\unittie\hbox{fm}}
\def\fminverse{\unittie\hbox{fm}^{-1}}          \let\infm=\fminverse
\def\eV{\unittie\hbox{eV}}    \let\ev=\eV
\def\MeV{\unittie\hbox{MeV}}    \let\Mev=\MeV  \let\mev=\MeV
\def\GeV{\unittie\hbox{GeV}}    \let\Gev=\GeV  \let\gev=\GeV
\def\MeVfm{\hbox{MeV--fm}}  \let\mevfm=\MeVfm

%
%
%

\def\undervec#1{\setbox0\hbox{$\tildesymbol$}\setbox1\hbox{$#1$}#1%
                \dimen0=\wd0\advance\dimen0by\wd1\divide\dimen0by2
                 \kern-\dimen0\lower1.3ex\hbox{$\tildesymbol$}}
\def\sundervec#1{\setbox0\hbox{$\ssize\tildesymbol$}
                 \setbox1\hbox{$\ssize #1$}#1%
                \dimen0=\wd0\advance\dimen0by\wd1\divide\dimen0by2
                 \kern-\dimen0\lower.85ex\hbox{$\ssize\tildesymbol$}}
\def\ssundervec#1{\setbox0\hbox{$\sssize\tildesymbol$}
                 \setbox1\hbox{$\sssize #1$}#1%
                \dimen0=\wd0\advance\dimen0by\wd1\divide\dimen0by2
                 \kern-\dimen0\lower.6ex\hbox{$\sssize\tildesymbol$}}

\def\uvec#1{{\mathchoice{\undervec#1}%
     {\undervec#1}{\sundervec#1}{\ssundervec#1}}}

\def\longvec{\overrightarrow}

\def\boldvector#1{\ifcat#1A {{\bf #1}} \else \mbf{#1} \fi}

\def\mbf#1{{
\mathchoice
{\hbox{\boldmath$\displaystyle{#1}$}}
{\hbox{\boldmath$\textstyle{#1}$}}
{\hbox{\boldmath$\scriptstyle{#1}$}}
{\hbox{\boldmath$\scriptscriptstyle{#1}$}}
}}

\def\vectorsunder{\def\vector{\uvec}}
\def\vectorsbold{\def\vector{\boldvector}}      
\def\vectorsarrow{\def\vector{\mathaccent"017E }}  
\vectorsbold

%
%
%
\def\Tc{T_c}
\def\kfermi{k_{\sssize {\rm F}}}    
\def\epsfermi{\epsilon_{\sssize {\rm F}}}    
     \let\efermi=\epsfermi
\def\Mstar{M^{\ssize\ast}}       
\def\Mstarsq{M^{\ssize\ast}{}^2}
\def\Estar{E^{\ssize\ast}}
\def\EFstar{E^{\ssize\ast}_{\sssize {\rm F}}}
     \let\Efstar=\EFstar
\def\kstar{k^{\ssize\ast}}
\def\rhosmall#1{\rho_{\sssize {#1}}}
\def\rhoB{\rhosmall{\rm B}}
\def\rhozero{\rhosmall{0}}
\def\rhothree{\rhosmall{3}}
\def\rhos{\rho_{{\rm s}}}
\def\rhov{\rho_{{\rm v}}}
\def\Ylzero{Y_{l0}}     \let\Ylzer=\Ylzero
\def\Ylm{Y_{lm}}       \let\ylm=\Ylm
\def\Ghartree{G_{{\ssize \rm H}}}   \let\GHartree=\Ghartree
\def\Msun{M_{\odot}}
%
%
%
%
%
\def\mn{{\mu\nu}}         \let\munu=\mn
\def\gmunu{g^\mn}         \def\etamunu{\eta^\mn}
\def\gmunul{g_\mn}        \def\etamunul{\eta^\mn}
\def\sigmamunu{\sigma^\mn}
\def\sigmamunul{\sigma_\mn}
\def\sigmai{\sigma^i}
\def\sigmaj{\sigma^j}
\def\sigmaij{\sigma^{ij}}
\def\sigmavec{\vector\sigma}
\def\tauthree{\tau_{{\sssize 3}}}
\def\tauvec{\vector\tau}
\def\gammamu{\gamma^{\mu}}       \let\gammu=\gammamu
\def\gammamul{\gamma_{\mu}}      \let\gammul=\gammamul
\def\gammafive{\gamma^5}         \let\gamfiv=\gammafive
\def\gammafivel{\gamma_5}        \let\gamfivl=\gammafivel
\def\gammazero{\gamma^{0}}       \let\gamzer=\gammazero
\def\gammazerol{\gamma_{0}}      \let\gamzerl=\gammazerol
\def\gammavec{\vector\gamma}        \let\gamvec=\gammavec
\def\gammavecl{\vector\gamma}       \let\gamvecl=\gammavecl
\def\alphavec{\vector\alpha}         \let\alphvec=\alphavec
\def\slash#1{\rlap/{#1}}
\def\kslash{\slash{\mkern-1mu k}}    
\def\dmu{\partial^{\mu}}
\def\dmul{\partial_{\mu}}
\def\dnu{\partial^{\nu}}
\def\dnul{\partial_{\nu}}
\def\dmudmu{\dmul\dmu}        \let\dalamb=\dmudmu
\def\Dmu{D^{\mu}}
\def\Dmul{D_{\mu}}
\def\contract#1#2{\gamma^{#1}{#2}_{#1}}
\def\del{\vector\nabla}
\def\delsq{\del^2}
\def\deldot#1{\del\mbf{\cdot}\vector #1}
%
%
%
%
%
%
\def\lagrangian{{\cal L}}        \let\lag=\lagrangian
\def\hamiltonian{{\cal H}}
\def\Amu{A^\mu}
\def\Amul{A_\mu}
\def\Fmunu{F^\mn}
\def\Fmunul{F_\mn}
\def\Gmunu{G^\mn}
\def\Gmunul{G_\mn}
\def\Vzero{V_{\sssize 0}}
\def\Vmu{V^{\mu}}
\def\Vmul{V_{\mu}}
\def\Vvec{\vector V}      \let\Vvector=\Vvec
\def\bzero{b_{\sssize 0}}
\def\bmu{b^{\mu}}
\def\bmul{b_{\mu}}
\def\Bvec{\vector B}
\def\psibar{\overline\psi}
\def\psidag{\psi^{\dagger}}         \let\psidagger=\psidag
\def\Gnk{G_{n\kappa}(r)}
\def\Fnk{F_{n\kappa}(r)}
\def\phizero{\phi_{\sssize 0}}
\def\Azero{A_{\sssize 0}}
\def\xzero{x_{\sssize 0}}
\def\ubar{\bar{u}}
\def\dbar{\bar{d}}
\def\cbar{\bar{c}}
\def\qbar{\bar{q}}
\def\qbarq{\qbar q}
\def\Tmunu{T^\mn}
\def\Tnumu{T^{\nu\mu}}
%
%
%
%
\def\intback{\kern-.1em}
\def\dthree#1{\intback{\rm d}^3\intback #1}
\def\dfour#1{\intback{\rm d}^4\intback #1}
\def\dthreepi#1{\intback {\tsize{\rm d}^3\intback #1 \over
                                 \mathstrut\tsize (2\pi)^3}}
\def\dfourpi#1{\intback {\tsize{\rm d}^4\intback #1 \over
                                 \mathstrut\tsize (2\pi)^4}}
\def\dn#1#2{\intback{\rm d}^{#1}\intback #2}
\def\dnpi#1#2{\intback {\tsize{\rm d}^{#1}\intback #2 \over
                                 \mathstrut\tsize (2\pi)^{#1}}}
\def\intdn#1#2{\int\dn#1#2}
\def\intdnpi#1#2{\int\dnpi#1#2}
\def\dfourq{\dfour{q}}
\def\dfourk{\dfour{k}}
\def\dfourx{\dfour{x}}
\def\dthreeq{\dthree{q}}
\def\dthreek{\dthree{k}}
\def\dthreex{\dthree{x}}
\def\intdfourkpi{\int\dfourpi{k}}
\def\intdfourqpi{\int\dfourpi{q}}
\def\intdnkpin{\int\intback {\tsize{\rm d}^n\intback k \over
                   \mathstrut\tsize (2\pi)^n}}
\def\dhpi#1{\intback {\tsize{\rm d}^4 \intback #1 \over
         \mathstrut\tsize (2\pi)^3}}
\def\partder#1#2{{\partial #1\over\partial #2}}
                    \let\partialderiv=\partder  \let\partderiv=\partder
\def\deriv#1#2{{{\rm d} #1\over {\rm d}#2}}
\def\dbyd#1{{{\rm d}\over {\rm d}#1}}
\def\partialdbyd#1{{\partial \over\partial #1}}   \let\partdbyd=\partialdbyd
%
%
%
%
\def\threej#1#2#3#4#5#6{\left(\matrix{#1 & #2 & #3 \cr
                                      #4 & #5 & #6 \cr}\right)}
\def\sixj#1#2#3#4#5#6{\left\{\matrix{#1 & #2 & #3 \cr
                                      #4 & #5 & #6 \cr}\right\}}
\def\ninej#1#2#3#4#5#6#7#8#9{\left\{\matrix{#1 & #2 & #3 \cr
                                           #4 & #5 & #6 \cr
                                           #7 & #8 & #9 \cr}\right\}}
\def\alphas{\alpha_{{\rm s}}}
\def\gv{g _{\rm v}}        \let\gvec=\gv
\def\gvsq{\gv^2}
\def\grho{g_\rho}
\def\grhosq{\grho^2}
\def\gs{g_{\rm s}}         \let\gsca=\gs
\def\gssq{\gs^2}
\def\gpi{g_{\pi}}
\def\gpisq{\gpi^2}
\def\fpi{f_{\pi}}
\def\fpisq{\fpi^2}
\def\Fpi{F_{\pi}}
\def\GF{G_{{\sssize \rm F}}}
\def\gpiNN{g_{\pi{\sssize NN}}}
%
%
%
%
\def\mq{m_{{\rm q}}}         \let\mquark=\mq
\def\mv{m _{\rm v}}      \let\mvec=\mv
\def\mvsq{\mv^2}
\def\mrho{m_\rho}
\def\mrhosq{\mrho^2}
\def\ms{m_{\rm s}}       \let\msca=\ms
\def\mssq{\ms^2}
\def\mpi{m_{\pi}}
\def\mpisq{\mpi^2}
\def\Mnucleon{M_{{\sssize N}}}  \let\mnucleon=\Mnucleon
%
%
%
%
%
\def\aspac{\noalign{\vskip 2pt}}
\def\bspac{\noalign{\vskip 4pt}}
\def\cspac{\noalign{\vskip 6pt}}
\def\dspac{\noalign{\vskip 8pt}}
\def\espac{\noalign{\vskip 10pt}}
\def\fspac{\noalign{\vskip 12pt}}
\def\gspac{\noalign{\vskip 14pt}}
\def\hspac{\noalign{\vskip 16pt}}
%
%
%
%
%
\def\nucleus#1#2{$^{#2}${\rm #1}}
%
%
%
%
%
\def\etc{{\it etc.}}
\def\etal{{\it et al.}}
\def\eg{{\it e.g.},\ }
\def\ie{{\it i.e.},\ }
\def\cf{{\it cf.}}
%
%
%
%
%
\def\Tav#1{\langle \mkern-3mu\langle \mkern2mu%
              #1 \mkern2mu \rangle \mkern-3mu\rangle}
     \let\ensavg=\Tav
\def\abs#1{| #1 |}
\def\overlap#1#2{\langle #1 | #2 \rangle}
\def\matrixelement#1#2#3{\bral{#1}#2\ketl{#3}}
   \let\me=\matrixelement
\def\commutator#1#2{[#1,#2]}
\def\anticommutator#1#2{\{#1,#2\}}
\def\sTav#1{\left\langle \mkern-3mu\left\langle \mkern2mu%
              #1 \mkern2mu \right\rangle \mkern-3mu\right\rangle}
\def\sabs#1{\left| #1 \right|}
\def\soverlap#1#2{\left\langle #1 \right| \left. #2 \right\rangle}
\def\smatrixelement#1#2#3{\bra{#1\vphantom{#2#3}}#2\ket{#3\vphantom{#1#2}}}
   \let\sme=\smatrixelement
\def\scommutator#1#2{\left[#1,#2\right]}
\def\santicommutator#1#2{\left\{#1,#2\right\}}
\def\VEV#1{\left\langle #1 \right\rangle}    \let\vev=\VEV
\def\VEVl#1{\langle #1 \rangle}
\def\VEVgl#1{\bigl\langle #1 \bigr\rangle}
\def\VEVGl#1{\Bigl\langle #1 \Bigr\rangle}
\def\VEVggl#1{\biggl\langle #1 \biggr\rangle}
\def\bra#1{\left\langle #1 \right|}
\def\ket#1{\left| #1 \right\rangle}
\def\bral#1{\langle #1 |}
\def\ketl#1{| #1 \rangle}
\def\bragl#1{\bigl\langle #1 \bigr|}
\def\ketgl#1{\bigl| #1 \bigr\rangle}
\def\Bragl#1{\Bigl\langle #1 \Bigr|}
\def\Ketgl#1{\Bigl| #1 \Bigr\rangle}
\def\braggl#1{\biggl\langle #1 \biggr|}
\def\ketggl#1{\biggl| #1 \biggr\rangle}
\def\dubdots{}
\def\dubbral#1{\langle #1 \Vert}
\def\ddubbral#1{\langle #1 \dubdots}
\def\dubketl#1{\Vert #1 \rangle}
\def\ddubketl#1{\dubdots #1 \rangle}
\def\dubbragl#1{\bigl\langle #1 \bigr\Vert}
\def\ddubbragl#1{\bigl\langle #1 \dubdotsgl}
\def\dubketgl#1{\bigl\Vert #1 \bigr\rangle}
\def\ddubketgl#1{\dubdotsgl #1 \bigr\rangle}
\def\dubBragl#1{\Bigl\langle #1 \Bigr\Vert}
\def\ddubBragl#1{\Bigl\langle #1 \dubDotsgl}
\def\dubKetgl#1{\Bigl\Vert #1 \Bigr\rangle}
\def\ddubKetgl#1{\dubDotsgl #1 \Bigr\rangle}
\def\dubbraggl#1{\biggl\langle #1 \biggr\Vert}
\def\ddubbraggl#1{\biggl\langle #1 \dubdotsggl}
\def\dubketggl#1{\biggl\Vert #1 \biggr\rangle}
\def\ddubketggl#1{\dubdotsggl #1 \biggr\rangle}
\def\dubbra#1{\left\langle #1 \left\Vert}
\def\dubket#1{\right\Vert #1 \right\rangle}
%
%

%
%
%
%
\def\Phiket#1{\ket{\Phi_{#1}}}
\def\Phibra#1{\bra{\Phi_{#1}}}
\def\Psiket#1{\ket{\Psi_{#1}}}
\def\Psibra#1{\bra{\Psi_{#1}}}
%
%
\newcommand\beq{\begin{equation}}
\newcommand\eeq{\end{equation}}
\newcommand\beqa{\begin{eqnarray}}
\newcommand\eeqa{\end{eqnarray}}

\def\eqref#1{(\ref{#1})}

\def\Famp#1{{\cal F}^{#1}}
\def\uspinor#1{{\cal U}_{#1}({\bf x})}
\def\vspinor#1{{\cal V}_{#1}({\bf x})}
%

%
%
\begin{symbols}
\symdef\Akl{A^{\vphantom{\dagger}}_{{\bf k}\lambda}}
\symdef\Adagkl{A^{\dagger}_{{\bf k}\lambda}}

\symdef\bcdens{{\pmb{$\cal B$}}}
\symdef\Bhat{\hat B}
\symdef\Bkl{B^{\vphantom{\dagger}}_{{\bf k}\lambda}}
\symdef\Bdagkl{B^{\dagger}_{{\bf k}\lambda}}

\symdef\cf{{\it cf.}}
\symdef\coords{t,{\bf x}}
\symdef\currmu{B^{\mu}}

\symdef\d{{\rm d}}
\symdef\degen{{\gamma \over (2\pi)^3}}

\symdef\edens{{\cal E}}
\symdef\eg{{\it e.g.},\ }
\symdef\Estark{\Estar (k)}

\symdef\Fa{F_{a}(r)}
\symdef\frho{f_{\rho}}
\symdef\fv{f_{\rm v}}

\symdef\Ga{G_{a}(r)}

\symdef\Hhat{\hat H}

\symdef\ie{{\it i.e.},\ }

\symdef\kboltz{k_{\sssize {\rm B}}}    
\symdef\ksub{{\bbox{k}}}
\symdef\kvec{{\bf k}}

\symdef\lagrang{{\cal L}}
\symdef\lambdat{{\widetilde \lambda}}

\symdef\momdens{{\pmb{$\cal P$}}}
\symdef\momhat{{\bf{\hat P}}}
\symdef\msigma{m_\sigma}
\symdef\msigmasq{m^2_\sigma}

\symdef\Nbar{\bar{\mkern-2muN}}

\symdef\pcoords{t,-{\bf x}}
\symdef\phizero{\phi_0}

\symdef\scattamp{{\cal F}}
\symdef\sumkl{\sum_{{\bf k}\lambda}}

\symdef\UoptqE{U_{\rm opt}(q,E)}

\symdef\ua{\b{$a$}}
\symdef\uh{\b{$h$}}
\symdef\uL{\b{$L$}}
\symdef\ulambda{\b{$\lambda$}}
\symdef\uN{\!\b{$\mkern4muN$}}
\symdef\uNbar{\b{$\mkern6mu\bar{\mkern-2muN}$}}
\symdef\upi{\b{$\mkern2mu\pi$}}
\symdef\uQ{\b{$Q$}}
\symdef\uR{\b{$R$}}
\symdef\urho{\b{$\mkern2mu\rho$}}
\symdef\uU{\!\b{$\mkern4muU$}}
\symdef\uv{\b{$v$}}
\symdef\uxi{\b{$\xi$}}

\symdef\Vmunu{V^{\mu\nu}}
\symdef\Vmunul{V_{\mu\nu}}

\symdef{\vecalpha}{{\bbox{\alpha}}}
\symdef{\veccdot}{\,{\bbox{\cdot}}\,} 
\symdef{\veccross}{\,{\bbox{\times}}\,} 
\symdef{\vecgamma}{{\bbox{\gamma}}}
\symdef{\vecnabla}{{\bbox{\nabla}}}
\symdef{\vecpi}{{\bbox{\pi}}}
\symdef{\vecrho}{{\bbox{\rho}}}
\symdef{\vecsig}{{\bbox{\sigma}}}
\symdef{\vectau}{{\bbox{\tau}}}

\symdef\Vzero{V_0}
\end{symbols}
%
%





\preprint{\vbox{\hfill IU/NTC\ \ 96--17}}

\title{Recent Progress in Quantum Hadrodynamics}

\author{Brian D. Serot}
\address{Department of Physics and Nuclear Theory Center \\
         Indiana University, Bloomington, Indiana\ \ 47405}
\author{John Dirk Walecka}
\address{Department of Physics \\
    The College of William and Mary, Williamsburg, Virginia\ \ 23185 \\
     {\rm and} \\
    Thomas Jefferson National Accelerator Facility \\
    12000 Jefferson Avenue, Newport News, Virginia\ \ 23606\\ }
%
%

\maketitle
\vspace{0.25in}
\centerline{(\today )}
%
\begin{abstract}
Quantum hadrodynamics (QHD) is a framework for describing the nuclear 
many-body problem as a relativistic system of baryons and mesons.
Motivation is given for the utility of such an approach and for the
importance of basing it on a local, Lorentz-invariant lagrangian density.
Calculations of nuclear matter and finite nuclei in both renormalizable and
nonrenormalizable, effective QHD models are discussed.
Connections are made between the effective and renormalizable models, as well
as between relativistic mean-field theory and more sophisticated treatments.
Recent work in QHD involving nuclear structure, electroweak interactions
in nuclei, relativistic transport theory, nuclear matter under extreme
conditions, and the evaluation of loop diagrams is reviewed.
\end{abstract}
\vspace{0.8in}
\centerline{To be published in {\em International Journal of Modern
Physics E}}


\narrowtext


\newpage


\section{Introduction}
\label{sec:intro}
 
The study of atomic nuclei plays an important role in the
development of many-body theories.  
Early experimental probes of the nucleus were limited to energy scales 
considerably less than the nucleon mass $M \approx 939\,{\rm MeV}/c^2$, 
and the nucleus has traditionally been described as a collection of
nonrelativistic nucleons interacting through an instantaneous
two-body potential, with the dynamics given by the Schr\"odinger equation.  
The two-body potential is fitted to
the empirical properties of the deuteron and to low-energy
nucleon--nucleon (NN) scattering data, and one then attempts to
predict the properties of many-nucleon systems.  This is a
difficult problem, because the NN potential is strong, short ranged 
($R \approx 1\,\rm fm$), spin dependent, and has a very repulsive
central core.  Nevertheless, over a period of many years and
with the advent of more and more powerful computers, reliable
methods have been developed for solving the nonrelativistic
nuclear many-body problem.  

A new generation of accelerators will
allow us to study nuclei at higher energies, at shorter
distances, and with greater precision than ever before.  For
example, electron--nucleus scattering at CEBAF (now known as the Thomas
Jefferson National Accelerator Facility) will sample distance scales 
down to tenths of a Fermi, and ultra-relativistic heavy-ion collisions
at RHIC may produce nuclear densities of 10 times equilibrium density and
temperatures of several hundred MeV.  
These experiments will clearly
involve physics that goes beyond the Schr\"odinger equation,
such as relativistic motion of the nucleons, dynamical meson
exchanges, baryon resonances, modifications of hadron
structure in the nucleus, and the dynamics of the quantum
vacuum, which will include the production of a quark-gluon
plasma.  The challenge to theorists is to develop techniques
that describe this new physics, while maintaining the important
general properties of quantum mechanics, Lorentz covariance,
electromagnetic gauge invariance, and microscopic causality.
 
Since quantum chromodynamics (QCD) of quarks and gluons is
the fundamental theory of the strong interaction, it
is natural to look to QCD as the means to describe this new
physics.  While this may be desirable in principle, there are
many difficulties in practice, primarily because the QCD
coupling is strong at distance scales relevant for the vast
majority of nuclear phenomena.  Although significant progress
has been made in performing strong-coupling
lattice calculations, actual QCD predictions at nuclear length
scales with the precision of existing (and anticipated) data are
not presently available.  This situation will probably persist
for some time, particularly with regard to many-nucleon systems.
 
In contrast, a description based on hadronic degrees of freedom
is attractive for several reasons.  First, these variables are
the most efficient at normal densities and low temperatures, and
for describing particle absorption and emission, because these are the 
degrees of freedom actually observed in experiments.
Second, hadronic calculations can be calibrated by requiring
that they reproduce empirical nuclear properties and scattering
observables; we can then extrapolate to the extreme situations
mentioned earlier.  
As an example, accurate microscopic
meson-exchange models have been constructed to describe the NN
interaction \cite{Na73,Br76,Pa79,Zu81,Ma87,Te87,Ma89,Pl94,St94}.  
These contain several mesons, the most important
of which are the $\pi (0^-,1)$, $\sigma (0^+,0)$, $\omega
(1^-,0)$, and $\rho (1^-,1)$, where the indicated quantum
numbers denote spin, parity, and isospin, together with both N
and $\Delta (1232)$ degrees of freedom.  
Moreover, relativistic field theories of hadrons have been 
successful in describing the bulk and single-particle properties of nuclei
and nuclear matter in the mean-field and 
Dirac--Brueckner--Hartree--Fock approximations \cite{Se86,Te87,Ma89,Se92}.
 
Our basic goal is therefore to formulate a consistent
microscopic treatment of nuclear systems using hadronic (baryon
and meson) degrees of freedom.  
In principle, one could derive the form of the hadronic theory
from the underlying QCD lagrangian, using the ideas of modern Effective
Field Theory \cite{We90a,We91,We92a,Ge93}.
For example, one could define the low-energy effective theory by requiring
scattering amplitudes computed using hadrons to ``match'' the
corresponding amplitudes derived from the underlying theory.
Unfortunately, unlike certain cases where this matching can be done
(the Standard Model of weak interactions, QED descriptions of atomic
physics, and interactions in heavy-quark systems) \cite{Ma96}, 
the derivation of a 
low-energy hadronic theory directly from QCD is intractable at present.
Thus we must rely on other properties of QCD to constrain hadronic 
lagrangians.

One constraint is that the effective hadronic theory should embody the 
symmetries of QCD: Lorentz invariance, parity conservation, isospin symmetry, 
and spontaneously broken chiral symmetry.\footnote{%
The final two symmetries are only approximate.
We also want to maintain the usual gauge invariance in electromagnetic
interactions.}
These symmetry constraints severely restrict the forms of the allowed
interactions, but they are insufficient to completely specify the low-energy
dynamics.
Thus we will use existing phenomenology to guide us in the
construction of the effective lagrangian, in order to find the relevant 
degrees of freedom and the most efficient ways to structure the interactions.

For example, meson-exchange models of the NN interaction tell us which
mesons and baryons are the most important; these models have been useful
both for scattering and for the nuclear matter problem.
Moreover, experiments on
the electrodisintegration of the deuteron show unambiguously the
presence of pion-exchange currents, which arise when the
incoming (virtual) photon couples to a pion being exchanged
between two nucleons.
Medium-energy pion--nucleus
and proton--nucleus scattering indicate the importance of baryon
resonances, such as the $\Delta$, in nuclear reactions. 
In addition, a decomposition of the low-energy NN scattering amplitude 
using Lorentz invariants reveals that the empirical Lorentz
scalar, vector, and pseudoscalar amplitudes are much larger than
the amplitudes deduced from a nonrelativistic decomposition containing
Galilean invariants.
These large amplitudes have important consequences for the
spin, velocity, and density dependence of the NN interaction
and are at the heart of relativistic descriptions of
nucleon--nucleus scattering that reproduce spin observables in a
very economical fashion \cite{Mc83,Sh83,Cl83,Wa87}.
Finally, relativistic mean-field models using classical Lorentz scalar
and vector fields show that these are useful degrees of freedom for
describing bulk and single-particle nuclear properties and for elucidating
the important density dependence in the NN interaction. 

A Lorentz-covariant description is important for extrapolation to 
astrophysical objects and for describing processes at large 
energy-momentum transfer, as will be observed with the new accelerators.
On the other hand, recasting the nuclear many-body
problem in nonrelativistic form (\ie with two-component rather than
four-component nucleon spinors) leads to
interactions (both two-body and few-body) that are similar to those used 
in successful calculations of few-nucleon systems.
Many years of study within the nonrelativistic framework have produced
a quantitative description of the structure of light nuclei (here both NN
and NNN potentials are important) \cite{Wi92,Fr95,Pu95}.
Moreover, we have a qualitative understanding 
of some general features of nuclear structure, such as the importance of the
Pauli principle in reducing NN correlations \cite{Wa95}, which justifies
both the shell model and the single-nucleon optical potential,
and the interplay of single-particle and collective degrees of
freedom, which determines the shape of the nucleus \cite{Ei87}.  
Although the equilibrium properties of nuclear matter cannot
be precisely reproduced with modern NN potentials \cite{Da83,Da85,Ja92a}, 
the addition of a phenomenological, density-dependent interaction to the 
free NN potential leads to excellent results for the single-particle
structure and charge and mass densities of a large number of
nuclei \cite{Go79}. 
One of the goals of the effective hadronic theory is to provide a deeper
understanding of these successes and a more concrete link between these
results and the underlying QCD.

We desire a microscopic treatment of the nuclear many-body problem that is
consistent with quantum mechanics, special relativity, unitarity,
causality, cluster decomposition, and the intrinsic symmetries mentioned
earlier.
The modern viewpoint is that relativistic quantum field theory based on a
local, Lorentz-invariant lagrangian density is
simply the most general way to parametrize an $S$ matrix (or other
observables) consistent with these desired properties \cite{We95}.
Thus there is no reason that relativistic quantum field theory should
be reserved for ``elementary'' particles only.
We will refer to relativistic field theories based on hadrons as
{\it quantum hadrodynamics} or QHD \cite{Se86,Se92}.  
In principle, the field-theoretic formulation allows for the construction
of ``conserving approximations'' \cite{Ba61,Ba62} that maintain the
general properties mentioned above.
Analyses based on QHD, as defined here, can hopefully provide
a correct description of many-baryon systems at large
distances and at energy scales that are not too high.
One must examine, however, the limitations of hadronic field theory when
one attempts to extrapolate away from this regime.

Although spontaneously broken chiral symmetry implies that $\pi\pi$
and $\pi$N interactions are weak for small momenta, the NN interaction is 
too strong to be treated perturbatively in the couplings.
(This is also true in nonrelativistic formulations of the nuclear
many-body problem.)
It is therefore necessary to develop reliable nonperturbative
approximations, so that unambiguous comparisons between theory
and experiment can be made.  The formulation of practical,
reliable techniques for finite-density calculations in
strong-coupling relativistic quantum field theories is still basically
an unsolved problem \cite{Da86,Ka87,Fu89,Ko95d,Ka96a}. 
The development of such
tools in a hadronic field theory is not only useful in its own
right, but it may also provide insight into similar approaches
for QCD.  Nuclear many-body theory has had such influence on
other areas of physics in the past.

Historically, QHD models \cite{Wa74,Se79} were confined to the class of
renormalizable field theories, which can be characterized by a finite
number of coupling constants and masses \cite{Ca82,Co84}. 
The motivation was that these parameters could be calibrated to observed
nuclear properties, and one could then extrapolate into regions of
extreme density, temperature, or energy-momentum transfer without the
appearance of new, unknown parameters \cite{Wa74}.
The hope was that nonrenormalizable and vacuum effects would be small
enough that they could be adequately described by the long-distance
structure of a renormalizable hadronic theory, through the systematic
evaluation of quantum loops \cite{Se92}.
These models had several successes, some of which will be described below,
but also difficulties in carrying out the renormalization program, as well
as concrete indications that renormalizability is too restrictive.
Thus it is important to generalize these models to the more modern
viewpoint of nonrenormalizable, effective field theories, 
which can {\em still\/} provide
a consistent treatment of the nuclear many-body problem.
Reviewing the progress made toward this generalization in the past few
years is a central theme of this article.

Another goal of this article is to review the developments in QHD that
have transpired since 1992.\footnote{%
We try to be as complete as possible with references from early 1992 through
1995.}
We begin by discussing simple renormalizable models: {QHD--I} \cite{Wa74},
which contains neutrons, protons, and the isoscalar, Lorentz
scalar and vector mesons $\sigma$ and $\omega$;
the linear $\sigma$ model \cite{Sc57,Ge60,Le72}, which contains
neutrons, protons, pions, and neutral scalar mesons interacting
in a chirally invariant fashion; and the extension of these models to
include the isovector $\vecrho$ meson, which we generically call
{QHD--II} \cite{Se79,Ma82}.
These models serve as pedagogical tools for introducing
the relativistic mean-field and Dirac--Hartree approximations, and
their application to both infinite nuclear matter and the ground
states of atomic nuclei.  
At the relativistic mean-field theory (RMFT) level with classical,
isoscalar, Lorentz scalar and vector fields, {QHD--I} provides a
simple picture of the equilibrium properties of nuclear matter.
We focus on the new features that arise in a relativistic framework 
and emphasize the important concepts of Lorentz covariance and
self-consistency.  
When extended to finite nuclei through Dirac--Hartree calculations with 
a classical, isovector, Lorentz vector field and a few parameters fitted to 
the properties of nuclear matter, one derives the nuclear shell model 
and can obtain, through the relativistic impulse approximation (RIA), an 
efficient description of the scattering of medium-energy nucleons 
from nuclei.
We also show that the bulk and single-particle properties of nuclei
provide stringent enough constraints to distinguish between various
RMFT models.
These successful QHD results are discussed in Section~\ref{sec:qhd}.

In Section~\ref{sec:chiral}, we examine the linear sigma model.
We show how the constraints of linear chiral symmetry and spontaneous
symmetry breaking, when realized in the usual fashion, {\em preclude\/}
a successful description of nuclei at the mean-field level.
We then discuss how the exchange of two correlated pions between nucleons
produces a strong mid-range attraction in the scalar-isoscalar channel;
this allows for an alternative (nonlinear) implementation of the chiral
symmetry, with an additional scalar field to simulate correlated two-pion
exchange, that produces successful mean-field results.
We conclude that the scalar field in \nobreak{QHD--I} should be regarded 
as an {\em effective\/} field that includes the pion dynamics that is most
important for describing bulk nuclear properties.

In Section~\ref{sec:few}, we consider the application of
effective hadronic theories to the two- and few-nucleon problems.
Several meson-exchange models of the NN interaction have been
developed and applied to nuclear systems, both relativistically
and nonrelativistically.
These models focus primarily on the precise reproduction of the
two-nucleon data, and they are typically not renormalizable.
We discuss the relativistic quasipotential formalism and its successful
application to two-nucleon systems, as well as its extension to the study
of two-nucleon correlations in nuclear matter.
We then examine how the RMFT results are modified when short-range
correlations are included and also gain some insight into the relationship
between relativistic and nonrelativistic approaches to the nuclear matter
problem.
The successes of these nonrenormalizable models, as well as the introduction 
of an effective scalar field in Section~\ref{sec:chiral}, give strong 
empirical hints that one must really view QHD as arising from an effective, 
nonrenormalizable lagrangian that yields an economical description of the 
underlying theory of QCD in the nuclear domain.  

Therefore, in Section~\ref{sec:eft}, we present a formulation of QHD in the
modern context of effective field theory.
Since a nonrenormalizable QHD lagrangian generally contains an infinite
number of unknown constants, one must identify suitable expansion 
parameters and define a meaningful truncation scheme.
Based on the results in Section~\ref{sec:qhd}, we argue that although
the mean meson fields are large on the scale of nuclear energies (several 
hundred MeV versus tens of MeV), and one has all the complexities of a 
strong-coupling theory, the {\em ratios\/} of the mean fields to the 
nucleon mass are small at moderate densities.
These ratios thus provide suitable expansion parameters.
This observation leads naturally into Section~\ref{sec:rmft}, 
where a wide range of applications of RMFT to nuclear structure throughout 
the periodic table and to a variety of nuclear phenomena is discussed.
The numerous successful calculations since early 1992
are considered within the
general context of effective field theory, and the importance of both
renormalizable and nonrenormalizable couplings is considered.
We also introduce some basic ideas of density functional 
theory \cite{Dr90} to illustrate how RMFT models can include many-body
effects that go beyond the simple Hartree approximation.
When combined with the systematics of two-nucleon correlations described
in Section~\ref{sec:few}, these ideas help to explain why RMFT models are
so successful.

The RMFT has also been widely applied to relativistic transport theory
and heavy-ion reactions through the Vlasov--Uehling--Uhlenbeck 
equations for the phase-space distribution function.  
We briefly review the basic theoretical background
and discuss developments since 1992 in Section~\ref{sec:matter}.
We also consider the extrapolation of RMFT results for the nuclear equation 
of state to high baryon density and high temperature.

To have a consistent effective field theory, one must also consider
loop corrections to the RMFT.
These loops contain both familiar many-body contributions \cite{Fe71} and
quantum vacuum effects.
The underlying vacuum dynamics of QCD is implicitly contained in the
parameters of the effective hadronic lagrangian; nevertheless, these vacuum
effects are modified in the presence of valence nucleons at finite density.
Although chiral perturbation theory \cite{Ga84,Me93a}
provides a systematic framework for including pion loops in meson--meson
and meson--nucleon scattering amplitudes, the
inclusion of vacuum loops that arise in the many-body theory is still
an unsolved problem. 
We review the present status of loop calculations in QHD in 
Section~\ref{sec:loops}.
Finally, Section~\ref{sec:summ} contains a summary.

The reader is assumed to have a working knowledge of
relativistic quantum field theory, canonical quantization, and
the use of path integrals at zero temperature.  For general
background on relativistic field theory, the reader can turn to a number of 
texts\cite{Bj64,Bj65,Na78,It80,Ra81,Ma84,Ry85,Ka89,St93,Um93,Gr93,We95,We96};
for background on quantum hadrodynamics, two recent reviews exist that
cover the basic formulation and developments up to 1992 \cite{Se86,Se92}.  
A recent text develops many of the theoretical tools in more detail 
\cite{Wa95}.



\section{Quantum Hadrodynamics (QHD)}
\label{sec:qhd}

The purpose of these next three sections is to develop some insight into the
ingredients necessary for a realistic relativistic description of few- and
many-body nuclear systems.
In this section, we review the basic ideas of QHD and some of its 
applications.  
Most of the concepts are developed in \cite{Se86,Se92}, which discuss
the literature up to 1992, and a recent text provides a more thorough 
treatment of the theoretical tools and background \cite{Wa95}. 
We will not repeat all of that material here.
Moreover, since we are concentrating on basic results, we postpone 
until later the discussion of recent progress, where we will try to be as 
complete as possible with developments and references from early 1992
through 1995. 
While the literature in this area prior to 1992 is detailed in 
\cite{Se86,Se92}, we would like to single out for special mention the 
important background papers \cite{Sc51,Jo55,Du56,We67,Mi72}.

Quantum hadrodynamics as defined above is a general framework
for the relativistic nuclear many-body problem.  The detailed
dynamics must be specified by choosing a particular lagrangian density.  
As we will discuss later,
there is increasing evidence that QHD must be regarded as an
effective field theory, where all types of hadronic couplings
satisfying general symmetry requirements are to be included.
Nonetheless, the original motivation for using a renormalizable
lagrangian to formulate a consistent relativistic nuclear
many-body theory based on hadronic degrees of freedom remains
valid.  Furthermore, as we shall see, there is now convincing
empirical evidence that whatever the form of the effective theory, it must
be dominated by strong isoscalar, Lorentz scalar and vector interactions.

To introduce the relativistic formalism, we consider a simple
model called QHD--I \cite{Wa74}, which contains fields for
baryons $\Big[\psi =
\pmatrix{\psi_{\rm p}\cr
\psi_{\rm n}\cr}\Big]$ 
and neutral scalar ($\phi$)
and vector ($\Vmu$) mesons.
The lagrangian density for this model (with $\hbar = c = 1$)
is given by \cite{Se86,Se92}

\beqa
    \lagrang &=& \psibar \big[ \gammamul (i\dmu - \gv\Vmu )
         - (M - \gs\phi)
         \big] \psi + \tover12 (\dmul\phi\dmu\phi - \mssq \phi^2)
         \nonumber \\[3pt]
     &  &\qquad {}
       -\tover{1}{3!}\kappa\phi^3 - \tover{1}{4!}\lambda\phi^4
       - \tover14 F_{\mu\nu}F^{\mu\nu} + \tover12 \mvsq\Vmul\Vmu
            + \delta\lagrang   \ ,
         \label{eq:qhdone}
\eeqa
where $F^{\mu\nu} \equiv \dmu V^{\nu} - \dnu\Vmu$ and
$\delta\lagrang$ contains counterterms.
The parameters $M$, $\gs$, $\gv$, $\ms$, $\mv$, $\kappa$, and $\lambda$
are phenomenological constants that may be determined (in principle) from 
experimental observables.
This lagrangian resembles massive QED with an
additional scalar interaction, so the resulting
relativistic quantum field theory is
{\em renormalizable\/} \cite{Bo70}.
The inclusion of the scalar self-interactions proportional to $\phi^3$ and
$\phi^4$ make this the most general lagrangian consistent with 
renormalizability (for these degrees of freedom) \cite{Bo77}.
The counterterms in $\delta\lagrang$ are used for renormalization.
 
The motivation for this model has evolved considerably since it was
introduced.
As discussed in the Introduction, when the empirical NN scattering amplitude
is described in a Lorentz-covariant fashion, it contains
large isoscalar, scalar and four-vector
pieces \cite{Mc83,Sh83,Cl83,Wa87}, and the simplest way to reproduce these
is through the exchange of neutral scalar and vector mesons.
The neutral scalar and vector
components are the most important for describing bulk nuclear
properties, which is our main concern here.
Other Lorentz components of the NN interaction average essentially to zero
in spin-saturated nuclear matter and may be incorporated as refinements to the
present model; we will discuss pion dynamics in the following section.
The important point is that even in more complete models, the dynamics 
generated by scalar and vector mesons will remain; thus, it is important 
to first understand the consequences of these degrees of freedom for 
relativistic descriptions of nuclear systems.
 
The field equations for this model follow from the Euler--Lagrange
equations and
can be written as
\beqa
(\dmul\dmu + \mssq ) \phi + \tover12 \kappa \phi^2
        + \tover16 \lambda \phi^3 &=& \gs \psibar\psi   \ ,
         \label{eq:kg}\\[3pt]
\dnul F^{\nu\mu} + \mvsq \Vmu &=& \gv \psibar \gammamu \psi  \ ,
             \label{eq:qed}\\[3pt]
[\gammamu (i\dmul - \gv\Vmul ) - (M - \gs\phi )]\psi &=& 0   \ .
             \label{eq:diracone}
\eeqa
(The counterterms have been suppressed.)
Equation \eqref{eq:kg}\ is a Klein--Gordon equation with a scalar 
source term and nonlinear scalar self-interactions.
Equation \eqref{eq:qed}\
looks like massive QED with the conserved baryon current
\beq
   \currmu
    \equiv (\rhoB , \bcdens )
    =\psibar \gammamu \psi
      , \qquad
   \dmul B^{\mu} = 0 \ ,   \label{eq:barone}
\eeq
rather than the (conserved) electromagnetic current as the source.
Finally, Eq.~\eqref{eq:diracone}\
is the Dirac equation with scalar and vector fields
entering in a minimal fashion.
These field equations imply that the canonical energy-momentum tensor
$\Tmunu$
is conserved ($\dmul\Tmunu = \dnul\Tmunu = 0$).
 
When quantized, Eqs.~\eqref{eq:kg}--\eqref{eq:diracone}\ become
{\em nonlinear quantum field equations}, whose exact solutions (if they
exist) are very complicated.  In particular, they describe mesons and
baryons {\em that are not point particles}, but rather objects
with structure due to the implied (virtual) meson and
baryon-antibaryon loops.  
Here the dynamical input of renormalizability is apparent, since we are 
assuming that the intrinsic structure (or at least the long-range part of 
it) can be described using hadronic degrees of freedom.  
Strictly speaking, the validity of this input and its limitations have
yet to be tested conclusively within the framework of QHD;
nevertheless, as we shall see, there are now strong indications that this
assumption is too optimistic.
 
We also expect the coupling constants in
Eqs.~\eqref{eq:kg}--\eqref{eq:diracone}\ to be large, so
perturbative solutions are not useful.  Fortunately, there is an
approximate nonperturbative solution that can serve as a starting point
for studying the implications of the lagrangian in Eq.~\eqref{eq:qhdone}.
Consider a system of $B$ baryons in a large box of volume $V$ at zero
temperature.  Assume that we are in the rest frame of the
matter, so that the baryon flux $\bcdens = 0$.  As the baryon
density $B/V$ increases, so do the source terms on the
right-hand sides of Eqs.~\eqref{eq:kg}\ and \eqref{eq:qed}.
If the sources are large enough, the meson field operators can be
approximated by their expectation values, which are classical
fields: %
\beq
    \phi \rightarrow \langle \phi \rangle \equiv \phizero   ,\qquad
    \Vmu \rightarrow
   \langle \Vmu \rangle \equiv (\Vzero,0) \ .
          \label{eq:mfappx}
\eeq
For our stationary,
uniform system, $\phizero$ and $\Vzero$ are {\em constants\/} that
are independent of space and time, and since the matter is at rest,
the classical three-vector field $\Vvec = 0$.
 
It is important to emphasize that the preceding
``mean-field'' theory (MFT) serves only as a starting point for
calculating corrections within the framework of QHD, using
Feynman diagrams and path-integral methods, as discussed in 
\cite{Se86,Se92,Wa95}.
We will return later to decide at which densities this starting point
is actually useful.

\subsection{The Nuclear Matter Equation of State}  
When the meson
fields in Eq.~\eqref{eq:qhdone}\ are approximated by the
constant classical fields of Eq.~\eqref{eq:mfappx}, we arrive at
the mean-field lagrangian density %
\beq
    {\cal L}_{\sssize {\rm MFT}} =
              \psibar [ i \dmu\gammamul - \gv\Vzero\gammazerol
              - (M - \gs\phizero)]\psi
              -\tover12\mssq\phizero^2
              -\tover{1}{3!}\kappa\phizero^3
              -\tover{1}{4!}\lambda\phizero^4
              +\tover12\mvsq\Vzero^2   \ .
    \label{eq:lagrang}
\eeq
(The counterterms have been suppressed.)
The conserved baryon four-current remains as in Eq.~\eqref{eq:barone},
and the canonical energy-momentum tensor becomes
\beq
    \Tmunu_{\sssize {\rm MFT}}
         = i\psibar\gammamu\dnu\psi - (\tover12 \mvsq \Vzero^2
                  -\tover12 \mssq\phizero^2
        -\tover{1}{3!}\kappa\phizero^3 - \tover{1}{4!}\lambda\phizero^4
         )\gmunu    \ .
    \label{eq:enmomt}
\eeq
As discussed by Freedman \cite{Fr78}, there is no need to symmetrize
$\Tmunu$ if we consider only uniform nuclear matter.
This follows because the additional terms in the symmetrized tensor enter
as a total four-divergence, whose diagonal matrix elements vanish in a 
uniform system.
 
Since the meson fields are classical, only the fermion field must be
quantized.
The Dirac field equation follows from $\lagrang_{\sssize {\rm MFT}}$:
\beq
[i\gammamul\dmu - \gv\gammazerol\Vzero - (M - \gs\phizero)]\psi(\coords)
                             = 0  \ ,
    \label{eq:Diracf}
\eeq
and since this equation is {\em linear}, it can be solved exactly.
The scalar field $\phizero$ shifts the baryon mass from $M$ to
$\Mstar \equiv M - \gs\phizero$, while the vector field $\Vzero$ shifts the
energy spectrum.
We look for normal-mode solutions
with both positive and negative energies, as is natural for
the Dirac equation.  These solutions can be used to define
quantum field operators $\psi$ and $\psidagger$ in the usual fashion,
and by imposing the familiar equal-time anticommutation relations,
we can construct the baryon number operator
$\Bhat \equiv
\int\dthreex\, (:\psibar\gammazero\psi :)$ and the four-momentum operators
${\hat P}^{\mu} = (\Hhat,\momhat)\equiv \int\dthreex\, {\hat T}^{0\mu}$,
with the results
\beqa
    \Hhat - \me{0}{\Hhat}{0}  &\equiv & 
      \Hhat_{\sssize {\rm MFT}} + \delta H \ ,   
         \label{eq:hamone}\\[3pt]
    \Hhat_{\sssize {\rm MFT}} &=&
    \sumkl (\kvec^2 + \Mstarsq )^{1/2}
                   (\Adagkl\Akl + \Bdagkl\Bkl)
                   +\gv\Vzero\Bhat  \nonumber \\
        &  & \qquad\qquad {}+(\tover12\mssq\phizero^2
               +\tover{1}{3!}\kappa\phizero^3
               +\tover{1}{4!}\lambda\phizero^4
                   - \tover12\mvsq\Vzero^2)V    \ ,
              \label{eq:hammft} \\[3pt]
    \delta H &=& -\sumkl \Big[ (\kvec^2 + \Mstarsq )^{1/2} -
                         (\kvec^2 + M^2)^{1/2} \Big]   \ ,
              \label{eq:deltah}\\[3pt]
    \Bhat &=& \sumkl   (\Adagkl\Akl - \Bdagkl\Bkl)    \ ,
              \label{eq:bop}\\
    \momhat &=& \sumkl   \kvec   (\Adagkl\Akl + \Bdagkl\Bkl) \ .  
              \label{eq:momop}
\eeqa
Here $\Adagkl$, $\Bdagkl$, $\Akl$, and $\Bkl$ are creation and
destruction operators for (quasi)baryons and (quasi)antibaryons
with shifted mass and energy, and $\Bhat$ is the ``normal-ordered'' baryon 
number operator, which clearly counts the number of baryons minus the
number of antibaryons.  
(The index $\lambda$ denotes both spin and isospin projections.) 
The correction term $\delta H$ arises
from placing the operators in $\Hhat_{\sssize {\rm MFT}}$ in
normal order and includes the contribution to the energy
from the filled Dirac sea, where the baryon mass has been
shifted by the uniform scalar field $\phizero$\cite{Se86}.
Since all energies are measured relative to the vacuum, we must subtract the
total energy of the Dirac sea in the vacuum state $\ket{0}$, where the
baryons have their free mass $M$.
We will return later to discuss this ``zero-point energy''
correction; for now, let us concentrate on the MFT hamiltonian
defined by Eq.~\eqref{eq:hammft}.
 
Since $\Hhat_{\sssize {\rm MFT}}$ is diagonal, this model
mean-field problem has been solved {\em exactly\/} once the
meson fields are specified; their determination is discussed
below.  The solution retains the essential features of QHD:
explicit mesonic degrees of freedom, consistency with relativistic 
covariance \cite{Fu90}, and
the incorporation of antiparticles.  
Since $\Bhat$ and $\momhat$ are also diagonal, the baryon number and
total momentum are constants of the motion, as are their
corresponding densities $\rhoB$ and $\momdens$, since the volume
is fixed.
 
For uniform nuclear matter, the ground state is obtained by
filling energy levels with spin-isospin degeneracy $\gamma$ up
to the Fermi momentum $\kfermi$.  (The generalization to finite
temperature will be discussed at the end of this section.) The
Fermi momentum is related to the baryon density $\rhoB \equiv B/V$ by %
\beq
    \rhoB = {\gamma\over (2\pi)^3} \int_0^{\kfermi} \dthree{k} =
       {\gamma\over 6\pi^2}  \kfermi^3   \ ,
         \label{eq:olddens}
\eeq
where the degeneracy factor is 4 for symmetric ($N=Z$) matter
and 2 for pure neutron matter ($Z=0$).
The constant vector field $\Vzero$ can be expressed in terms of conserved
quantities from the expectation value of the vector meson field equation
\eqref{eq:qed}:
\beq
   \Vzero = \frac{\gv}{\mvsq} \rhoB \ .    \label{eq:Visrho}
\eeq

The expressions for the energy density and pressure now take the simple
forms\cite{Se86}
\beqa
    \edens &=&
    {\gvsq\over 2\mvsq}  \rhoB^2 + {\mssq\over 2\gssq}  (M - \Mstar)^2
    + {\kappa\over 6\gs^3} (M - \Mstar)^3
    + {\lambda\over 24\gs^4} (M - \Mstar)^4 \nonumber\\[3pt]
  & & \qquad {}
    + {\gamma\over (2\pi)^3} \int_0^{\kfermi} \dthree{k} \ \Estark   \ ,
         \label{eq:endens} \\[3pt]
    p  &=&
    {\gvsq\over 2\mvsq}  \rhoB^2 - {\mssq\over 2\gssq}  (M - \Mstar)^2
    - {\kappa\over 6\gs^3} (M - \Mstar)^3
    - {\lambda\over 24\gs^4} (M - \Mstar)^4 \nonumber\\[3pt]
  & & \qquad {}
    +\frac{1}{3} 
    {\gamma\over (2\pi)^3} \int_0^{\kfermi} \dthree{k} \
         \frac{\kvec^2}{\Estark}   \ ,
         \label{eq:oldpress}
\eeqa
where $\Estark \equiv (\kvec^2 + \Mstarsq )^{1/2}$.  The first
four terms in Eqs.~\eqref{eq:endens}\ and \eqref{eq:oldpress}\
arise from the classical meson fields.  The final terms in these
equations are those of a relativistic gas of baryons of mass
$\Mstar$.  These expressions give the nuclear matter equation of
state at zero temperature in parametric form: $\edens (\rhoB )$
and $p (\rhoB )$.
 
The constant scalar field $\phizero$, or equivalently, the
effective mass $\Mstar$, can be determined thermodynamically at
the end of the calculation by minimizing $\edens (\Mstar )$ with
respect to $\Mstar$.  This produces the {\em self-consistency
condition} 
\beq
    \Mstar
          = M - {\gssq \over \mssq } \rhos
            + {\kappa\over 2\gs\mssq} (M - \Mstar)^2
            + {\lambda\over 6\gs^2 \mssq} (M - \Mstar)^3 \ ,
                          \label{eq:mftsc}
\eeq  
where the scalar density $\rhos$ is defined by
\beq
   \rhos \equiv \langle :\psibar\psi :\rangle
    =  {\gamma\over (2\pi)^3} 
            \int_0^{\kfermi} \dthree{k}\ {\Mstar\over\Estark }\ .
            \label{eq:rhosdef}
\eeq
Equation \eqref{eq:mftsc}\ is equivalent to the MFT scalar
field equation for $\phizero$.
Note that the scalar density is smaller than the baryon density
[Eq.~\eqref{eq:olddens}]
due to the factor $\Mstar / \Estark$, which is an effect of Lorentz
contraction.
Thus the contribution of rapidly moving baryons to the scalar source is
significantly reduced.
Most importantly, Eq.~\eqref{eq:mftsc}\ is a {\em transcendental
self-consistency equation\/}
for $\Mstar$ that must be solved at each value of $\kfermi$.
This illustrates the {\em nonperturbative\/} nature of the mean-field
solution.
 
\begin{figure}[t]
 \setlength{\epsfxsize}{5.0in}
 \centerline{\epsffile{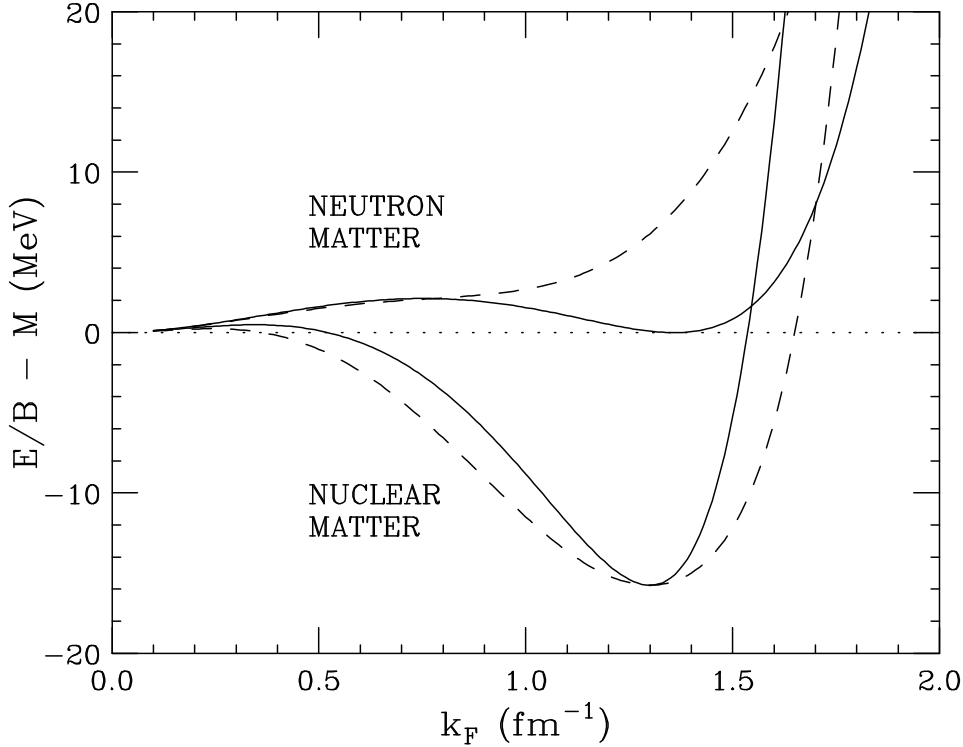}}
  \vspace{13pt}\renewcommand{\baselinestretch}{1.0}
    \caption{\protect\label{fig:mftsat}%
Saturation curves for nuclear matter.
These results are calculated in the relativistic mean-field theory with
baryons and neutral scalar and vector mesons (QHD--I).
The coupling constants are chosen to fit the value and position of the
minimum.
The solid curve uses the couplings in Eq.~\protect\eqref{eq:coupls},
while the dashed curve uses the parameter set NLC, as described 
later in the text.
The predictions for neutron matter ($\gamma = 2$) are also shown.}
\end{figure}
 
To analyze these results, we initially set $\kappa = \lambda = 0$, as in
the original version of the model \cite{Wa74}.
An examination of the analytic expression \eqref{eq:endens}\ for
the energy density shows that the system is unbound ($\edens /
\rhoB > M$) at either very low or very high densities. 
At intermediate densities, the attractive
scalar interaction will dominate if the coupling constants are
chosen properly.  The system then {\em saturates}.  Nuclear
matter with an equilibrium Fermi wavenumber $\kfermi^0 = 1.30\infm$
and an energy/nucleon $e_0 \equiv (\edens / \rhoB - M) =
-15.75\MeV$ is obtained if the couplings are chosen
as\footnote{The values $C^2_{\rm s} = 267.1$ and $C^2_{\rm v} =
195.9$ used in \protect\cite{Se86} yield $\kfermi^0 = 1.42\infm$.} %
\beq
   C^2_{\rm s} \equiv \gssq \Big({M^2\over \mssq}\Big) = 357.4   , \qquad
   C^2_{\rm v} \equiv \gvsq \Big({M^2\over \mvsq}\Big)
                          = 273.8 \ .
         \label{eq:coupls}
\eeq
In this approximation, the nuclear compression modulus $K$ is $545 \MeV$.
Note that the meson masses enter only through the {\em ratios\/} 
$g^2_i / m^2_i$ in Eqs.~\eqref{eq:endens},
\eqref{eq:oldpress}, and \eqref{eq:mftsc}.
(For a more complete discussion of the relevant dimensional coupling
parameters and their specification from nuclear matter properties,
see \cite{Fu96}.)
The resulting saturation curve is shown in Fig.~\ref{fig:mftsat}.
In this approximation, the relativistic properties of the scalar and vector
fields are responsible for saturation; a Hartree--Fock variational estimate
built on the nonrelativistic (Yukawa) potential limit of the interaction shows
that such a system is unstable against collapse \cite{Fe71,Wa95}.
 
\begin{figure}[t]
 \setlength{\epsfxsize}{4.0in}
 \centerline{\epsffile{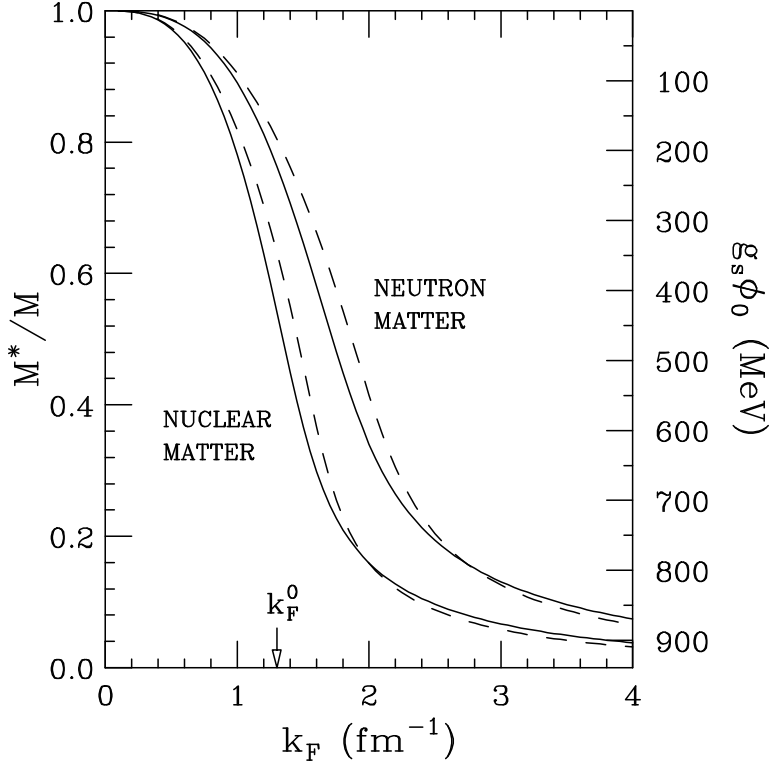}}
  \vspace{13pt}\renewcommand{\baselinestretch}{1.0}
    \caption{\protect\label{fig:mftmstar}%
Effective mass as a function of density for nuclear ($\gamma = 4$) and neutron
($\gamma = 2$) matter based on Fig.~\protect\ref{fig:mftsat}.}
\end{figure}
 
The solution of the self-consistency condition \eqref{eq:mftsc}\
for $\Mstar$ yields an effective mass that is a decreasing
function of the density, as illustrated in
Fig.~\ref{fig:mftmstar}.  Evidently, $\Mstar /M$ becomes small
at high density and is significantly less than unity at ordinary
nuclear densities ($\Mstar / M = 0.541$ at $\kfermi^{\vphantom{0}}
= \kfermi^0$).
This is a consequence of the large scalar
field $\gs\phizero$, which is approximately 400 MeV (at 
$\kfermi^{\vphantom{0}} = \kfermi^0$) and which
produces a large attractive contribution to the energy/baryon.
There is also a large repulsive energy/baryon from the vector
field $\gv\Vzero
\approx 350\MeV$.
Thus {\em the Lorentz structure of the interaction leads to a new energy scale
in the problem}, and the small nuclear binding energy ($\approx 16 \MeV$)
arises from the cancellation between the large scalar attraction and vector
repulsion.
As the nuclear density increases, $\Mstar$ decreases,
the scalar source $\rhos$ becomes smaller
than the vector source $\rhoB$, and the attractive forces saturate,
producing the minimum in the binding curve.\footnote{%
Figure 15 in \cite{Se86} shows the decrease in $\rhos$ relative to
$\rhoB$ for $0 < \kfermi^{\vphantom{0}} \leq \kfermi^0$.}
Clearly, because of the sensitive cancellation involved near the equilibrium
density, corrections to the MFT must be calculated before the importance of
this saturation mechanism can be assessed.
Nevertheless, the Lorentz structure of the interaction provides {\em
a new saturation mechanism that is not present in the nonrelativistic
potential limit}, as this limit ignores the distinction between Lorentz
scalar and vector fields.
We will see later that part of this new saturation mechanism can be 
expressed in terms of repulsive {\em many-body\/} forces in a nonrelativistic
formulation.
 
\begin{figure}[t]
 \setlength{\epsfxsize}{5.0in}
 \centerline{\epsffile{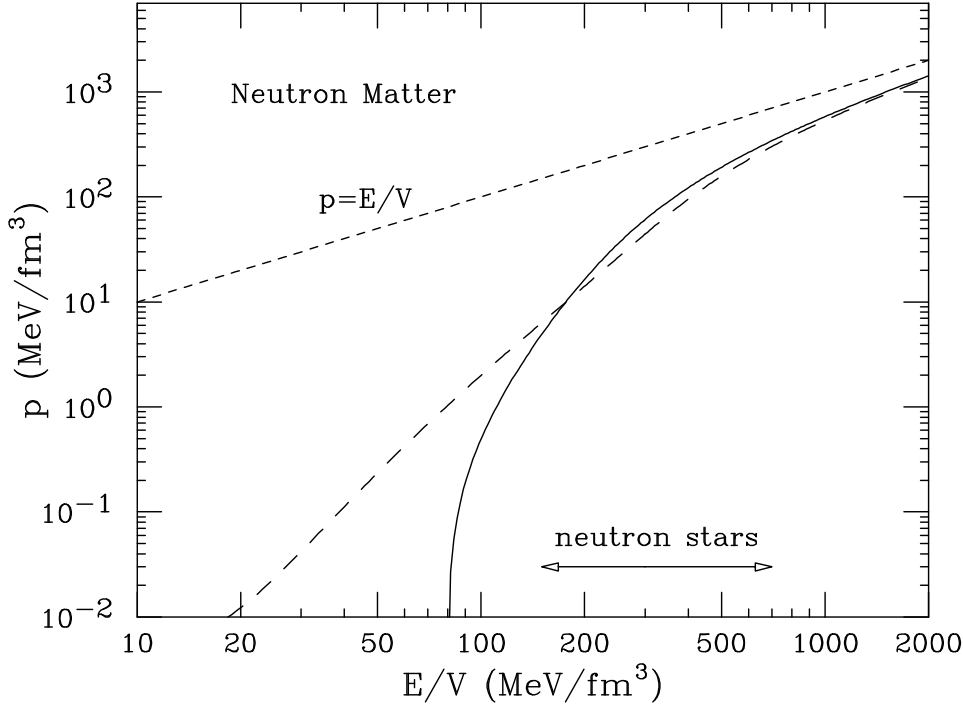}}
  \vspace{13pt}\renewcommand{\baselinestretch}{1.0}
    \caption{\protect\label{fig:neutroneos}%
Predicted equation of state for neutron matter at all densities.
The solid and dashed curves show the result for QHD--I based on
Fig.~\protect\ref{fig:mftsat}.  
The dotted line represents the causal limit $p = \edens$.
The density regime relevant for neutron stars is also shown.}
\end{figure}
 
The corresponding curves for neutron matter obtained by setting
$\gamma = 2$ are also shown in Figs.~\ref{fig:mftsat}\ and
\ref{fig:mftmstar}, and the equation of state (pressure {\em
vs}.  energy density) for neutron matter at all densities is
given in Fig.~\ref{fig:neutroneos}.  
At high densities, the system approaches the
``causal limit'' $p = \edens$ representing the stiffest possible
equation of state.  Thus we have a simple, two-parameter model
that is consistent with the equilibrium point of normal nuclear
matter and that allows for a covariant, causal extrapolation to
any density.
 
Nevertheless, the two-parameter (linear) model is fit only to the 
equilibrium point of nuclear matter; the values of $\Mstar$ and $K$ at 
equilibrium are predictions.
For example, for a given Fermi wavenumber $\kfermi^0$ and energy/nucleon
$e_0$ at equilibrium, $\Mstar$ at equilibrium must satisfy \cite{Fu96}
\beq
\bigl(e_0 + M + \sqrt{(\kfermi^0 )^2 + \Mstarsq}\bigr) \rhoB^0
  - (M - \Mstar )\rhos^0
    - {2\gamma\over (2\pi)^3} 
            \int_0^{\kfermi^0} \dthree{k}\ \Estark = 0 \ .
  \label{eq:JDWmstar}
\eeq
As will be shown in the next subsection, the properties of finite nuclei
place significant constraints on both $\Mstar$ and $K$, and the
linear model predicts too small a value for the former and too large a value
for the latter.
Moreover, the absence of isovector mesons in QHD--I leads to 
a bulk symmetry energy that is too small, and consequently, the repulsive 
forces in neutron matter are underestimated.
This can be corrected by introducing a mean field for the isovector
$\vecrho$ meson, as we describe shortly.
Thus, for comparison, we also show in 
Figs.~\ref{fig:mftsat} through \ref{fig:neutroneos}
the more realistic MFT results obtained in a nonlinear model (NLC), with
parameters fit to the properties of nuclei and given in 
Table~\ref{tab:params}, below.
The important point is that the additional parameters allow for small
adjustments of the nuclear properties near equilibrium, {\em but the
basic features implied by the large scalar and vector fields remain intact\/}
\cite{Wa74,Se86,Bo89,Bo91,Fu96}. 
 
\subsection{Finite Nuclei}
\label{sec:finite}  
We now generalize the results of the preceding subsection to study atomic 
nuclei.  
We continue to work in the mean-field approximation to QHD--I, but since the
system now has finite spatial extent, these fields are spatially
dependent.\footnote{%
As usual in discussions of nuclear
structure, calculations will be carried out in a frame where the
nucleus is at rest.} 
If we initially restrict consideration to spherically symmetric nuclei, 
the meson fields depend only on the radius, and since the baryon current is 
conserved, the spatial part of the vector field $\Vvec$ again
vanishes\cite{Se86}.
Thus the mean-field QHD--I lagrangian of Eq.~\eqref{eq:lagrang}\ becomes 
\beqa
    {\cal L}_{\sssize {\rm MFT}}^{\rm (I)} &=&
              \psibar [ i \gammamul  \dmu - \gv\gammazerol\Vzero -
              (M - \gs\phizero)]\psi
             -\tover12[ (\del \phizero )^2 + \mssq\phizero^2]
                        \nonumber \\[2pt]
    & &\qquad  {}
              -\tover{1}{3!}\kappa\phizero^3
              -\tover{1}{4!}\lambda\phizero^4
              +\tover12[(\del \Vzero )^2 + \mvsq\Vzero^2]   \ ,
    \label{eq:newlagrang}
\eeqa
and the Dirac equation for the baryon field is
\beq
\big\{ i \gammamul \dmu - \gv \gammazerol \Vzero (r) - [M - \gs \phizero (r) ]
         \big\} \psi (x) = 0  \ .      
              \label{eq:baryon}
\eeq
Appropriate values for the scalar and vector couplings ($\gs$ and $\gv$),
masses ($\ms$ and $\mv$), and nonlinear parameters ($\kappa$ and
$\lambda$) will be given below.
 
Although the baryon field is still an operator, the meson fields are
classical; hence Eq.~\eqref{eq:baryon}\
is linear, and we may again seek normal-mode solutions
of the form $\psi (x) = \psi ({\bf x}) \e^{-iEt}$.
This leads to the eigenvalue equation
\beq
         h\psi ({\bf x})
         \equiv \{ {-i} \alphavec\veccdot \del
         + \gv \Vzero (r) + \beta
              [M - \gs \phizero (r)]\}\psi ({\bf x})
              = E \psi ({\bf x}) \ ,      
                   \label{eq:hamil}
\eeq
which defines the single-particle Dirac hamiltonian $h$,
with $\alphavec$ and $\beta$ the usual Dirac matrices.
Equation \eqref{eq:hamil}\ has both positive- and negative-energy solutions
$\uspinor{}$ and $\vspinor{}$, which allow
the field operators to be constructed
in the Schr\"odinger picture.
The positive-energy spinors can be written as
\beq
    \uspinor{\alpha} \equiv {\cal U}_{n k m t} ({\bf x})
         =       \pmatrix{i \big[ G_{n k t}(r)/r \big]
                          \Phi_{k m}        \cr\bspac
                          - \big[ F_{n k t}(r)/r \big]
                          \Phi_{-k m}        \cr}
           \zeta_t \ ,   
         \label{eq:spinordef}
\eeq
where $n$ is the principal quantum number, $\Phi_{k m}$ is a spin-1/2
spherical harmonic \cite{Ed57}, and $\zeta_t$ is a two-component
isospinor labeled by the isospin projection $t$.
(We take $t=\frac{1}{2}$ for protons and $t=-\frac{1}{2}$ for neutrons.)
The phase choice in Eq.~\eqref{eq:spinordef}\
produces real bound-state wave functions
$F$ and $G$ for real potentials $\phizero$ and $\Vzero$,
and the normalization is given by
\beq
    \int_0^{\infty} \intback \d r \big( |G_{\alpha}(r)|^2
                                     +|F_{\alpha}(r)|^2 \big)
         = 1  \ ,       
              \label{eq:norm}
\eeq
which ensures unit probability to find each nucleon somewhere in space.
 
The classical meson field equations follow from Eq.~\eqref{eq:newlagrang}\
and resemble Eqs.~\eqref{eq:kg}\ and \eqref{eq:qed} restricted to static,
spherically symmetric fields.
With the general form for the spinors in Eq.~\eqref{eq:spinordef},
we can evaluate the
nuclear densities, which serve as source terms in the meson field equations.
Assume that the nuclear ground state consists of filled shells up to some
value of $n$ and $k$, which may be different for protons and neutrons;
this is appropriate for doubly magic nuclei.
In addition, assume that all bilinear products of baryon operators are
normal ordered, which removes contributions from the
negative-energy spinors $\vspinor{\alpha}$.
This amounts to neglecting the filled Dirac sea of baryons
and defines the mean-field approximation.
The contributions from the Dirac sea to nuclear matter will be considered in
Section~\ref{sec:eft}.
 
With these assumptions, the local baryon ($\rhoB$) and scalar ($\rhos$)
densities become
\beq
    \pmatrix{\rhoB ({\bf x}) \cr
           \rhos ({\bf x}) \cr}  
       = \sum_{\alpha}^{\rm occ}\,
            {\overline{\cal U}}_{\alpha}({\bf x})
         \pmatrix{\gammazero \cr 1 \cr}
            \uspinor{\alpha}
       = \sum_{a}^{\rm occ} \Big( {2j_a + 1 \over 4 \pi r^2}\Big)
            \big( |\Ga |^2 \pm |\Fa |^2 \big) \ ,  
                   \label{eq:densities}
\eeq
which holds for filled shells, as appropriate for spherically symmetric
nuclei.
The remaining quantum numbers are denoted
by $\{\alpha\} = \{a;m\} \equiv \{n, k, t ; m \}$, and the nonzero
integer $k$ determines $j$ and $\ell$ through $k = (2j + 1)(\ell - j)$.
Notice
that since the shells are filled, the sources are spherically symmetric.
 
The sources produce the meson fields, which satisfy static
Klein--Gordon equations:
\beqa
{\d^2 \over \d r^2}  \phizero (r) + {2 \over r} {\d \over \d r}  \phizero (r)
         - \mssq \phizero (r) - {\kappa\over 2} \phizero^2 (r)
         - {\lambda\over 6} \phizero^3 (r)
          &=& - \gs \rhos (r)   \ ,
                   \label{eq:scalarfield} \\[3pt]
{\d^2 \over \d r^2}  \Vzero (r) + {2 \over r} {\d \over \d r}  \Vzero (r)
         - \mvsq \Vzero (r) &=& - \gv \rhoB (r) \ .
                                        \label{eq:vectorfield}
\eeqa
The equations for the baryon wave functions follow upon
substituting Eq.~\eqref{eq:spinordef}\
into Eq.~\eqref{eq:hamil}, which produces
\beqa
{\d \over \d r}  \Ga + {k \over r}  \Ga -
       \big[ E_a - \gv \Vzero (r) + M - \gs \phizero (r) \big] \Fa &=& 0 \ ,
                      \label{eq:Geqn} \\[4pt]
{\d \over \d r}  \Fa - {k \over r}  \Fa +
       \big[ E_a - \gv \Vzero (r) - M + \gs \phizero (r) \big] \Ga &=& 0 \ .
                      \label{eq:Feqn}
\eeqa
Thus the spherical nuclear
ground state is described by coupled, ordinary
differential equations that may be solved by an iterative procedure, as
discussed in \cite{Ho81,Fu87}.
They contain all information about the static ground-state nucleus in this
approximation.
 
The mean-field hamiltonian can be computed just as for infinite
matter, and after normal ordering, the ground-state energy
is given by
\beqa
E & = & \int\dthree{x}\ \Big\{ \frac{1}{2} \big[ (\del \phizero )^2 + \mssq
              \phizero^2 \big] - \frac{1}{2} \big[ (\del \Vzero )^2 +\mvsq
              \Vzero^2 \big]
          + {\kappa\over 3!} \phizero^3 + {\lambda\over 4!}\phizero^4
           \nonumber \\[3pt]
    &  &\qquad\qquad {}+
              \sum_\alpha^{\rm occ} {\cal U}_{\alpha}^{\dagger}({\bf x})
              \big[ -i\alphavec\veccdot\del + \beta (M-\gs\phizero) 
              + \gv \Vzero \big]\, \uspinor{\alpha}\Big\} \ .   
                   \label{eq:Efunc}
\eeqa
Here the meson fields are functions of the radial coordinate. 
Notice that if we interpret this expression as an {\em energy functional\/}
for the Dirac--Hartree ground state, extremization with respect to the meson
fields reproduces the field equations \eqref{eq:scalarfield}\ and
\eqref{eq:vectorfield}, with the densities from Eq.~\eqref{eq:densities}.
Moreover, extremization with respect to the baryon wave functions
${\cal U}_{\alpha}^{\dagger} ({\bf x})$, subject to the constraint
\beq
    \int\dthree{x}\  {\cal U}_{\alpha}^{\dagger} ({\bf x})
         {\cal U}_{\alpha}^{\vphantom{\dagger}} ({\bf x}) = 1          
    \label{eq:morenorm}
\eeq
for all occupied states (which is enforced by Lagrange multipliers
$E_\alpha$), leads to the Dirac equation \eqref{eq:hamil}.
This alternative derivation of the Dirac--Hartree equations from an energy
functional is useful for extensions of the simple model discussed 
above \cite{Fu96}.
 
Once the solutions to the Dirac--Hartree equations have been found, the
ground-state energy can be computed by using the
Dirac equation \eqref{eq:hamil} to  introduce the eigenvalues $E_a$ and by
partially integrating the meson terms to introduce the densities.
In the end,
the total energy of the system is given by
\beq
E = \sum_a^{\rm occ} E_a (2j_a +1) - {1 \over 2} \int\dthreex
         \big[ {-\gs} \phizero (r) \rhos (r) + \gv \Vzero (r) \rhoB (r) \big]
%
          - {1\over 12} \int\dthreex \big[
           \kappa \phizero^3 (r) + \tover12 \lambda \phizero^4 (r) \big] \ .
                         \label{eq:DHenergy}
\eeq

Before discussing the Dirac--Hartree solutions, let us generalize the
equations to include some additional degrees of freedom and couplings.
Although the isoscalar meson fields are the most important for describing
general properties of nuclear matter, a quantitative comparison with actual
nuclei requires the introduction of some additional dynamics.
 
For example, it is necessary to include the electromagnetic
interaction to account for the Coulomb repulsion between
protons.  Moreover, since hadronic interactions exhibit an
almost exact SU(2) isospin symmetry, the nucleons can couple to
isovector mesons in addition to the isoscalar (neutral) mesons
of QHD--I.  These isovector mesons, for example the
$\vecrho$ and $\vecpi$, come in three charge states
($+$, 0, $-$) and couple differently to the proton and neutron.
Thus they affect the nuclear symmetry energy, which arises when
there are unequal numbers of neutrons and protons.
 
The construction of a renormalizable lagrangian containing
charged, massive vector fields is somewhat complicated and is
discussed at length in Abers and Lee \cite{Ab73}; applications to
the present model can be found in
\cite{Se86}.  For our purposes, we require only the classical
contributions from these fields, and in this case, the
lagrangian simplifies considerably.  In particular, since the
nuclear ground state has well-defined charge, only the neutral
rho meson field (denoted by $b_0$) enters, and since the
ground state is assumed to have well-defined parity and
spherical symmetry, there is no classical pion field.  Thus the
mean-field lagrangian for this extended model, which we call
QHD--II, is given by %
\beqa
    {\cal L}_{\sssize {\rm MFT}}^{\rm (II)} &=&
              \psibar [ i \gammamul  \dmu - \gv\gammazerol\Vzero
              -\grho\tover12\tau_3\gammazerol b_0 -e \tover12 (1+\tau_3)
              \gammazerol A_0
              -(M - \gs\phizero)]\psi
                        \nonumber \\[2pt]
    & &\qquad
             {}-\tover12[ (\del \phizero )^2 + \mssq\phizero^2]
             -\tover{1}{3!}\kappa\phizero^3 - \tover{1}{4!}\lambda\phizero^4
             +\tover12[(\del \Vzero )^2 + \mvsq\Vzero^2]
                        \nonumber \\[2pt]
    & &\qquad
             {}+\tover12 (\del A_0)^2
              +\tover12[(\del b_0 )^2 + \mrhosq b_0^2]   \ .
    \label{eq:QHDII}
\eeqa
Here $A_0$ is the Coulomb potential, $e$ is the proton electromagnetic charge,
$\grho$ is the rho-nucleon coupling constant, and $\tau_i$ are the usual
isospin Pauli matrices.
For now, all of
the boson fields are assumed to be functions of the radial coordinate
only.
 
The Dirac--Hartree equations for this extended model can be
derived just as before.  The Dirac equations for the baryon wave
functions now contain $b_0$ and $A_0$, and because of the
structure of the $\tau_3$ matrix, $b_0$ couples with opposite
sign to protons and neutrons, and $A_0$ couples only to the
protons.  In addition to the source terms in
Eq.~\eqref{eq:densities}, which sum over both proton and neutron
occupied states, the source term for the $\rho$ meson involves
the {\em difference\/} between proton and neutron densities,
while the Coulomb source involves only protons.  These different
types of couplings allow for a more accurate reproduction of
real nuclei, where the proton and neutron wave functions are not
identical.  The full set of equations are presented in
\cite{Se86} and are used to compute the results discussed below.
 
\begin{table}[t]
\caption{Dirac--Hartree Parameter Sets.
Note that $\ms$ and $\kappa$ are in MeV.
\protect\label{tab:params}}
\vspace{.1in}
    \newdimen\digitwidth
    \setbox0=\hbox{\rm 0}
    \digitwidth=\wd0
    \catcode`?=\active
    \def?{\kern\digitwidth}
        \begin{tabular}{lcccccc}
Set &$\gssq$ &$\gvsq$ &$\grhosq$ &$\ms$ &$\kappa$ &$\lambda$ \\[2pt]
\hline
L2     &109.63&190.43 &65.23  &520.??&???0  &???0?    \\
NLB??    &?94.01&158.48 &73.00  &510.0?&?800  &??10?  \\
NLC    &?95.11&148.93 &74.99  &500.8?&5000  &$-$200?   \\
        \end{tabular}
\end{table}
 
\subsubsection{Spherical Nuclei}
The solutions of the preceding equations
depend on the parameters $\gs$, $\gv$, $\ms$, and $\grho$ (when
the $\rho$ meson is included); $\kappa$ and $\lambda$ will be set to
zero in this subsection.
We take the experimental values
$M = 939\MeV$, $\mv = m_{\omega} = 783\MeV$, $\mrho = 770\MeV$,
and $e^2/4 \pi =\alpha = 1/137.036$ (which determines the
Coulomb potential) as fixed.  The free parameters can be chosen
by requiring that when the Dirac--Hartree equations are solved
in the limit of infinite nuclear matter, the empirical
equilibrium density ($\rhoB^0 = 0.1484\ {\rm fm}^{-3}$), energy/nucleon 
($e_0 = -15.75\ {\rm MeV}$), and bulk symmetry energy (35 MeV) are 
reproduced.\footnote{The number of significant digits in the empirical
input values is not intended to indicate how accurately these quantities
are known.
We are merely reporting the precise values used in \protect\cite{Ho81}
to determine the model parameters.}
The empirical equilibrium density is determined here from the
density in the interior of \nucleus{Pb}{208} and corresponds to
$\kfermi^0 = 1.30 \infm$.  We also fit the empirical rms
charge radius of \nucleus{Ca}{40} ($r_{\rm rms} = 3.482$\ fm), which
is determined primarily by $\ms$.  This procedure produces the
parameters in the row labeled L2 in Table~\ref{tab:params},
which are taken from \cite{Ho81}.  This parameter set yields the
same values for $C^2_{\rm s}$ and $C^2_{\rm v}$ as in
Eq.~\eqref{eq:coupls}, so that $\Mstar /M = 0.541$ and $K\approx
545 \MeV$ at equilibrium.

\begin{figure}[t]
 \setlength{\epsfxsize}{4.5in}
 \centerline{\epsffile{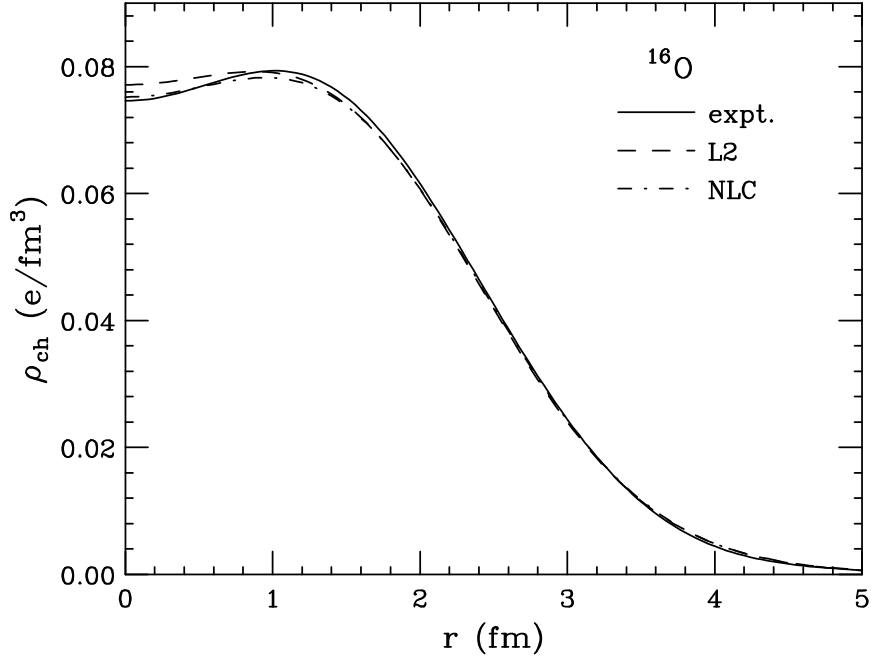}}
  \vspace{13pt}\renewcommand{\baselinestretch}{1.0}
    \caption{\protect\label{fig:16O}%
Charge density distribution for \protect\nucleus{O}{16}.
The experimental curve is from \protect\cite{De87}.
The Dirac--Hartree calculations for parameter set L2 yield the long-dashed 
curve, while those from set NLC yield the dot-dashed curve.}
\end{figure}
 
\begin{figure}[p]
 \setlength{\epsfxsize}{4.5in}
 \centerline{\epsffile{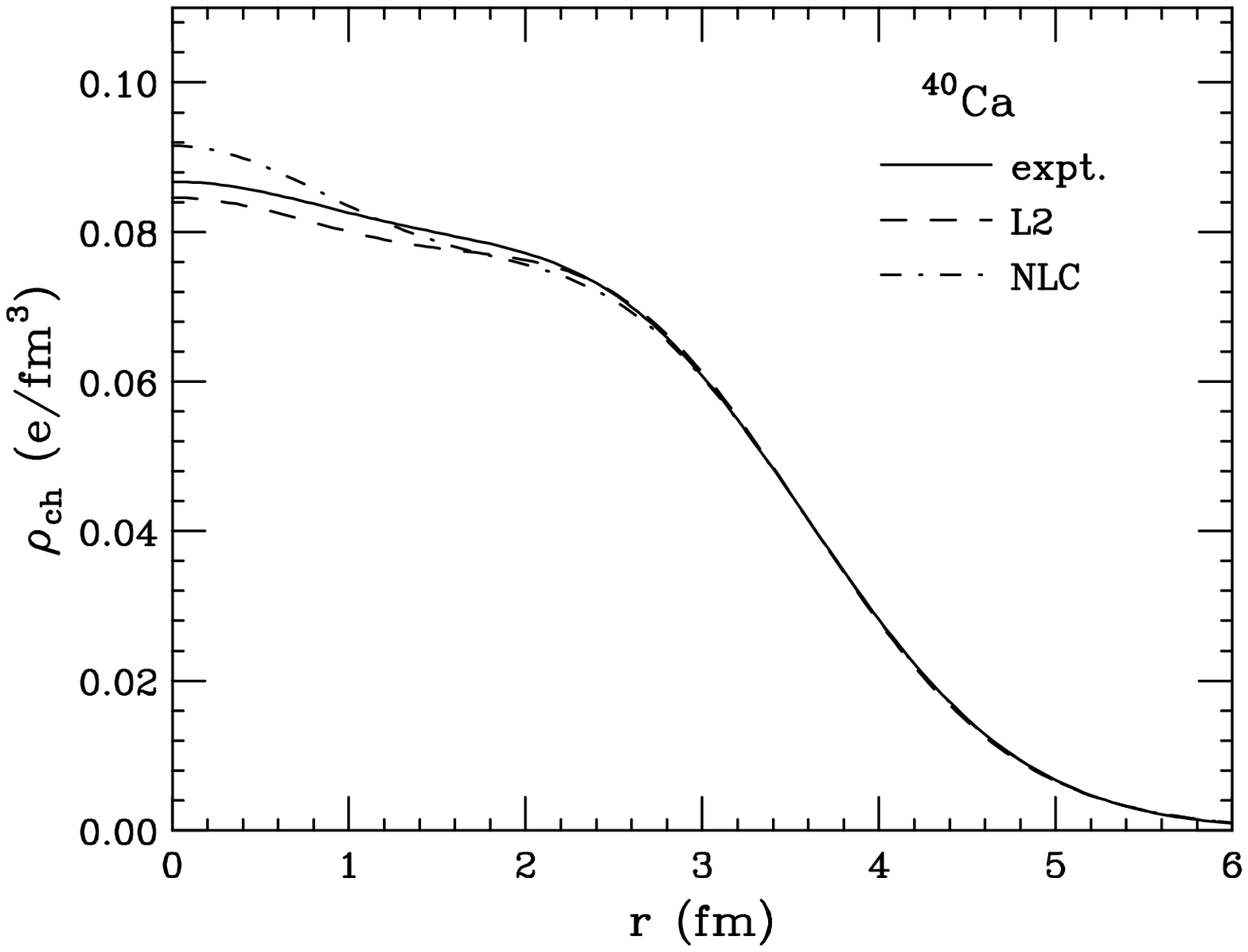}}
  \vspace{13pt}\renewcommand{\baselinestretch}{1.0}
    \caption{\protect\label{fig:40Ca}%
Charge density distribution for \protect\nucleus{Ca}{40}.
The experimental curve is from \protect\cite{De87}.
The Dirac--Hartree calculations for parameter set L2 yield the long-dashed 
curve, while those from set NLC yield the dot-dashed curve.}
%
%
 \setlength{\epsfxsize}{4.5in}
 \centerline{\epsffile{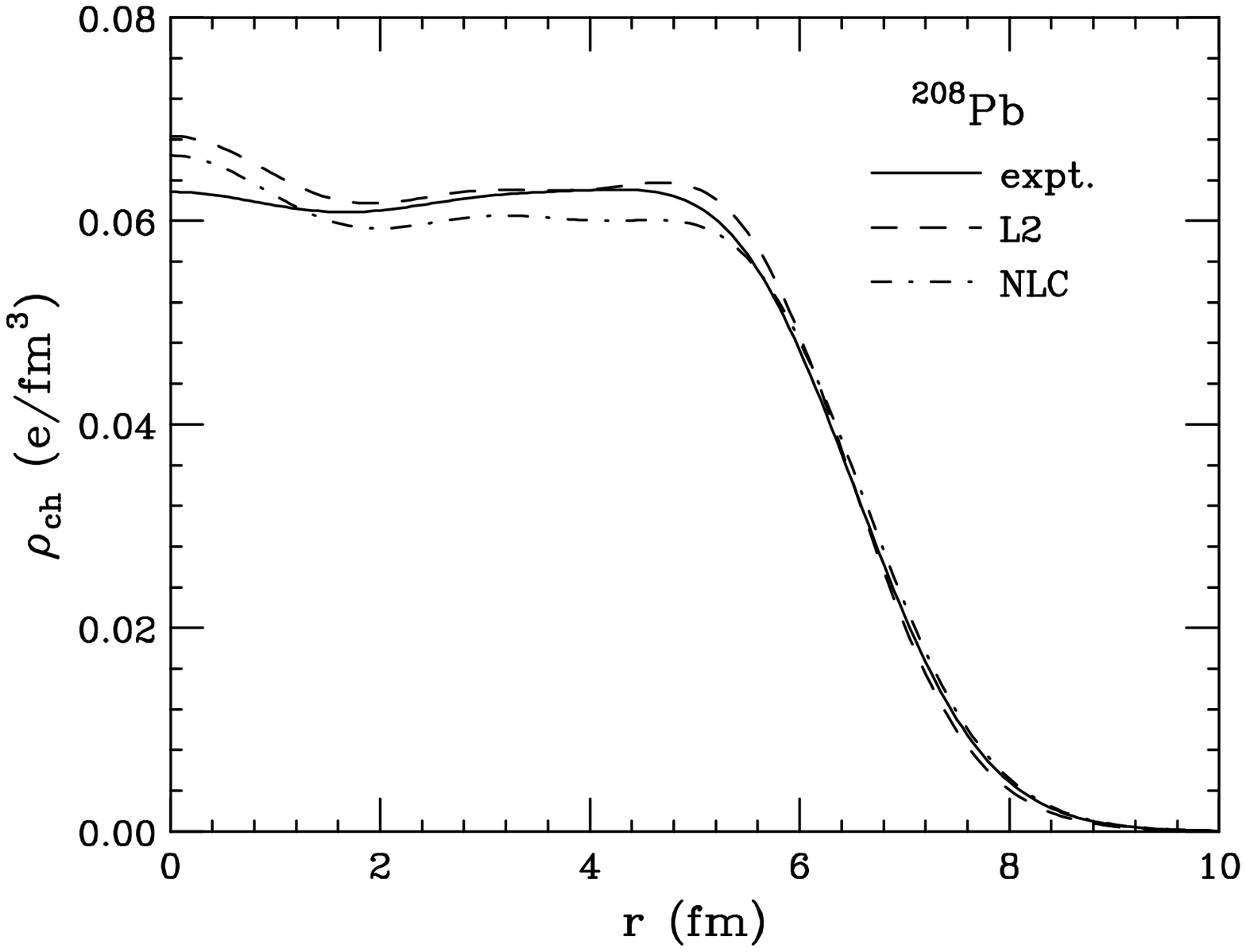}}
  \vspace{13pt}\renewcommand{\baselinestretch}{1.0}
    \caption{\protect\label{fig:208Pb}%
Charge density distribution for \protect\nucleus{Pb}{208}.
The solid curve is from \protect\cite{De87}.
Dirac--Hartree results are indicated by the long-dashed curve (set L2)
and the dot-dashed curve (set NLC).}
\end{figure}
 
Once the parameters have been specified, the properties of all closed-shell
nuclei are determined in this approximation.
For example, Figs.~\ref{fig:16O}\ through \ref{fig:208Pb}\ show the
Dirac--Hartree charge densities of \nucleus{O}{16}, \nucleus{Ca}{40},
and \nucleus{Pb}{208} compared with
the empirical distributions determined from electron scattering \cite{De87}.
The empirical proton charge form factor has been folded with the
calculated ``point proton'' density to determine the charge density.
 
\begin{figure}[t]
 \setlength{\epsfxsize}{5in}
 \centerline{\epsffile{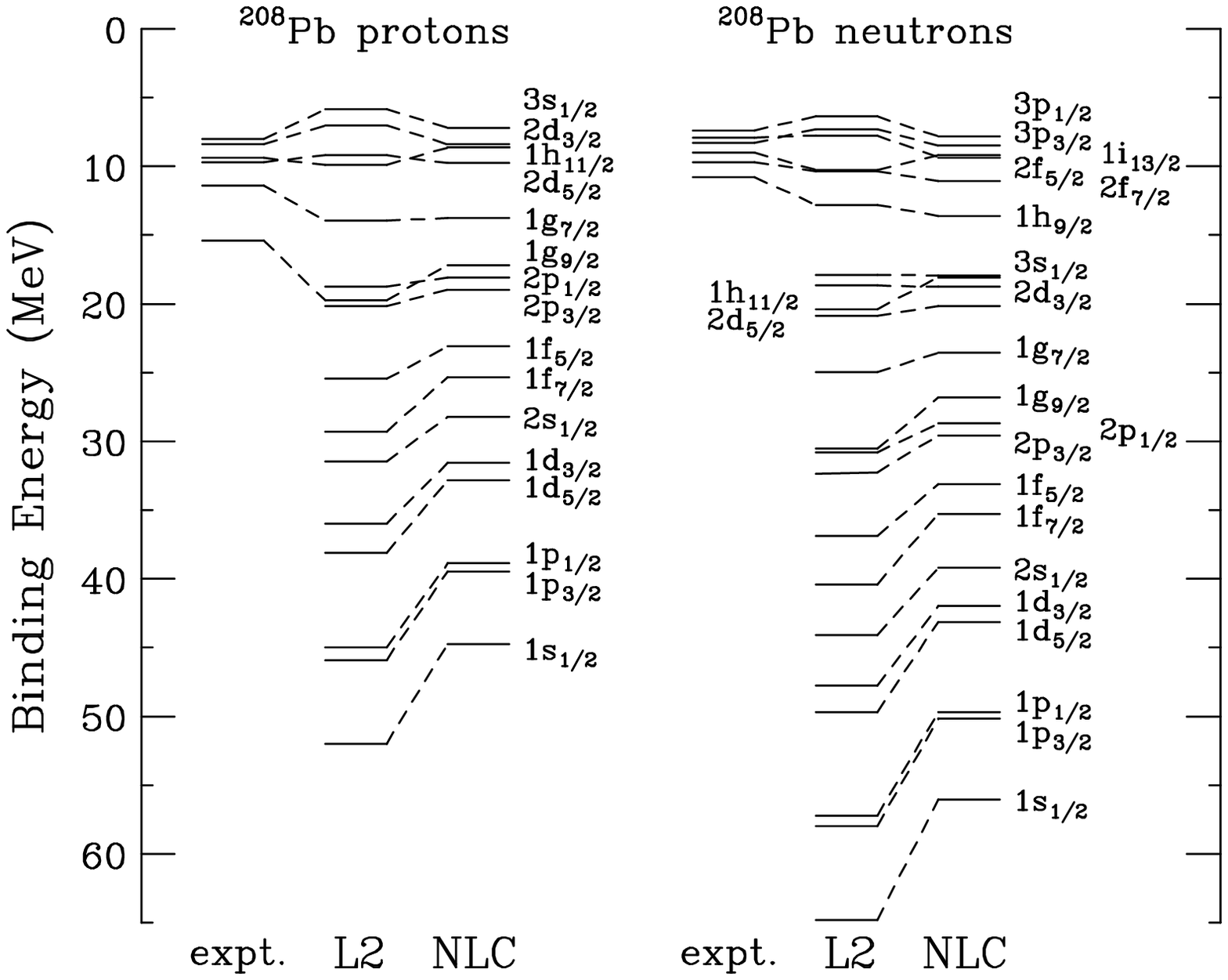}}
  \vspace{13pt}\renewcommand{\baselinestretch}{1.0}
    \caption{\protect\label{fig:shells}%
Predicted spectrum for occupied single-particle levels
in \protect\nucleus{Pb}{208}.
Experimental values are taken from neighboring nuclei.}
\end{figure}
 
In Fig.~\ref{fig:shells}, the predicted energy levels in
\nucleus{Pb}{208} are compared with experimental values derived from
neighboring nuclei \cite{Bo69,Ra79}. The relativistic
calculations clearly reveal a shell structure; the level orderings and major
shell closures of the nuclear shell model are correctly reproduced.
This successful result arises from
the spin-orbit interaction that occurs naturally when a Dirac
particle moves in large, spatially varying classical scalar and vector
fields \cite{Fu36,Kr73,Kr74,Mi74,Mi75}. Note that whereas $\gs\phizero$ and
$\gv\Vzero$ tend to {\em cancel\/} in the central potential that
determines saturation, they {\em add\/} constructively in the spin-orbit
potential.  We emphasize that no parameters are adjusted
specifically to produce the spin-orbit interaction, as is
usually the case in nonrelativistic calculations.  Thus, with a
minimal number of phenomenological parameters determined from
bulk nuclear properties, {\em one derives the level structure of the
nuclear shell model}.

We emphasize that in this relativistic model of nuclear structure,
the calculation of the ground state is {\em self-consistent}.
The scalar and vector fields follow directly from the scalar and baryon
densities, which are in turn determined by the solutions to the Dirac
equation~\eqref{eq:hamil}\ in the classical fields.
Moreover, this relativistic shell model arises from a simple approximation
to the underlying QHD lagrangian.
Thus one has a consistent many-body framework to systematically investigate 
corrections.
 
\subsubsection{Deformed Nuclei} 
To study the systematics of this
relativistic model of nuclear structure, we extend the preceding
equations to deal with deformed, axially symmetric nuclei.  This
allows us to calculate not only the ground states of nuclei with
fully closed shells, but also those for even-even nuclei between
closed shells.  We will concentrate here on nuclei with $12 \leq
B \leq 40$, which includes the $1p$ and $2s$--$1d$
shells \cite{Fe71}. The restriction to azimuthal and
reflection-symmetric deformations is reasonable for light,
even-even nuclei.
 
These assumed symmetries of the ground state,
together with the assumption of well-defined parity, imply that the
nonvanishing meson fields are the same ones that appear in
spherical nuclei \cite{Pr87}.
Thus the Dirac--Hartree equations are essentially the same as those written
earlier, except that all fields now depend on both a radial and
an angular coordinate [for example, $\phizero (r,\theta )$], and
all the differential equations become partial differential
equations.  The source densities are still computed as in
Eq.~\eqref{eq:densities}, but they now depend on $r$ and
$\theta$.  There are several methods for solving the resulting
set of coupled partial differential equations, and the
interested reader is directed to the literature for a
discussion \cite{Le86,Pa87,Fu87,Zh91a}. 
The equilibrium deformation is obtained by choosing the occupied 
single-particle states to minimize the energy.
 
The results of Furnstahl, Price, and Walker \cite{Fu87}
for quadrupole moments in the $2s$--$1d$ shell show how this
observable can constrain the properties of nuclear matter near
equilibrium density.
These authors use the full complement of parameters in the present
model and constrain them to produce nuclear saturation at 
$\kfermi^0 = 1.30 \infm$  with a binding energy of 
15.75 MeV and a bulk 
symmetry energy of 35 MeV, as well as the correct rms radius for
\nucleus{Ca}{40}, just as for the calculations of spherical nuclei in the
preceding subsection. 
An infinite number of parameter sets will satisfy these constraints, and
two examples (NLB and NLC) are shown in Table~\ref{tab:params}. 
Both of these sets produce roughly equal
values for $\Mstar$ at nuclear matter equilibrium ($\Mstar /M = 0.61$ for
NLB and $\Mstar / M = 0.63$ for NLC), so they generate similar
spin-orbit splittings and deformations.
As seen in Figs.~\ref{fig:16O}--\ref{fig:shells}, NLC accurately 
reproduces the nuclear charge densities and splittings for spherical
nuclei (results for NLB are similar).
Moreover, the agreement with experimental quadrupole moments is 
excellent, particularly the systematic trends and the oscillation between 
oblate and prolate shapes around $B=32$ (see Fig.~7 in \cite{Fu87}).  
Thus the successful description of spherical nuclei in this relativistic 
model can be extended to reproduce the observed systematics of light 
deformed nuclei {\em with the same parameters}.

These authors also observe, however, that if one uses the set L2 to
compute the quadrupole moments, the smaller value of $\Mstar / M = 0.54$
at equilibrium leads to a prediction of {\em spherical\/} shapes for
\nucleus{C}{12}, \nucleus{Si}{28}, and \nucleus{S}{32}, which are
inconsistent with B(E2; $0^+ \rightarrow 2^+$) values derived from
experiment \cite{Le75}.
Thus, the sensitive dependence of the deformation on the level density
near the Fermi surface (which is determined essentially by the inverse of
$\Mstar$) allows one to conclude that the small adjustments in
equilibrium  properties afforded by the nonlinear scalar couplings 
$\kappa$ and $\lambda$ are significant.
A similar analysis of the $2s$--$1d$ shell binding energies 
(see Fig.~10 in \cite{Fu87}) shows that
compression moduli greater than roughly 350 MeV are too large, which
is true of sets L2 ($K \approx 545\MeV$) and NLB ($K \approx 420\MeV$).
Note that the favored parameter set NLC (with $K \approx 225\MeV$) has a 
negative value for the quartic coupling $\lambda$, which may lead to 
problems for large values of the scalar field, since the energy is no 
longer bounded from below.
As we will see later, however, the mean-field energy is actually valid
only as an expansion in powers of the fields, and thus global questions
about stability are not particularly relevant.



\subsection{Nucleon--Nucleus Scattering}
\label{sec:RIA}
The scattering of
medium-energy nucleons from nuclei can provide information about
both nuclear structure and the NN interaction.  Since the NN
interaction has complex spin, isospin, momentum, and density
dependence, nucleon--nucleus scattering exhibits a wide variety
of phenomena.  As a starting point for describing these
phenomena, we use the Dirac--Hartree description of the nucleus,
together with the relativistic impulse approximation (RIA),
which assumes that the interaction between the projectile and
target nucleons has the same form as the interaction
between two nucleons in free space.  This interaction is used to
produce a nucleon--nucleus optical potential that incorporates
the leading term in a multiple-scattering series.
 
Although the simple QHD models discussed above are useful for
studying the average properties of the nuclear interaction, they
are less useful for describing the detailed quantitative
features (such as spin dependence) of the full NN scattering
amplitude.  These quantitative features are important for any
reasonable description of the nucleon--nucleus scattering
observables.  The RIA allows us to combine the empirical
free-space scattering amplitude with a relativistic calculation
of the nuclear ground state.
 
The RIA as originally formulated \cite{Mc83,Sh83,Cl83} involves
two basic procedures.  First, the experimental NN scattering
amplitude is represented by a particular set of five 
Lorentz-covariant functions \cite{Mc83a} that multiply the so-called
``Fermi invariant'' Dirac matrices.  The Lorentz covariant
functions are then folded with the Dirac--Hartree target
densities to produce a first-order optical potential for use in
the Dirac equation for the projectile \cite{Ho91}.
Here we briefly summarize the formalism and the results; a more complete
discussion is given in \cite{Se86}.
 
The constraints of Lorentz covariance, parity conservation,
isospin invariance, and that free nucleons are on their mass
shell imply that the invariant NN scattering operator
$\scattamp$ can be written in terms of five complex functions
for $pp$ scattering and five for $pn$ scattering.  In the
original RIA, $\scattamp$ was taken as %
\beq
\scattamp = \Famp{S}
          + \Famp{V} \gammamu_{\sssize (0)}
                \gamma^{\phantom\mu}_{\sssize (1)}{}_{\mu}
          + \Famp{P} \gammafive_{\sssize (0)}\gammafive_{\sssize (1)}
          + \Famp{T} \sigma^{\mu \nu}_{\sssize (0)}
                     \sigma^{\phantom\mu}_{\sssize (1)}{}_{\mu \nu}
          + \Famp{A}\gammafive_{\sssize (0)} \gammamu_{\sssize (0)}
                    \gammafive_{\sssize (1)}
                    \gamma^{\phantom\mu}_{\sssize (1)}{}_{\mu}\ ,    
                        \label{eq:mrwamp}
\eeq
where the subscripts (0) and (1) refer to the incident and
struck nucleons, respectively.  Each amplitude $\Famp{L}$ is a
complex function of the Lorentz invariants $t$ (four-momentum
transfer squared) and $s$ (total four-momentum squared), or
equivalently, of the momentum transfer $q$ and incident energy
$E$.  It is found empirically that the amplitudes $\Famp{S}$,
$\Famp{V}$, and $\Famp{P}$ are much larger than any amplitudes
obtained in a nonrelativistic decomposition, which uses
Galilean-invariant operators.
 
The RIA optical potential $\UoptqE$ is defined as
\beq
\UoptqE = -{4 \pi i p \over M} 
         \langle \Psi | \sum_{n=1}^A
         \e^{i {\bf q}\cdot {\bf x}(n)}
         \scattamp (q,E; n) | \Psi \rangle \ ,
            \label{eq:oldopt}
\eeq
where $\scattamp$ is the scattering operator of Eq.~\eqref{eq:mrwamp}, 
$p$ is the magnitude of the projectile three-momentum in the 
nucleon--{\it nucleus\/} c.m. frame (where the scattering observables are
calculated), $| \Psi \rangle$ is the $A$-particle nuclear ground state, 
and the sum runs over all nucleons in the target.  
$\scattamp$ is a function of the momentum transfer $q$ and collision 
energy $E$, which we take to be the proton--nucleus c.m. energy; this
amounts to neglecting nuclear recoil.
 
With these simplifications, the Dirac optical potential is
local, and only diagonal nuclear densities are needed.  For a
spin-zero nucleus, the only nonzero densities are the baryon and
scalar densities of Eq.~\eqref{eq:densities}, plus a tensor term
computed by inserting $\sigma^{0i}$ between the spinors in
Eq.~\eqref{eq:densities}.  Thus the optical potential takes the
form %
\beq
U_{\rm opt} = U^S + \gammazero U^V - 2 i \vecalpha\veccdot {\hat r}
                   U^T  \ ,    \label{eq:optnew}
\eeq
where $U^L \equiv U^L (r;E)$ for each component.
The tensor contribution $U^T$ is small and is neglected in
what follows.  The RIA optical potential then has only scalar
and vector contributions, and the Dirac equation for the
projectile has precisely the same form as in
Eq.~\eqref{eq:hamil}, with $U^V$ replacing $\gv \Vzero$ and
$U^S$ replacing $(-\gs \phizero)$: %
\beq
h \, \uspinor{0} = \Big\{ {-i \vecalpha} \veccdot \vecnabla +
       U^V (r;E) + \beta \big[ M + U^S (r;E) \big] \Big\} \, \uspinor{0}
                   = E  \,\uspinor{0} \ .       \label{eq:scattD}
\eeq
In practice, one includes in $U^V$ the Coulomb potential computed from the
empirical nuclear charge density; other electromagnetic contributions arising
from the proton anomalous magnetic moment are of similar size to the tensor
term $U^T$.
 
Since representative RIA results have been shown many times in the
literature (see, for example, Figs.~25--28 in \cite{Se86}), we will
not reproduce the figures here.\footnote{%
Computer codes for performing RIA calculations are also readily
available \protect\cite{Ho91}.}
The target densities in these calculations are taken from the results for 
spherical nuclei discussed above, {\em with no further adjustment of
parameters}, and the RIA calculations agree remarkably well with the data.  
Moreover, when compared with nonrelativistic impulse-approximation
calculations, the relativistic results are superior,
particularly for the spin observables.
Although the nonrelativistic results improve when higher-order corrections
are included \cite{Co90,Ra90}, 
one concludes that the important spin- and density-dependent
effects are already contained in the  relativistic impulse-approximation
framework \cite{Pi84,Hy85,Lu87}.
 
The spin dynamics is inherent in the relativistic formalism and
arises naturally from the large Lorentz scalar and vector
potentials in the Dirac equation \eqref{eq:scattD}.  This is
precisely the same spin dynamics that produces the observed
spin-orbit splittings in the bound single-particle levels.  {\em
Thus the relativistic Hartree calculations provide a minimal
unifying theoretical basis for both the nuclear shell model and
medium-energy proton--nucleus scattering---two essential aspects
of nuclear physics}.
 
\subsection{Nuclear Excited States}
\label{sec:RPA}
  
We now turn to the calculation of
nuclear response functions and the properties of nuclear excited
states, as described in the random-phase approximation (RPA)
built on the MFT ground state.  These excitations arise from the
consistent linear response of the ground state, in which the
nucleons move coherently in varying classical meson fields that
are in turn determined by oscillatory nuclear sources.  Many
studies of the relativistic nuclear response have been carried
out for both infinite matter and finite nuclei, beginning with
the pioneering work of Chin \cite{Ch77}, and here we will focus
on two basic issues.  
The first is the importance of consistency,
which implies that the particle-hole interaction in the excited
states must be the same (and use the same parameter values) as
the interaction in the ground state.
This ensures that the occupied and unoccupied single-particle states
are indeed orthogonal.
Second, the relativistic response involves not only the familiar
positive-energy particle-hole configurations, but also
configurations that mix positive- and negative-energy states.
These new configurations are crucial for the conservation of the
electromagnetic current and the separation of the ``spurious''
$J^{\pi} = 1^-$ state.  This emphasizes that the Dirac
single-particle basis is complete only when both positive- and
negative-energy states are included.
 
The calculation of the linear response is basically the same as
in nonrelativistic many-body theory \cite{Fe71}. The principal
idea is to compute the particle-hole (polarization) propagator
and to extract the collective excitation energies and transition
amplitudes from the poles and residues of this propagator.
Several methods have been developed and applied to finite
nuclei \cite{Fu85,Fu85a,Ni86,We87,We88,Fu88a,Bl88,Sh89,Da90,Ho90,Pi90,Pr92a}.
Initial calculations were restricted primarily to isoscalar excitations, 
because fitting bulk nuclear properties constrains only the isoscalar
particle-hole interaction significantly, and because the
isovector response depends critically on pion dynamics, with the
associated complexity that we discuss later.  

An excellent discussion of the role of consistency is contained in the work
of Dawson and Furnstahl \cite{Da90}, where results for the low-lying,
negative-parity, isoscalar states in \nucleus{C}{12}, \nucleus{O}{16},
and \nucleus{Ca}{40} are
compared to several empirical levels that might be reasonably
described as particle-hole excitations. 
The full RPA eigenvalues agree favorably with the empirical values,
which is a nontrivial result because of the large cancellations
between scalar and vector contributions.  
Moreover, it is found that the negative-energy states play an important 
role in determining the RPA spectrum, particularly for the spurious
$1^-$ state.  
Lorentz covariance implies that in a consistent RPA calculation, this 
state should appear at zero excitation energy, which occurs only when the 
full Dirac basis is maintained.

Moreover, RPA calculations by Furnstahl \cite{Fu89b} of
the ratio of the transition charge density to the longitudinal
current for a particular ($3^-,0$) excitation in \nucleus{O}{16} show that 
if only the positive-energy (particle-hole) configurations are retained, 
the electromagnetic current is not conserved.  
In contrast, with the full RPA calculation, including the contributions 
from negative-energy states, current conservation is restored.
The conclusion is that it is {\em essential\/} to include all states in 
the Dirac basis to maintain the conservation of the current.
 
\subsection{Nuclear Matter at Finite Temperature}
The preceding discussion of the nuclear matter equation of state was 
restricted to zero temperature.
The extension to finite temperature is straightforward in the
MFT, since the hamiltonian is diagonal and the mean-field
thermodynamic potential $\Omega$ can be calculated
exactly.\footnote{%
We neglect the zero-point corrections from
the Dirac sea in this section (see \cite{Fr78}\ and
\cite{Fu91}), as well as thermal contributions from the massive
isoscalar mesons.} 
The results for the scalar density, baryon density, energy density, and 
pressure are given by (here $\kappa = \lambda = \grho = 0$) \cite{Se86} 
\beqa
        \rhos &=& \degen \int\dthree{k}\,   {\Mstar \over \Estar(k)}\,
                 (n_k + {\overline n}_k) \ ,
                    \\[4pt]
        \rhoB &=& \degen \int\dthree{k}\,   (n_k -
                   {\overline n}_k)  \ ,  \\[4pt]
        \edens &=& {\gvsq \over 2\mvsq}\rhoB^2 + {\mssq \over 2\gssq}
                    (M - \Mstar)^2 + \degen\int\dthree{k}\,  \Estar(k)
                    (n_k + {\overline n}_k) \ ,
                       \label{eq:edensT} \\[4pt]
        p &=& {\gvsq \over 2\mvsq}\rhoB^2 -
              {\mssq \over 2\gssq} (M - \Mstar)^2
         + {1 \over 3}\degen\int\dthree{k}\,  
         {{\bf k}^2 \over \Estar(k)}\, (n_k
         +{\overline n}_k)  \ ,  \label{eq:pT}
\eeqa
where the baryon and antibaryon distribution functions are
\beq
              \pmatrix{n_k(T,\nu)\cr
              {\overline n}_k(T,\nu)
              \cr} \equiv
              \Big\{1+\e^{[\Estar(k)\mp \nu  ]/T}
              {\Big\}}^{-1} \ ,  \label{eq:dist}
\eeq
and the reduced chemical potential is $\nu \equiv \mu - \gv\Vzero$.
(We set Boltzmann's constant $k_{\scriptscriptstyle{\rm B}} =1$.)
The appropriate value of $\Mstar$ is determined by minimizing the 
thermodynamic potential with respect to that parameter:
\beq
   \left( {\partial\Omega \over \partial\Mstar}\right)_{\mu , V, T}
      = 0 \ .
\eeq

The nuclear matter equation of state at all densities and temperatures for
this hadronic MFT model (QHD--I) is shown in \cite{Se92,Wa95}.
The model has also been combined with a simple description of quark-gluon
matter to describe the hadron/quark phase transition as a function of
temperature and density \cite{Se86}.
A more extensive examination of the equation of state, as well as a 
discussion of the covariance of the finite-temperature results, is contained
in \cite{Fu90}.



\section{Pion Dynamics and Chiral Symmetry}
\label{sec:chiral}

The relativistic neutral scalar and vector fields are the most important 
for determining the bulk properties of nuclear systems.  
Nevertheless, the lightest and most accessible meson is the pion, whose 
interactions with nucleons and nuclei have been extensively studied at 
the meson factories.  
It is therefore impossible to formulate a complete and quantitative hadronic 
theory without including pionic degrees of freedom.  
Here we briefly review some important aspects of pion dynamics in QHD.
More complete discussions, together with references to the original 
literature, can be found in \cite{Se86,Se92a,Wa95}.

In the limit of massless $u$ and $d$ quarks, QCD possesses a global, 
chiral ${\rm SU}(2)_L \times {\rm SU}(2)_R$ symmetry.
Chiral transformations may be written in terms of a set of vector
($V$) and axial-vector ($A$) generators, which produce corresponding
isospin rotations.
The symmetry is spontaneously broken, leading to the existence of
pseudoscalar Goldstone bosons (pions).
The vector (isospin) symmetry, which forms an ${\rm SU}(2)_V$ subgroup
of the original chiral group, remains unbroken.
Our effective hadronic theories should respect these underlying symmetries,
which have important consequences for the way mesons interact
with themselves and with each other; a thorough discussion is contained
in \cite{Wa95}.
In nature, electromagnetic interactions and finite quark masses imply that
these symmetries are only approximate, but the symmetry-violating terms
can be added as small perturbations.

\subsection{The Linear Sigma Model}
\label{sec:sigma}

The simplest model illustrating these ideas is the linear sigma model
\cite{Sc57,Ge60,Le72}, which contains a pseudoscalar ($\gammafivel$)
coupling between pions and nucleons, plus an auxiliary scalar field
(denoted here by $s$) to implement the symmetry.
A small symmetry-violating (SV) term is included to generate a finite pion
mass.
We will also include a massive, neutral, isoscalar vector field to supply
a repulsive interaction, as in QHD--I.

By demanding that the theory be Lorentz covariant, parity invariant, isospin
and chiral invariant, and renormalizable, one is led to the form
\beqa
    \lagrang_{\sigma\omega} & = &
              \lagrang_{\rm chiral} + \lagrang_{\rm SV} \ ,
              \label{eq:lsw}\\[3pt]
    \lagrang_{\rm chiral} & = & \psibar \big[ \gammamul (i\dmu - \gv\Vmu )
              - \gpi (s + i\gammafivel \vectau\veccdot\vecpi ) \big] \psi
              +\tover12 (\dmul s \dmu s + \dmul \vecpi\veccdot \dmu\vecpi)
               \nonumber\\[3pt]
    & &\qquad
              -\tover14 \lambdat (s^2 + \vecpi{}^2 - v^2)^2
              -\tover14 F_{\mu\nu}F^{\mu\nu} +\tover12 \mvsq\Vmul\Vmu
              +\delta\lagrang\ ,
        \label{eq:lchiral} \\[3pt]
    \lagrang_{\rm SV} & = & \epsilon s \ .  
              \label{eq:lsb}
\eeqa
Here $\psi$, $\vecpi$, and $s$ are the nucleon, isovector pion, and neutral
scalar meson fields, respectively, and $\gpi$ is the pion--nucleon coupling
constant.  
The parameters $\lambdat$ and $v$ describe the strength of the meson
self-interactions, and $\epsilon$ is a small chiral-symmetry-violating
parameter related to the pion mass; the exact chiral limit is obtained by
setting $\epsilon = 0$.  
The form of the meson self-interactions allow for spontaneous symmetry
breaking, which is used to give the nucleon a finite mass.

The lagrangian \eqref{eq:lchiral} is invariant under global vector and 
axial-vector isospin
transformations, which imply (by Noether's theorem) the conserved isovector
currents\footnote{The (isoscalar) baryon current 
$B^\mu = \psibar \gammamu \psi$ is also conserved.}
\beqa
    {\bf T}^\mu &=& \tover12 \psibar \gammamu \vectau \psi
                   + \vecpi \veccross \dmu \vecpi \ , 
           \label{eq:conscurrV}\\
    {\bf A}^\mu &=& \tover12 \psibar \gammafivel\gammamu\vectau \psi
              + \vecpi\dmu s - s \dmu\vecpi 
    \label{eq:conscurrA}
\eeqa
in the chiral limit $\epsilon = 0$.  
When $\epsilon\not= 0$, we obtain instead the PCAC relation
\beq
    \dmul {\bf A}^\mu = \epsilon\vecpi\ , \label{eq:PCAC}
\eeq
which follows from the field equations.
Note that the chiral symmetry is realized {\em linearly}, which means that
under a general chiral transformation, neutrons mix with protons, and
the scalar mixes with the pions.
(The vector field is a chiral scalar.)
Moreover, the linear symmetry requires that the scalar and pion couple to
the nucleon with equal strength $\gpi \approx 13.4$.
In applications, one often relaxes this condition (with no
justification) and sets $\gs = \gpi /g_A$, where the axial coupling
$g_A \approx 1.26$; this {\em ad hoc\/} procedure allows the 
Goldberger--Treiman relation to be satisfied at the tree
level \cite{We67,Le72}.

The baryon mass is generated by spontaneous symmetry 
breaking \cite{Le72,Se86}, which implies that the scalar field $s$ has a 
nonzero vacuum expectation value $s_0$, and
that the pion is a massless Goldstone boson (in the limit $\epsilon = 0$).  
In terms of the shifted scalar field 
\beq
\sigma \equiv s_0 - s
  \label{eq:sshift}
\eeq
and the physical masses defined by
\beq
    M = \gpi s_0 \ , \qquad
    \epsilon = \frac{M}{\gpi}\, \mpisq \ , \qquad
    \lambdat = \frac{\msigmasq - \mpisq}{2M^2}\, \gpisq \ ,
         \label{eq:mdefs}
\eeq
the lagrangian $\lagrang_{\sigma\omega}$ reads
\beqa
    \lagrang_{\sigma\omega} & = & \psibar \big[ i\gammamul\dmu - \gv\gammamul
                   \Vmu - (M - \gpi\sigma ) -i\gpi \gammafivel
                 \vectau\veccdot\vecpi \big] \psi
         +\tover12 (\dmul\sigma\dmu\sigma - \msigmasq\sigma^2 )
         \nonumber\\[4pt]
    & & \qquad
         -\tover14 F_{\mu\nu}F^{\mu\nu} + \tover12 \mvsq\Vmul\Vmu
         +\tover12 (\dmul\vecpi\veccdot\dmu\vecpi - \mpisq\vecpi{}^2)
              \nonumber\\[4pt]
    & & \qquad
         + \gpi\,\frac{\msigmasq - \mpisq}{2M}\, \sigma (\sigma^2
                             +\vecpi{}^2)
         -\gpisq\,\frac{\msigmasq - \mpisq}{8M^2}\, (\sigma^2
                             +\vecpi{}^2)^2 \ ,
    \label{eq:lswafter}
\eeqa
where the counterterms $\delta\lagrang$ will here and henceforth be
suppressed.  
Note that the explicit chiral symmetry violation is now contained entirely in
the parameter $\mpisq$, the square of the pion mass.  
Moreover, if we consider the (renormalized) parameters
$M$, $\mpi$, $\mv$, and $\gpi$ as experimentally known, then
apart from the vector meson coupling $\gv$, there is only one free
parameter in this model, namely, the chiral scalar mass $\msigma$.  
Both the signs and magnitudes of the nonlinear meson interactions 
proportional to $\sigma^3$ and $\sigma^4$ are determined 
by the chiral symmetry and its spontaneous breaking in the vacuum.  

A vector-isovector $\vecrho$ meson can be added in a renormalizable
manner to produce QHD--II, as discussed in \cite{Se86}.\footnote{%
Including the isovector $\vecrho$ meson destroys the chiral symmetry.
A fully chiral, renormalizable model (QHD--III)
containing both the $\vecrho$ and
${\bf a}_1$ mesons is derived in \protect\cite{Se92b}.
See also \protect\cite{Ko93a,Fo95a}.}
At the mean-field level of interest here, the rho meson enters as in
Eq.~\eqref{eq:QHDII}.
Given the similarity between the lagrangians in Eqs.~\eqref{eq:lswafter}
and \eqref{eq:qhdone}, it is only natural (but incorrect!)
to identify the $\sigma$ field with the scalar field $\phi$ studied earlier.
Since the pion field vanishes at the mean-field level (we assume our
systems possess the same symmetries as those in Section~\ref{sec:qhd}),
the dynamical equations are exactly the same as before,
except that the nonlinear scalar interactions are now determined by the
symmetry:
\beq
    \frac{\kappa_{\sigma\omega}}{M} = -\frac{3\gpi}{M^2}(\msigmasq - \mpisq )
     \ , \qquad
    \lambda_{\sigma\omega} = \frac{3\gpisq}{M^2} (\msigmasq - \mpisq )
         \ . \label{eq:swNLs}
\eeq

Unfortunately, the identification of $\sigma$ with $\phi$ leads to serious
difficulties.
First, even though there are still two free parameters in symmetric nuclear
matter, it is impossible to reproduce the empirical equilibrium point, 
as documented long ago by Kerman and Miller \cite{Ke74}.
(Their result is reproduced in Fig.~5 of \cite{Fu96}.)
The basic problem is that the nonlinear scalar interactions are large and
have the wrong signs.
[Compare $\kappa_{\sigma\omega} < 0$ and $\lambda_{\sigma\omega} >0$ in
Eq.~\eqref{eq:swNLs} with the favored NLC values in Table \ref{tab:params}.]
This problem was solved in \cite{Bo83,Sa85}, where interactions between
the scalar and vector fields were introduced by making the replacement
\beq
  \tover12 \mvsq\Vmul\Vmu \longrightarrow 
     \tover12 \eta^2 \gvsq \Vmul\Vmu (s^2 + \vecpi^2)
\eeq
in Eq.~\eqref{eq:lchiral}.
Now the vector field also acquires its mass through spontaneous symmetry
breaking, when the scalar field is shifted as in Eq.~\eqref{eq:sshift}.
This results in the replacement
\beq
  \tover12 \mvsq\Vmul\Vmu \longrightarrow {1\over 2}\,\mvsq\Vmul\Vmu
     - {\gpi \mvsq \over M}\, \Vmul\Vmu\sigma
     + {1\over 2}\, {\gpisq\mvsq \over M^2}\, 
        \Vmul\Vmu (\sigma^2 + \vecpi^2)
          \label{eq:bogfix}
\eeq
in Eq.~\eqref{eq:lswafter}.
The signs of the scalar--vector cubic and quartic interactions are 
{\em opposite\/} to those in Eq.~\eqref{eq:lswafter},
which now allows the nuclear matter equilibrium
point to be reproduced.\footnote{%
The new scalar--vector interactions destroy the renormalizability of the
model, but we will overlook that for the moment.}

Nevertheless, an extensive mean-field analysis shows that
it is impossible to generate realistic results for nuclear densities, 
single-particle spectra, and total binding energies 
within this framework \cite{Fu93,Fu93a,Fu96}.
The primary problem is still the existence of scalar nonlinearities with 
incorrect systematics; in addition, the large value of the (known) scalar 
coupling leads to a scalar mass that is much larger than that required by the
phenomenology ($\msigma \approx 800\MeV$ compared to a desired value of
$\msigma \approx 500\MeV$).

We conclude that the standard form of spontaneous chiral symmetry breaking,
implemented in a model with a linear realization of the symmetry,
{\em cannot\/} produce successful
nuclear phenomenology at the mean-field level when we identify the chiral
scalar $\sigma$ with the scalar in QHD--I.\footnote{%
It has recently been shown \protect\cite{He94a,Ca96} that realistic nuclear 
systematics can be achieved in a linear realization by using a 
{\em logarithmic\/}, nonrenormalizable scalar potential to generate the
spontaneous symmetry breaking.}
Note that the interaction potential (the term proportional to $\lambdat$)
in Eq.~\eqref{eq:lchiral} is the only form consistent with both chiral
symmetry and renormalizability.
This failure of the sigma model is therefore evidence, even at the mean-field
level, that the simultaneous constraints of linear chiral symmetry and
renormalizability are too restrictive.

A different approach to the chiral symmetry is more compatible with observed 
nuclear properties and with successful relativistic mean-field models.
Here one takes the chiral scalar mass $\msigma$ to be {\em large\/} to 
eliminate the unphysical scalar nonlinearities and then generates the 
mid-range attractive force between nucleons {\em dynamically\/} through 
correlated two-pion exchange.\footnote{One can see that the scalar
self-interactions become small by writing the lagrangian
\protect\eqref{eq:lswafter} in terms of the re-scaled field $\chi \equiv
\msigma \sigma$.
Note, however, that the $\sigma \vecpi^2$
coupling becomes large when $\msigma$ becomes large.}
Numerous calculations \cite{Ja75,Du77,Du80,Li89,Li90,Ki94} have
shown that the exchange of two interacting pions in the scalar-isoscalar
channel produces a strong attractive force with a range comparable to that
of a scalar meson with a mass of 500 to 600 MeV.
In principle, this approach can be realized within the linear models 
discussed above, by rejecting the mean-field approximation (which is
inadequate) and by retaining contributions from correlated two-pion exchange
between nucleons.
This clearly introduces much greater complexity, since one must
first construct a boson-exchange kernel containing correlated two-pion
exchange and then allow this kernel to act to all orders (for example, in a
ladder approximation) to determine the NN interaction and the resulting
nuclear matter energy density \cite{Ma89}.
This calculation is further complicated by the need to maintain
chiral symmetry at finite density, which is difficult to do when one uses
the non-derivative (``pseudoscalar'' or PS) $\pi$N coupling implied by the
linear realization of the symmetry \cite{Ma82,Ho82,Wa95}.
These complexities motivate us to look for a more efficient way to implement
the chiral symmetry.

\subsection{A Nonlinear Realization of Chiral Symmetry}
\label{sec:nlsigma}

The chiral dynamics of pions and the successful mean-field picture described
in Section~\ref{sec:qhd} can be combined if the chiral symmetry is realized
in a {\em nonlinear\/} fashion.
In a nonlinear realization, chiral transformations mix the nucleon and pion
fields, and a scalar-isoscalar field is unchanged.
There are many ways to implement a nonlinear realization; here we follow
the approach of Weinberg \cite{We67} and discuss a different method in
Section~\ref{sec:eft}.

First notice that a chiral rotation matrix can be written as
\beq
 \exp (\tover{i}{2} {\vector \omega}\veccdot\vectau \gammafivel )
         = {\mbf 1} \cos (\omega /2) + i {\hat n}\veccdot\vectau
              \gammafivel \sin (\omega /2) \ ,
         \label{eq:chiralmat}
\eeq
where $\mbf 1$ denotes the unit matrix in spin-isospin space
and ${\vector\omega}\equiv {\hat n}\omega$ is real.  
The key observation is that the linear terms in $\sigma$ and $\vecpi$ 
in Eq.~\eqref{eq:lswafter}\ can be written as a chiral rotation matrix:
\beqa
    M - \gpi\sigma + i\gpi \vectau\veccdot\vecpi \gammafivel & = &
         [(M-\gpi\sigma)^2 + \gpisq \vecpi{}^2]^{1/2} \exp (i{\vector\theta}
              \veccdot\vectau \gammafivel )
              \nonumber\\[3pt]
    & = & [(M-\gpi\sigma)^2 + \gpisq \vecpi{}^2]^{1/2} [{\mbf 1} \cos\theta
              +i{\hat n}\veccdot\vectau \gammafivel \sin\theta ]
              \ , \label{eq:amazing}
\eeqa
where $\vecpi = {\hat n}\pi$ and ${\vector\theta} = {\hat n}\theta$.  
It now follows immediately that
\beq
    \cos\theta = \frac{M-\gpi\sigma}{[(M-\gpi\sigma)^2
              + \gpisq \vecpi{}^2]^{1/2}} \ , \qquad
    \sin\theta = \frac{\gpi\pi}{[(M-\gpi\sigma)^2
              + \gpisq \vecpi{}^2]^{1/2}} \ .  
         \label{eq:thetamess}
\eeq
One can then eliminate the linear pion--nucleon interaction 
in the lagrangian
$\lagrang_{\sigma\omega}$ by defining a new baryon field $N$ as a unitary
transformation of the old baryon field $\psi$:
\beq
    N \equiv \exp (\tover{i}{2} {\vector\theta}\veccdot\vectau 
        \gammafivel ) \psi
      = \sqrt{\tover12 (1 + \cos\theta)}\ ({\mbf 1}+
         i{\vector\xi}\veccdot\vectau
         \gammafivel )\psi \ ,
         \label{eq:newbaryon}
\eeq
where
\beq
    {\vector\xi} \equiv {\hat n}\frac{\sin\theta}{1 + \cos\theta} \ ,
         \label{eq:xidef}
\eeq
and by defining new meson fields $\vecpi'$ and $\sigma'$ as
\beq
    {\vector\xi} \equiv \frac{\gpi}{2M}\,\vecpi' \equiv
              \frac{f}{\mpi}\, \vecpi' \ , \qquad
    M - \gpi\sigma' \equiv \frac{M - \gpi\sigma}{\cos\theta} \ .  
    \label{eq:newmesons}
\eeq

Although the algebra is tedious, it is straightforward \cite{Wa95} to rewrite
$\lagrang_{\sigma\omega}$ in terms of the new fields as
%
\beqa
    \lagrang_{\sigma\omega} &=& \Nbar \Big\{ i\gammamul\dmu
         - \gv \gammamul \Vmu
         - (M - \gpi\sigma') 
                   \nonumber\\[4pt]
%
    & & \qquad {}
         + \frac{1}{1 + (f/\mpi )^2\vecpi'{}^2} \, \bigl[
         (f/\mpi )\gammamu\gammafivel \vectau\veccdot \dmul\vecpi'
         - (f/\mpi )^2 \gammamu \vectau\veccdot\vecpi' \veccross\dmul \vecpi'
         \bigr] \Big\}N
              \nonumber\\[4pt]
    & & \qquad {}
         + \tover12 (\dmul\sigma'\dmu\sigma' - \msigmasq\sigma'{}^2 )
         + \tover12 R[R\,\dmul\vecpi'\veccdot\dmu\vecpi' -\mpisq\vecpi'{}^2]
              \nonumber\\[4pt]
    & & \qquad {}
         + (\msigmasq - \mpisq) \big[ (f/\mpi )\sigma'{}^3
              -\tover12 (f/\mpi )^2 \sigma'{}^4 \big] 
         -\tover14 F_{\mu\nu}F^{\mu\nu} + \tover12 \mvsq\Vmul\Vmu
      \ ,
    \label{eq:nonlinear}
\eeqa
where we have defined the ratio
\beq
    R \equiv R(\sigma',\vecpi') \equiv \frac{1-2(f/\mpi )\sigma'}{1 +
                   (f/\mpi )^2 \vecpi'{}^2} \ .  
    \label{eq:Rdef}
\eeq
This somewhat imposing lagrangian has several important advantages.
First, the linear PS coupling between the nucleon and pion has
been replaced by a derivative or pseudovector (PV) coupling and a so-called
``sea gull'' term, where the nucleon couples to two pions simultaneously.  
In fact, there are an infinite number of such couplings, when one expands the
prefactor $[1 + (f/ \mpi )^2\vecpi'{}^2]^{-1}$ as a power series in
$\vecpi'{}^2$.  
Nevertheless, all of these couplings involve at least one derivative acting 
on the pion field; thus, in the limit of vanishing pion momenta, the pions
and nucleons {\em decouple}, as required by the chiral soft-pion theorems
that follow from the ${\rm SU}(2)_L \times {\rm SU}(2)_R$ current algebra
and PCAC \cite{Sa69,Do92}. 

Second, the new pseudovector coupling constant is
\begin{eqnarray}
    f^2  =  g_{\pi}^2 \left( \frac{m_{\pi}}{2 M} \right)^2
        \approx  1.0 \ , 
                                          \label{eq:ccf}
\end{eqnarray}
which is much smaller than the pseudoscalar coupling constant 
$g_{\pi}^2 / 4 \pi \approx 14.4 $.
Moreover, the explicit derivative couplings eliminate the sensitive
cancellations between Feynman diagrams that are necessary in the linear
realization.
The large coupling and sensitive cancellations show that the linear
realization is an {\em inefficient\/} way to implement the symmetry.

As noted earlier, $\msigma$ should be taken to be large, to avoid the
unwanted nonlinear interactions.
If desired, $\msigma$ can be kept finite, so that it plays the role of a 
regulator that maintains the renormalizability of  the model \cite{Ma82}, 
or $\msigma$ can be taken to infinity, so that the chiral scalar field 
{\em decouples}, resulting in the effective nonlinear model of 
Weinberg \cite{We67,Wa95}.\footnote{%
In contrast to the lagrangian \eqref{eq:lswafter}, if one rewrites the
{\em nonlinear\/} lagrangian \eqref{eq:nonlinear} in terms of $\chi \equiv
\msigma \sigma'$, it is easy to see that {\em all\/} interactions involving
the scalar field vanish in the $\msigma \rightarrow \infty$ limit.}
If the process $\pi + \pi \rightarrow \pi + \pi$ is investigated in
this model with a heavy $\sigma$, and the
chiral-invariant Born amplitude is unitarized, one observes
a broad, low-mass, near-resonant amplitude in the $(0^+,0)$
channel, even though the chiral $\sigma$ has a large mass.
When this model $\pi\pi$ scattering amplitude is included in the
two-pion-exchange part of the NN interaction, the result is a dynamically
generated, broad, low-mass ($\approx 600 \MeV$) peak that resembles
the exchange of a light scalar meson \cite{Li89,Li90}.

This strong scalar-isoscalar two-pion exchange can be simulated by 
introducing a low-mass, {\em effective\/} scalar field $\phi$
coupled directly to the nucleon, and scalar self-interactions can be added
to include a density dependence in the mid-range NN attractive force.
We emphasize that the purpose of the effective field is to parametrize the
NN attraction, so that we can avoid the complicated calculation of 
scalar-isoscalar pion loops.
All scalar propagation will be restricted to {\em spacelike\/} momenta,
and thus scalar particles are always virtual.\footnote{%
The effective scalar-isoscalar field is introduced to simplify the 
description of the attractive NN interaction.
Thus it plays a role in problems with $B \geq 1$ and cannot be 
directly interpreted as an on-shell particle.
The idea of introducing virtual degrees of freedom to describe fermionic
interactions is not revolutionary; phonon fields in metals and the
Ginzburg--Landau field in quantum liquids have played such a role for
many years with considerable success \protect\cite{Fe71}.}
At the mean-field level, the classical scalar field is an efficient way
to incorporate the effects of pion exchange that are the most important for
describing bulk nuclear properties.

If we now take $\msigma \rightarrow \infty$, the resulting nonlinear, chiral,
effective lagrangian is (primes on the fields are omitted)
\beqa
    \lagrang_{\rm eff} &=& \Nbar \Big\{ \gammamul
         (i\dmu - \gv \Vmu)
         - (M - \gs\phi) 
                   \nonumber\\[4pt]
%
    & & \qquad {}
         + \frac{1}{1 + (f/\mpi )^2\vecpi{}^2} \, \bigl[
         (f/\mpi )\gammamu\gammafivel \vectau\veccdot \dmul\vecpi
         - (f/\mpi )^2 \gammamu \vectau\veccdot\vecpi \veccross\dmul \vecpi
         \bigr] \Big\}N
              \nonumber\\[4pt]
     &  &\qquad {}
          + \tover12 (\dmul\phi\dmu\phi - \mssq \phi^2)
       -\tover{1}{3!}\kappa\phi^3 - \tover{1}{4!}\lambda\phi^4
       - \tover14 F_{\mu\nu}F^{\mu\nu} + \tover12 \mvsq\Vmul\Vmu
         \nonumber \\[3pt]
    & & \qquad {}
         + \frac{1}{2(1 + (f/\mpi )^2 \vecpi{}^2)}\,
     \Big[\frac{1}{1 + (f/\mpi )^2 \vecpi{}^2}\,
          \dmul\vecpi\veccdot\dmu\vecpi -\mpisq\vecpi{}^2 \Big]\ .
             \label{eq:Lww}
\eeqa
The explicit symmetry-violating term proportional to $\mpisq$ has been
included, and since $f/\mpi = \gpi / 2M$ from eq.~\eqref{eq:ccf}, the
exact chiral limit ($\mpi \rightarrow 0$) is sensible.

We retain the notation $N$ for the baryon field to remind us that it
transforms nonlinearly, in contrast to the field $\psi$ used earlier.
(Linear transformations are independent of the pion field, while nonlinear
transformations depend on the pion field.)
Although the nonlinear transformation law implied by Eq.~\eqref{eq:newbaryon}
is complicated, it has been shown \cite{We67,Co69} that under a general
chiral transformation,
\beq
N \longrightarrow N' = h(x) N \ ,
  \label{eq:Nmagic}
\eeq
where $h(x) \in {\rm SU(2)}_V$ is an {\em isospin\/} rotation matrix that
depends {\em locally\/} on the pion field.
Thus the inclusion of a Yukawa coupling
\beq
\lagrang_{\rm Y} = \gs \mkern2mu\Nbar N\phi
\eeq
(as well as scalar self-couplings) in $\lagrang_{\rm eff}$
leaves the chiral symmetry intact, because the light scalar is an 
isoscalar, and the nucleon field transforms as in Eq.~\eqref{eq:Nmagic}
\cite{Wa95}.\footnote{%
Note that one {\em cannot\/} add a Yukawa interaction $\psibar\psi\phi$ 
with an effective scalar field to the linear sigma model 
[Eq.~\protect\eqref{eq:lswafter}], as this destroys the chiral symmetry.}
Moreover, since the pion mean field vanishes, the chiral mean-field
theory obtained from $\lagrang_{\rm eff}$ 
produces {\em precisely\/} the same field equations and energy density as in
Section~\ref{sec:qhd}.
Finally, although the coupling strength $\gs$ of the light scalar is
comparable to $\gpi$ (as verified by explicit calculation of the correlated
two-pion exchange \cite{Li89,Li90}), we no longer require 
$\gs \! = \! \gpi$ and are free to adjust $\gs$ within a reasonable range.
The nonlinear scalar interactions, which parametrize the density dependence
of the correlated two-pion exchange, also contain adjustable parameters.

Thus, through these somewhat lengthy arguments, we conclude that {\em the
mean-field QHD--I model studied earlier is consistent with chiral symmetry}, 
provided we think in terms of a nonlinear realization of the symmetry.
The importance of the resulting scalar-isoscalar mean field and optical 
potential in producing a successful nuclear phenomenology was illustrated in 
Section~\ref{sec:qhd}.  
Although the one-pion-exchange interaction (and its iteration) produces a
relatively small contribution in nuclear matter due to its spin dependence,
the correlated two-pion-exchange contribution has a spin-independent,
isoscalar part that is large and that must be included from the 
outset.\footnote{%
In the few-nucleon problem, the situation is reversed.
Because of the more complicated spin dependence of the wave function, matrix
elements for one-pion exchange no longer average to zero.
Moreover, the existence of an almost-bound state at
essentially zero energy in the
${}^1S_0$ NN channel \protect\cite{Fe71} implies that there is a nearly
exact cancellation between the attractive (scalar) and repulsive (vector)
parts of the NN force in few-body systems.
Thus, the pion-exchange ``tensor force'' plays a more prominent role.}
The discussion in Section~\ref{sec:qhd} shows that the scalar-isoscalar
field in QHD--I is an efficient and successful way to incorporate these
important pionic effects.
{\em A profound change has occurred, however, because in contrast to 
the original proposal of QHD--I as a renormalizable field theory, we are
now forced to consider the scalar field as an effective degree of freedom
and the lagrangian in Eq.~{\rm \eqref{eq:Lww}} as a nonrenormalizable
effective lagrangian.}
We will consider this more general strategy for QHD in Section~\ref{sec:eft}.

\subsection{The $\Delta$ Isobar}
\label{sec:isobar}

The dominant phenomenological features of the low-energy $\pi$N interaction 
are that low-momentum pions interact weakly with nucleons (they decouple 
as $q_{\mu} \rightarrow 0$), and that the interaction is dominated by the
first pion--nucleon resonance, the $\Delta(1232)$.  
This resonance represents the first excited state of the baryon, with
$(J^{\pi}, T) = (\frac{3}{2}^+,\frac{3}{2})$.  
It is essential to include this degree of freedom in the theory to generate 
realistic results for pion--nucleus interactions and for few-nucleon systems.
It is impossible to put a field with these quantum numbers into a 
renormalizable lagrangian, but it has been shown that the $\Delta$ degree 
of freedom can be produced {\em dynamically\/} within the model.
The most efficient way to incorporate the $\Delta$ is
as an effective degree of freedom.

Here we simply summarize the results, as the arguments and references to the 
original literature are detailed in \cite{Se86,Se92a,Wa95}.  
The sum of $\pi$N ladder diagrams with nucleon exchange can be
investigated within the framework of the chiral $\pi$N theory
discussed above.  Partial-wave dispersion relations can be used,
with the one-baryon-exchange mechanism considered  as the driving term,
and the resulting integral equations solved with the $N/D$ 
method \cite{Li91}.
This is a relativistic extension of nonrelativistic Chew--Low theory. 
As in the Chew--Low theory, a resonance is found in the 
$(\frac{3}{2}^+,\frac{3}{2})$ channel.  
Evidently, the box diagram in the ladder sum involves a loop integral, 
which obtains significant contributions from large loop momenta
or short distances; thus, the predicted {\em position\/} of the
resonance is sensitive to the approximations made.
In contrast, the predicted resonance {\em width\/} is much less sensitive.
As a result of the work summarized in \cite{Se92a,Wa95}, it is clear that
the first excited state of the nucleon, the $\Delta$(1232) with
$(\frac{3}{2}^+, \frac{3}{2})$, which is the dominant feature of low-energy 
pion--nucleus interactions, is generated {\em dynamically\/} in QHD.
Thus it can also be included as an effective degree of freedom in a
nonrenormalizable hadronic lagrangian \cite{Bi82,De92a,We93,He96,Ta96}.



\section{Few-Nucleon Systems}
\label{sec:few}

The purpose of this section is to present some additional evidence of the
successes of hadronic field theory by considering the few-nucleon problem.
As before, we want to focus on basic ideas and phenomenology that will guide
us in the construction of an effective field theory for the many-body
problem.
We begin by summarizing modern models of the NN interaction
and show that when applied to two-nucleon systems, accurate results are
obtained both for electromagnetic observables and for threshold pion 
production.
We also consider the extension of the two-nucleon problem into nuclear
matter, by discussing the relativistic generalization of the 
independent-pair approximation \cite{Fe71,Wa95}, which is often called 
Dirac--Brueckner--Hartree--Fock (DBHF) theory.
This will allow us to compare nuclear matter calculations involving 
two-nucleon correlations to the mean-field results discussed earlier and
also to understand the relationship between the relativistic two-nucleon 
interaction and nonrelativistic many-body forces.

\subsection{The Nucleon--Nucleon Interaction}
\label{sec:NN}

Models of the NN interaction have been studied seriously for nearly 50 years.
An excellent historical review is contained in \cite{Ma89}.
Modern meson-exchange models provide excellent fits to experimental
observables up to and somewhat beyond pion-production threshold, which is at
a laboratory kinetic energy of roughly $300\MeV$.
Once the models are calibrated to the NN data, they can be used to study 
both the interaction of few-nucleon systems with experimental probes and the
nuclear many-body problem.
As mentioned in the Introduction, to achieve accurate reproduction of the
data, several mesons are needed [the most important are the $\pi (0^-,1)$, 
$\sigma (0^+,0)$, $\omega (1^-,0)$, and $\rho (1^-,1)$, although the
$\eta (0^-,0)$ and $\delta (0^+,1)$ are often included], and many models
contain both N and $\Delta$ degrees of freedom.
There are models based entirely on boson 
exchange \cite{Zu81,Ma87,Te87,Ma89,Gr90,Pl94},
as well as models that are supplemented by dispersion 
relations \cite{Br76,Pa79}
or by Regge theory \cite{Na73,St94}.
(A comparison of the different models is contained in \cite{Ma94e}.)
These models are not renormalizable and contain form factors
at the meson--nucleon vertices that are not usually expanded
in powers of momentum, as one would do in a strict implementation of
effective field theory \cite{We91,We92a,Or92,Va93,Or96}.
(In other words, the interactions are {\em nonlocal}.)
The range parameters of the vertex functions are treated as adjustable,
and they are important for producing realistic results for observables.

The NN system is usually studied with a {\em quasipotential\/}
approach, in which one starts with the two-nucleon Bethe--Salpeter equation 
and then reduces it to a three-dimensional integral equation by defining an
approximate two-nucleon propagator.
In free space, the Bethe--Salpeter equation can be written schematically
as \cite{Bj65}
\beq
  T = K + i\int K G G T \ ,  \label{eq:freeBS}
\eeq
where $K$ is the full two-body scattering kernel and $G$ is the exact baryon
propagator.
(For illustration, we consider only nucleons here.)
The four-dimensional integral implied in Eq.~\eqref{eq:freeBS} can be reduced
to a three-dimensional integral by defining a unitarized, two-particle
propagator $g_0$ and a quasipotential U:
\beqa
      T &=& U + \int U g_0 T \ , \label{eq:QBS}\\[3pt]
      U &=& K + \int K (iGG - g_0) U \ . \label{eq:QP}
\eeqa
These two equations are equivalent to \eqref{eq:freeBS}, and if one solves
both of them, the results should agree with the Bethe--Salpeter results
for any choice of $g_0$.
In practice, however, only Eq.~\eqref{eq:QBS} is retained, under the 
assumption that the corrections from \eqref{eq:QP} are small; moreover, the
quasipotential $U$ is approximated by keeping just the one-boson-exchange
``ladder'' kernel (which we call $V$), or the ladder terms plus some ``box''
and ``crossed-box'' amplitudes \cite{Ma89}.

To study the electromagnetic response of the two-nucleon system, the
quasipotential is fitted to reproduce the NN phase shifts and the properties
of the deuteron, and then the coupling of the electromagnetic current is
introduced.
Since the quasipotential can be defined in an infinite number ways, it is
important to test the sensitivity by utilizing different approaches;
currently, calculations using three different quasipotentials are
available \cite{Hu90,De93,Va95}.
Calculations of the deuteron charge and magnetic form factors are in good
agreement with experiment out to three-momentum transfers of roughly
$q^2 \approx 80\,{\rm fm}^{-2}$, provided that meson-exchange currents are 
included.
Moreover, calculated results for the tensor polarization ($T_{20}$) of the
deuteron agree with experiment out to $q^2 \approx 20\,{\rm fm}^{-2}$.
Significant effort has been spent in deriving meson-exchange currents that
guarantee electromagnetic gauge invariance, even when phenomenological form
factors are used at the vertices \cite{Gr87,Co94a}.
Nevertheless, there are still some open questions regarding the consistent
inclusion of realistic off-shell nucleon form factors and the model 
dependence of the transverse parts of the meson-exchange currents.

Meson-exchange currents are also important for describing pion production
near threshold in the reaction $ pp \rightarrow  pp \pi^0$.
A precise measurement of the total cross section for this process has
recently been performed \cite{Me90,Me92}, but calculations that include only
single-nucleon mechanisms severely underestimate the measured 
values \cite{Ko66,Sc69}.
In contrast, the inclusion of a scalar-meson exchange current (and smaller
contributions from other mesons) increases the cross section by about a
factor of five and leads to excellent agreement with the 
data \cite{Le92,Me92,Ho94b}.
The results are insensitive to changes in the potential
that generates the two-nucleon wave function and to different choices for
the phenomenological meson--nucleon form factors.
Although some questions have recently been raised about possible competing 
mechanisms, alternative calculations cannot reproduce the measured cross 
section \cite{Co96}.

\subsection{Two-Nucleon Correlations in Nuclear Matter}
\label{sec:DBHF}

The extension of the quasipotential approach to nuclear matter is known as
Dirac--Brueckner--Hartree--Fock theory \cite{An83,Ho84a,Br84,Ho87,Te87,Am92}.
The scattering matrix $T$ in Eq.~\eqref{eq:QBS} is replaced by a reaction
matrix $\Gamma$, which is determined by the quasipotential equation
\beq
      \Gamma = V + \int V g \Gamma \ , \label{eq:QDBHF}
\eeq
where $V$ is the (ladder-approximated) quasipotential and $g$ is a
unitarized, two-nucleon
propagator that includes interactions with the surrounding 
medium \cite{Ho84a,Se86,Ho87}.
In practice, one solves for matrix elements of $\Gamma$, so that the
driving term on the right-hand side
involves matrix elements of $V$; the new ingredient is that the Dirac wave 
functions must be determined {\em self-consistently}, since the
nucleons have large scalar and vector self-energies, analogous to the
mean fields studied earlier.

If one writes the nucleon self-energy $\Sigma (k)$ in (the rest frame of)
nuclear matter in terms of scalar ($\Sigma^{\rm s}$), timelike vector
($\Sigma^0$), and three-vector ($\Sigma^{\rm v}$) parts \cite{Se86}:
\beq
    \Sigma (k) = \Sigma^{\rm s} (k) - \gamma^0 \Sigma^0 (k) + 
                 \vecgamma\veccdot{\bf k}\, \Sigma^{\rm v} (k) \ ,
                     \label{eq:Sigdef}
\eeq
then the self-energy is determined by summing effective direct and exchange
interactions between nucleon pairs, which may be written schematically as
\beq
   \Sigma (k) = \sum_{E_{p} \leq E_{\rm F}} 
              [\matrixelement{{\bf k}{\bf p}}{\Gamma}{{\bf k}{\bf p}}
              -\matrixelement{{\bf k}{\bf p}}{\Gamma}{{\bf p}{\bf k}}]
       \ . \label{eq:DBHFsc}
\eeq
This is indeed a self-consistency condition, since the self-energy determines
the Dirac wave-functions and the reaction-matrix elements
that appear on the right-hand
side, and it is in turn determined by these quantities.
The $\Sigma^{\rm s}$ and $\Sigma^0$ components are analogous to the mean 
fields $(-\gs\phizero)$ and $(-\gv\Vzero)$ studied earlier;
$\Sigma^{\rm v}$ is a new contribution that arises because exchange 
diagrams are included.

The numerous approximations and procedures that go into solving the DBHF
equations are discussed at length 
in \cite{Ho84a,Br84,Se86,Ho87,Te87,Am92}.
The most important approximation is the use of a quasipotential, which
implies that the four-dimensional integral in the box diagram that
forms the fundamental unit of the ladder sum is reduced to a 
three-dimensional integral.
There are many possible reductions that can be identified as the relativistic
extension of the nonrelativistic Bethe--Goldstone equation \cite{Wa95};
unfortunately, numerical results for the nuclear matter binding energy are
sensitive to the reduction used.  
Even after this reduction, calculated results are sensitive to the 
high-momentum part of the loop integrals, implying important contributions 
from baryon transitions to states lying well above the Fermi surface.
Phenomenological form factors inserted at the vertices significantly reduce 
this sensitivity, but questions about the off-mass-shell and density 
dependence of these form factors remain to be studied.  
Moreover, since the self-energy $\Sigma$ enters explicitly in the 
self-consistent baryon spinors, the binding energy of
nuclear matter is sensitive to the self-consistency condition \cite{Le89a}.  
At present, the construction of a self-consistency condition
that leads to a conserving approximation when relativistic ladder
diagrams are summed is still a controversial topic \cite{Po88,De91,Hu93}.  

It must be emphasized that many of the conventional approximations used in 
DBHF theory have no counterparts in nonrelativistic Brueckner--Goldstone
theory, 
and thus the sensitivity to these approximations has never been 
systematically tested.
Nevertheless, it is still possible to make three important 
{\em qualitative\/}
statements about the effects of relativistic two-nucleon correlations in
nuclear matter:

\begin{enumerate}
\item
Although the correlation corrections produce changes in the RMFT nuclear 
matter binding energy that are of the same order as the binding energy 
itself, {\em the corrections to the large {\em RMFT} scalar and vector 
self-energies are small} \cite{Ho87}.
Thus the DBHF self-energies are essentially the same size as the scalar
and vector mean fields studied earlier.
(Full DBHF calculations yield typical results of $0.55 \lesssim \Mstar / M
\lesssim 0.65$ at equilibrium density \cite{Ma89,Br90a,De91,Am92}.)
Moreover, the momentum dependence (or state dependence) of $\Sigma^{\rm s}$
and $\Sigma^0$ is small; these self-energies are essentially constant for
occupied states in the Fermi sea.
Finally, the new term $\Sigma^{\rm v}$ is small.
The conclusion is that the successful RMFT picture presented above involving
large, constant, Lorentz scalar and vector fields persists when two-nucleon
correlations are included.\footnote{%
The large scalar and vector self-energies are also consistent with recent
analyses based on QCD sum rules \cite{Co91a,Co92a,Fu92a,Ji93a,Ji94a}.}
\item
The depletion of the Fermi sea due to correlation effects is considerably
smaller in the relativistic framework than in the nonrelativistic
framework \cite{Ja90}.
\item
The self-consistency condition \eqref{eq:DBHFsc} introduces a density 
dependence into the effective interaction $\Gamma$ that goes beyond what is 
included in nonrelativistic Brueckner--Goldstone theory.
Because of this extra density dependence, it is possible to 
{\em simultaneously\/} fit both the NN phase shifts and the nuclear matter 
equilibrium point at the two-hole-line level \cite{Ma89,Br90a,De91}.
Although an accurate calculation of the nuclear matter equilibrium point
is difficult, due to the sensitive cancellations, and
there are still open questions about three-hole-line corrections, true
many-body forces, and the role of the quantum vacuum, this successful
result {\em cannot be obtained\/} in a nonrelativistic framework 
with two-body potentials \cite{Da83,Da85}.
Further investigations are needed to make this result more conclusive.
\end{enumerate}  

\subsection{Relation to Nonrelativistic Calculations}

To close this section, we briefly examine why the DBHF theory produces
more favorable results for nuclear matter saturation at the two-hole-line
level than corresponding nonrelativistic Brueckner--Goldstone calculations.
It is well known that nonrelativistic calculations based on realistic NN
potentials predict equilibrium points that have either the correct
density but too little binding energy, or the correct binding energy
at too high a density; this produces the familiar ``Coester line''.
What is needed is an additional, density-dependent repulsive mechanism, 
and the new saturation
mechanism discussed in Section~\ref{sec:qhd} is precisely of this type.
Because of Lorentz covariance and self-consistency, as the nuclear density
increases, the nucleon effective mass $\Mstar$ decreases, and the nucleon
velocities increase.
This weakens the attractive force, and the net result is
an increased repulsion.

Although this additional repulsion is often described as a purely
``relativistic'' effect, it is easy to see that in a nonrelativistic 
framework, some of it can be attributed to three- and many-body 
forces \cite{Br87,Fo95}.
Rather than work with a complicated DBHF interaction, the basic point can
be illustrated using the RMFT model of Section~\ref{sec:qhd}; for
suitable choices of couplings and masses, simple scalar and vector exchange
resemble quite closely the scalar and vector components of the DBHF 
interaction $\Gamma$.
(We are not concerned here with numerous details like pi- and 
rho-exchange diagrams, the pion-exchange ``tensor force'', etc. 
We will also set the nonlinear scalar couplings $\kappa$ and $\lambda$ to 
zero to simplify the equations.)

The basic idea is to expand the RMFT nuclear matter energy density of 
Eq.~\eqref{eq:endens} in powers of the Fermi momentum $\kfermi$, and then to
group terms together to isolate various powers of $\rhoB$.
One can then identify contributions from increasing powers of $\rhoB$
as arising from two-body forces, three-body forces, etc., if one attempted
to reproduce the same energy/nucleon with nonrelativistic potentials.
We begin with the expansion of the self-consistency condition 
\eqref{eq:mftsc}, which is given through order $\kfermi^{11}$ 
by \cite{Se86,Fo95}
\beqa
   \Mstar &=& M - {\gssq\rhoB\over \mssq}\, \bigg[ 1 
            -  {3 \kfermi^2\over 10 M^2}
            +  {9 \kfermi^4\over 56 M^4}
            -  {5 \kfermi^6\over 48 M^6}
            +  {105 \kfermi^8\over 1408 M^8} \nonumber\\[3pt]
 & & \qquad\qquad\qquad
        {} - {3\over 5}\, {\gssq\over\mssq}\, \rhoB \bigg(
          {\kfermi^2\over M^3} - {48\kfermi^4\over 35 M^5}\Big)
          - {9\over 10}\, \Big({\gssq\over\mssq}\Big)^{\mkern-2mu 2}
            \Big({\rhoB^2\kfermi^2\over M^4}\Big) +\cdots \bigg] \ .
          \label{eq:mstarex}
\eeqa
Here $\rhoB = \gamma\kfermi^3 /6\pi^2$ as usual.

Substitution of this result into Eq.~\eqref{eq:endens} and expansion through
order $\kfermi^{11}$ yields, for the energy/nucleon,
\newpage
\beqa
  \edens /\rhoB &=& M + \bigg[ {3\kfermi^2\over 10 M}
            - {3\kfermi^4\over 56 M^3}
            + {\kfermi^6\over 48 M^5} - {15\kfermi^8\over 1408 M^7}
            + {21\kfermi^{10}\over 3328 M^9} + \cdots \bigg]
          \nonumber\\[4pt]
   & & \qquad
      +{\gvsq\over 2\mvsq}\, \rhoB - {\gssq\over 2 \mssq}\, \rhoB
      +{\gssq\over\mssq}\, {\rhoB\over M}\bigg[ {3\kfermi^2\over 10 M}
      -{36\kfermi^4\over 175 M^3} + {16\kfermi^6\over 105 M^5}
      -{64\kfermi^8\over 539 M^7} + \cdots \bigg]
          \nonumber\\[4pt]
  & & \qquad
      + \Big({\gssq\rhoB\over\mssq M}\Big)^{\mkern-2mu 2}
         \bigg[{3\kfermi^2\over 10 M}
          - {351\kfermi^4\over 700 M^3} + \cdots \bigg]
         + \Big({\gssq\rhoB\over\mssq M}\Big)^{\mkern-2mu 3}
            \bigg[{3\kfermi^2\over 10 M}
              - \cdots \bigg] \ .
        \label{eq:enex}
\eeqa
The first term is the baryon rest mass, followed by the nonrelativistic
Fermi-gas energy and the first few relativistic corrections, which are
essentially negligible at equilibrium density.
(These terms are the ones usually used to justify a nonrelativistic
treatment of the nuclear matter problem \cite{Ne85}.)
The next two terms (proportional to $\rhoB$)
give the nonrelativistic limit of the potential energy
coming from the vector and scalar mesons.
The following term in brackets (with overall factor $\rhoB$)
is a relativistic correction to the scalar potential energy that
arises from the Lorentz contraction factor in the scalar density,
evaluated for nucleons of mass $M$.
The final two terms (with overall factors of $\rhoB^2$ and $\rhoB^3$) are
also corrections to the scalar potential energy.
These arise from the self-consistency condition \eqref{eq:mstarex} on
$\Mstar$; self-consistency implies $\Mstar < M$, which increases the
velocities of the nucleons and thus also increases their energies.
These repulsive contributions are a signature of the velocity dependence
inherent in a Lorentz scalar interaction.
Nevertheless, the leading contributions to these terms (in powers of $\rhoB$)
can be reproduced in a nonrelativistic calculation by including
repulsive three-nucleon and four-nucleon potentials \cite{Br87,Fo95}.

There are several relevant points to be noted:
\begin{itemize}
\item
Important ``relativistic'' effects in the RMFT of nuclear 
matter are equivalent to many-body forces in a nonrelativistic framework.
\item
Although only the three- and four-body contributions are shown in
Eq.~\eqref{eq:enex}, {\em all\/} of the leading terms at each power
of $\rhoB$ arising from the self-consistency condition are {\em repulsive}.
Together with the Lorentz contraction contribution, they provide the 
new saturation mechanism discussed earlier.
\item
Since the scalar and vector mesons have relatively large masses
(compared to the pion), the nonrelativistic many-body potentials used to
generate these terms will be short-ranged \cite{Co95a}.
\item
With typical values for the RMFT parameters, one finds
$\gssq\rhoB / \mssq M \approx 0.5$ at equilibrium density.
It follows that the three- and four-body repulsive terms in 
Eq.~\eqref{eq:enex} are roughly 5 MeV and 2.5 MeV, respectively, which
are significant on the scale of the nuclear matter binding energy.
\item
These relativistic (or many-body, if you prefer) contributions
are inherent in the DBHF framework and allow a two-hole-line calculation
to (essentially) reproduce the equilibrium point of nuclear matter, with
an NN interaction fitted to phase shifts (see Fig.~10.13 in \cite{Ma89}).
Although, to our knowledge, a relativistic three-hole-line calculation of
nuclear matter has never been performed, one might hope that the smaller
depletion of the Fermi sea (as noted above) would lead to smaller three-body
corrections than in the nonrelativistic case, implying that the two-hole-line
results are fairly robust.
Relativistic three-body-cluster effects remain to be investigated.
\end{itemize}



\section{Effective Field Theory (EFT)}
\label{sec:eft}

\subsection{Introduction}
\label{sec:eftintro}

Quantum chromodynamics (QCD) is generally accepted as the underlying theory
of the strong interaction.
As we have seen, however, at energies relevant for most nuclear phenomena, 
hadrons are convenient and efficient degrees of freedom.
In particular, meson-exchange models of the NN interaction accurately
describe low-energy properties of the two-nucleon system, and
relativistic mean-field theory based on QHD--I and QHD--II provides a 
realistic description of the bulk and single-particle properties of nuclei.

Unfortunately, renormalizable QHD models have encountered 
difficulties due to large effects from loop integrals that incorporate
the dynamics of the quantum vacuum 
[Co87,
Pe87,Fu88,Fu89,We90,Li90a].
On the other hand, the ``modern'' approach to renormalization
\cite{Le89,Po92,Ge94,We95},
which makes sense of
effective, ``cutoff'' theories\footnote{Here ``cutoff'' means a 
regulator that maintains the appropriate symmetries.}
with  low-energy, composite degrees of freedom, provides an alternative. 
When composite degrees of  freedom are used, the structure of
the particles is described with increasing detail by
including more and more {\it nonrenormalizable\/} interactions
in a derivative expansion\cite{Le89}.
Moreover, nonrenormalizable interaction terms between the boson fields
allow us to describe the short-distance behavior of the
underlying theory of QCD.
So, based on the successful phenomenology we have seen so far, we would
like to generalize the description using
the modern viewpoint of effective field theory (EFT).

The strategy underlying EFT relies on two basic observations.
First, one argues that relativistic quantum field theory is
simply the most general way to parametrize an $S$ matrix (or other
observables) consistent with analyticity, unitarity, causality, cluster
decomposition, and symmetries (\eg Lorentz covariance, chiral
symmetry, \dots)\cite{We95}.
With this view, there is no reason that relativistic quantum field theory
should be reserved for ``elementary'' particles only.

Second, one observes that in most problems in physics, the relevant 
phenomena are confined to a specific length scale, and thus it is not
necessary to explicitly include dynamics at significantly shorter length
scales\cite{Ge93}.
This implies that, at least formally, one can construct an effective field
theory to be used at a given length scale by ``integrating out'' heavier
degrees of freedom corresponding to shorter length scales; the effects of
these heavier degrees of freedom will be implicitly contained in various
coupling parameters in the low-energy, effective 
theory \cite{Le89,Ka95a,Ma96}.
By fitting these parameters to experimental data, one can derive
relationships between different observables within the dynamical regime of
interest.

Note that this strategy is {\em opposite\/} to the strategy
of renormalizable field theory, where one argues that the {\em fixed\/}
number of unknown parameters can be determined at any convenient length scale
and then extrapolated to any other length scale using the equations of the 
``renormalization group''.
Indeed, the effective theory is expected to contain {\em numerous\/}
couplings of nonrenormalizable form when one integrates out the heavier
degrees of freedom.
These nonrenormalizable couplings incorporate the 
``compositeness'' of the low-energy degrees of freedom through their
implicit dependence on short-distance physics.

These considerations imply, of course, that the low-energy effective theory
will generally contain an infinite number of interaction terms, and thus
one needs an {\em organizing principle\/} to make sensible calculations.
First, one must find a suitable expansion parameter (or parameters) that
is small in the region of interest.
Second, one assumes ``naturalness'', which means that all of the unknown
couplings in the theory, when written in appropriate dimensionless form (as
discussed below), are of order unity.
Thus one can estimate contributions from various terms by counting powers
of the expansion parameter(s) and then truncate the lagrangian at the
desired level of accuracy.

Although the resulting framework is apparently structured less rigidly 
than the more familiar edifice of renormalizable quantum field theory,
EFT nevertheless has its own ``rules of the game'':
First, one cannot simply limit calculations to the tree level; loops
can and must be included, as this is the only way to correctly incorporate
unitarity.
Second, unknown parameters are to be determined either by explicitly
integrating out the short-distance physics and ``matching'' the low-energy
parameters to the results (if the underlying theory is tractable) or by
fitting to experiment (where, hopefully, the number of parameters is 
fewer than the number of data to be described).

In the case of the nuclear many-body problem, it is still impossible to 
compute the desired low-energy parameters by working directly with QCD.
Nevertheless, important constraints on the effective hadronic lagrangian are
established by maintaining the symmetries of QCD.
These include Lorentz invariance, parity conservation, 
isospin invariance, chiral symmetry, and electromagnetic gauge invariance.
These symmetries constrain the lagrangian most directly by restricting
the form of possible interaction terms;
if one is forced to fit unknown parameters to the data, one must
include in the effective lagrangian all (non-redundant) terms that are
consistent with the underlying symmetries.
Moreover, the nature of the symmetries may also dictate the appropriate
low-energy degrees of freedom, or give relationships between some of the
unknown parameters.

Redundancy may arise because there is significant freedom in choosing the
generalized coordinates of the effective lagrangian.
In contrast to renormalizable theories, where there is (usually) a
preferred choice of field coordinates in which the lagrangian is
manifestly renormalizable, the choice of field variables in an effective
theory is motivated by the desire to make the description of the interactions
as efficient as possible.
(This is similar to the situation in classical lagrangian mechanics.)
If a point transformation of the fields (subject to some mild
constraints \cite{Co69}) allows one to eliminate certain interaction terms,
they can be considered redundant \cite{Ba88,Ge91,Ba94b}.
The goal is then to find the best set of generalized
coordinates (fields), so that the inevitable truncations are as accurate
as possible.

A well-known application of EFT is chiral perturbation theory (ChPT), in 
which one observes that the spontaneous breaking of chiral symmetry in QCD 
implies that Goldstone bosons (pions, \dots ) are the relevant low-energy
degrees of freedom\cite{Le94}.
Chiral symmetry also implies that the pion interactions can be grouped 
order-by-order in the number of derivatives, so that there is a 
systematic expansion at low energies in powers of external 
momenta (and $\mpisq$) \cite{Ga84}.
Moreover, because the loop expansion also proceeds in powers of momenta, 
one can systematically compute loop corrections.
ChPT has been reasonably successful in describing scattering in the $B=0$
and $B=1$ sectors of low-energy QCD\cite{Do92,Me93a}.
Studies of two- and many-nucleon systems have been initiated and are
currently under active investigation\cite{Or92,Va93,Ly93,Or96,Ka96}.
However, the prospects for extending ChPT to many-body calculations at 
finite density are unclear at present.

Thus we are motivated to consider alternatives.
How can we exploit the ideas of EFT to develop a systematic, 
nonrenormalizable approach to nuclear structure?
Let us enumerate the relevant concepts:
\begin{enumerate}
\item Pions can be included by applying the framework of ChPT.
\item Nucleon fields are necessary, since these are the observed
fermionic degrees of freedom at low energies; that is, nucleons carry the
conserved baryon number $B$.
Moreover, nucleon compositeness is retained even at the ``tree'' level, if we
include nonrenormalizable interaction terms, as noted earlier.
\item QCD constraints will be imposed through {\em symmetries}; all
allowed (non-redun\-dant) terms must be included.
\item Since redefinitions of the field variables do not affect observables,
we must strive for the most efficient variables and parametrizations of
the interactions.
\item We must identify suitable expansion parameters.
Based on our earlier discussion, we know that the nuclear mean fields
(or self-energies) $\Phi\equiv\gs\phizero$ and $W\equiv\gv\Vzero$ are
roughly several hundred MeV at ordinary densities.
Thus we take as expansion parameters the ratios $\Phi /M$ and $W/M$, 
which are small at normal densities, and also the ratios of gradients of the 
fields to $M^2$, which are small in nuclei
($ |\bbox{\nabla}\Phi |/M^2 \approx 
|\bbox{\nabla} W |/M^2 \lesssim 0.1$)\cite{Fu95}.
If we then assume that the unknown coefficients are natural, we can
truncate the effective lagrangian.
Of course, after the calculations are finished and the parameters have been
fitted to empirical data, we must check that the naturalness criterion
is satisfied.
\item What about non-Goldstone bosons, like $\omega$, $\rho$, and our
effective scalar field $\phi$?
We argue that these are known to be useful degrees of freedom, since the
mid-range NN interaction can be efficiently described by the exchange of
these mesons\cite{Ma87,Ma89}.
Moreover, the coupling constants in ChPT at $O(E^4)$ in the meson sector can
be reproduced by a meson-resonance lagrangian applied at tree level,
with the vector mesons playing the leading 
role\cite{Me88,Do89,Ec89,Ec89a}; this is consistent with
the well-known hypothesis of vector-meson dominance\cite{Sa69}.
Finally, we have already noted that two-pion exchange in the scalar-isoscalar
channel can be efficiently simulated by a low-mass scalar field.\footnote{%
Recent results indicate that a low-mass scalar plays a similar role in
$\pi$N scattering \protect\cite{Me96}.}
Even though both the nucleon and scalar are composite objects, recent work
shows that the leading term [of $O(\Phi^2/M)$] in the interaction energy
of a nucleon in a classical scalar field is {\em model independent\/} and 
determined solely by Lorentz covariance \cite{Wa95a,Bi95}.
\item
An accurate description of pion interactions with few-body systems and
nuclei requires that the $\Delta$ resonance be included.
This can be done most efficiently by introducing the $\Delta$ as another
effective degree of freedom \cite{Pe68,Bi82,De92a,We93,Ta96}.
For simplicity, and because we are concerned primarily with bulk and
isoscalar properties of nuclei, we will omit $\Delta$ interactions in the
models discussed below.
\end{enumerate}

Thus our strategy will be to construct an EFT lagrangian containing nucleons,
pions, and low-mass scalar and vector mesons.
Heavier particles (with masses greater than roughly 1 GeV) will be 
integrated out.

\subsection{Naive Dimensional Analysis}
\label{sec:NDA}

There are still two important points to be addressed.
First, we must understand how to extract the dimensional scales of each
term in the lagrangian, so that the remaining dimensionless constants can
be checked for naturalness.
A naive dimensional analysis (NDA) for assigning a coefficient of the 
appropriate size to any term in the effective lagrangian has been proposed
by Manohar and Georgi\cite{Ma84a,Ge93a}. 
This allows for a determination of both the dimensional scales associated
with each term and for the inclusion of an overall dimensionless constant
that can be used to adjust the strength.
The basic assumption of naturalness is that once the appropriate
dimensional scales have been extracted, the overall dimensionless
coefficients should all be of order unity.
The NDA rules for a given term in the lagrangian density are:
\begin{description}
\item[(1)] Include a factor of $1/\fpi$ for each strongly interacting
        field.
\item[(2)] Assign an overall factor of $\fpisq M^2$.
\item[(3)] Multiply by factors of $1/M$ to achieve dimension (mass)$^4$.
\item[(4)] Include appropriate counting factors (such as $1/n!$ for
      $\phi^n$).
\end{description} 
Here $\fpi \approx 93\MeV$ is the pion-decay constant, and the nucleon mass
$M$ is taken as the generic large-momentum cutoff scale, which characterizes
the mass scale of physics beyond Goldstone bosons.
In some cases \cite{Fu96b}, it is more appropriate to use the (non-Goldstone)
meson masses rather than $M$, but we usually do not distinguish here between
the two.

As noted by Georgi\cite{Ge93a}, rule (1) simply assumes that the 
amplitude for producing any strongly interacting particle is proportional 
to the amplitude $\fpi$ for emitting a Goldstone boson.
This is a reasonable assumption, since $\fpi$ is the only natural scale.
Thus, by dividing each field by $\fpi$, we should arrive at a factor of
$O(1)$.
Rule (2) can be understood as an overall normalization factor that arises
from the standard way of writing the mass terms of non-Goldstone bosons. 
For example, one may
write the mass term of a scalar-isoscalar field $\phi(x)$ as
\beq
     {1\over 2}\mssq \phi^2 ={1\over 2}\fpisq M^2 
         {\mssq \over M^2} {\phi^2 \over \fpisq} \ , 
\eeq
where the scalar mass $\ms$ is treated as roughly the same
size as $M$.
By applying rule (1) and extracting the overall factor of $\fpisq M^2$, the
remaining ratios are of $O(1)$.
Since all terms will have the same overall scale factor $\fpisq M^2$,
higher-order terms  or terms with gradients of fields will be suppressed by
powers of $1/M$ relative to the leading mass terms, as a result of 
``integrating out'' physics above the scale $M$.
(A simple example is the low-momentum expansion of a tree-level propagator
for a heavy meson of mass $m_H$, which leads to terms with powers of
$\partial^2 / m^2_H$.)
It is exactly because of these $1/M$ suppression factors and dimensional 
analysis that one arrives at rule (3).
The origin of the combinatorial factors in rule (4) is discussed
in \cite{Fu96a}.

Applying these rules to a generic term in the effective lagrangian involving
the isoscalar fields and the nucleon field leads to (generalization
to include the pion, rho, and photon is straightforward) \cite{Fr96,Fu96a}
\beq
   \lagrang \sim g \frac{1}{m!}\,\frac{1}{n!}\,
     \bigg( {\psibar\Gamma\psi\over\fpisq M}
        \bigg)^{\mkern-2mu \ell} 
      \bigg( {\phi\over\fpi} \bigg)^{\mkern-2mu m}
      \bigg( {V\over\fpi} \bigg)^{\mkern-2mu n}
      \bigg( {\partial\ {\rm or}\ \mpi \over M}\bigg)^{\mkern-2mu p} 
      \fpisq M^2 
          \ , \label{eq:NDAgen}
\eeq
where $\psi$ is a baryon field, $\Gamma$ is any Dirac matrix, derivatives
are denoted generically by $\partial$,
and we have allowed for the possibility of 
chiral-symmetry-violating terms that contain the small parameter $\mpi /M$.
The product of all the dimensional factors then sets the scale in
terms of the pion-decay constant $\fpi$ and the nucleon mass $M$.
The overall coupling constant $g$ is dimensionless and of $O(1)$ if
naturalness holds.

These scaling rules imply that a general potential for the scalar meson can
be expanded as
\beq
      V_{\rm S} = \mssq\phi^2\bigg ({1\over 2}
              +{\kappa_3\over 3!}\,{\gs\phi \over M}
             + {\kappa_4 \over 4!}\,{\gssq\phi^2\over M^2}
             +\cdots \bigg ) \ . \label{eq:SP}
\eeq
Here we have included a factor of $1/\fpi$ for each power of
$\phi$; these factors are then eliminated in favor of 
$\gs\approx M/\fpi$, which is basically the Goldberger--Treiman 
relation \cite{Wa95}. 
Factorial counting factors are also included, since the NDA rules are
actually meant to apply to the tree-level scattering amplitude generated by 
the corresponding vertex \cite{We90a,Fu96a}.
 
The ``naturalness'' assumption states that after the dimensional factors 
and appropriate counting factors are extracted, the overall dimensionless 
coefficients [$g$ in Eq.~\eqref{eq:NDAgen}
and the $\kappa_3$, $\kappa_4$, \dots\ in Eq.~\eqref{eq:SP}]
should be of order unity.
It should be clear, however, that the preceding arguments are not a 
{\em proof\/} of naturalness, since we know of no physical law that forbids
large coefficients from appearing.
Nevertheless, without such an assumption, it is basically impossible to
construct an effective lagrangian with any predictive 
power.\footnote{The assumption of renormalizability also leads to
a finite number of parameters and well-defined predictions, but does so
by imposing {\em unnatural\/} restrictions on the lagrangian, namely, that
many parameters are identically zero in the absence of relevant symmetry
arguments.}
Until one can derive the effective hadronic lagrangian from QCD, the 
naturalness assumption must be checked by fitting to experimental data.

The second point to be addressed is that we are actually fitting the QHD
parameters based on calculations of finite-density observables (rather
than scattering observables).
One could, of course, fit the parameters to scattering observables, and
then perform corresponding calculations of nuclear properties, but we
must nevertheless understand how fitting parameters at the mean-field
level is to be interpreted in the context of EFT.
For this interpretation, we rely on the ideas of density functional
theory, which we discuss in Section~\ref{sec:rmft}.

\subsection{Nonlinear Chiral Symmetry Revisited}
Before applying the ideas of EFT to the construction of a model lagrangian,
we illustrate some basic features of nonlinear realizations of chiral
symmetry by building on our earlier results.
Return to the lagrangian of the chiral $\sigma\omega$ model, which (after
spontaneous symmetry breaking) is given in Eq.~\eqref{eq:lswafter}.
Recall that the nucleon mass $M$ is related to the vacuum expectation
value of the original scalar field (which we denote by $s_0$) through
\beq
s_0 = \frac{M}{\gpi} \ .
\eeq

We now introduce an $SU(2)$ matrix $\uU$ defined in terms of the pion fields
by
\beq
\uU \equiv \exp \Big( \frac{i}{s_0} \, \vectau\veccdot\vecpi \Big) \ ,
  \label{eq:firstU}
\eeq
as well as left- and right-handed baryon fields
\beq
\psi_L \equiv \frac{1}{2}\, (1 - \gammafivel )\psi \ , \quad
\psi_R \equiv \frac{1}{2}\, (1 + \gammafivel )\psi \ ,
\eeq
where $\psi$ is the baryon field in Eq.~\eqref{eq:lswafter}.
Consider now the following phenomenological, nonlinear generalization
$\lagrang$ of Eq.~\eqref{eq:lswafter}:\footnote{At this point, we free
ourselves from the constraint of renormalizability.}
\beqa
\lagrang &=& i\Big[\,\psibar_R \gammamul (\dmu + i\gv\Vmu ) \psi_R
              +\psibar_L \gammamul (\dmu + i\gv\Vmu ) \psi_L \Big]
        - \gpi s_0 \Big( 1 - \frac{\sigma}{s_0} \Big) 
         \Big[\, \psibar_R \uU^\dagger \psi_L 
             + \psibar_L \uU \psi_R \Big]
      \nonumber\\[4pt]
 & & \null
         +\tover12 (\dmul\sigma\dmu\sigma) + \tover14 s^2_0
          \tr\, (\dmul \uU \dmu \uU^\dagger) 
         - {\cal V} (\uU , \dmul \uU ; \sigma)
         + \tover14 \mpisq s^2_0 \tr\, (\uU + \uU^\dagger - 2)
      \nonumber\\[4pt]
 & & \null
         -\tover14 F_{\mu\nu}F^{\mu\nu} + \tover12 \mvsq\Vmul\Vmu \ .
   \label{eq:ljdw}
\eeqa
Here $\cal V$ is a meson potential built from the indicated fields and their
derivatives.
For $\mpisq = 0$, the new lagrangian is invariant under chiral
$SU(2)_{\rm L} \times SU(2)_{\rm R}$ transformations of the form
\beq
\psi_L \rightarrow \uL \psi_L \ , \quad
\psi_R \rightarrow \uR \psi_R \ , \quad
\uU \rightarrow \uL \uU \uR^\dagger \ ,
\eeq
where $\uL$ and $\uR$ are independent, {\em global} $SU(2)$ matrices, as
long as $\cal V$ is chosen to be invariant.
(The $\sigma$ and $\Vmu$ fields are unchanged.)

It is now a simple exercise to show that in the limit $s_0 \rightarrow
\infty$, in which the nucleon mass $M$ becomes very large, the two
lagrangians are identical:
\beq
\lagrang \ \longrightarrow \ \lagrang_{\sigma\omega} \ , \qquad
   s_0 \rightarrow \infty \ ,
\eeq
provided only that the potential in $\lagrang$ is chosen to take the
simple form
\beq
{\cal V} (\uU , \dmul \uU ; \sigma) \equiv  \frac{1}{2}\, \msigmasq
   \sigma^2 + O \Big( \frac{1}{s_0} \Big) \ .
\eeq
Inspection of Eq.~\eqref{eq:lswafter} shows that the meson interactions in
the $\sigma\omega$ model reduce to this form in the indicated limit.
More generally, different choices for $\cal V$ lead to different descriptions
of the spontaneous breaking of chiral symmetry, all of which preserve the
$SU(2)_{\rm L} \times SU(2)_{\rm R}$ invariance of the lagrangian.

To proceed further with $\lagrang$ of \eqref{eq:ljdw}, we implement the
following clever change of variables,
which leaves the particle content unchanged \cite{Do92}:
\beq
\uU \equiv \uxi \uxi \ , \quad
N_L \equiv \uxi^\dagger \psi_L \ , \quad
N_R \equiv \uxi \psi_R \ .
  \label{eq:clever}
\eeq
This change of variables takes the ``square root'' of $\uU$ and mixes
the pions into the nucleons in a manner reminiscent of the chiral
transformation in Eq.~\eqref{eq:newbaryon}.
Straightforward algebra now shows that the part of $\lagrang$ containing
the fermion fields becomes (we leave the remainder of $\lagrang$
in terms of $\uU$)
\beq
\lagrang_{\rm fermion} = 
         \overline N \Big[ i\gammamu (\dmul +i \uv_{\mu} + i\gv\Vmul)
             +\gammamu\gammafivel \ua_{\mu}
              -M +\gpi \sigma \Big] N  \ , \label{eq:lfnew}
\eeq
where
\beq
\uv_{\mu}      \equiv  
           -{i \over 2}(\uxi^{\dagger} \partial_{\mu} \uxi +
    \uxi \partial_{\mu} \uxi^{\dagger} ) \ , \quad
\ua_{\mu}      \equiv  -{i \over 2}(\uxi^{\dagger} \partial_{\mu} \uxi -
    \uxi \partial_{\mu} \uxi^{\dagger} ) \ .
   \label{eq:avdef}
\eeq

The new form of the lagrangian is invariant under the following nonlinear
chiral transformation \cite{Do92}
\begin{eqnarray}
\uxi(x) &\rightarrow& 
    \uL \uxi(x) \uh^{\dagger}(x) = \uh(x) \uxi(x) \uR^{\dagger}
               \ , \label{eq:xijdw} \\[4pt]
 N(x) &\rightarrow & \uh (x) N(x)  \ , \label{eq:Njdw}
\end{eqnarray}
where $\uh (x)$ is a {\em local\/} $SU(2)$ isospin transformation that
is defined by Eq.~\eqref{eq:xijdw} and that depends on the pion field
through $\uxi$.
This transformation implies
\beq
\ua_\mu   \rightarrow     \uh \ua_\mu \uh^{\dagger} 
               \ , \qquad\qquad
\uv_\mu   \rightarrow     \uh \uv_\mu \uh^{\dagger}
                   -i \uh\partial_\mu \uh^{\dagger} \ .
\eeq
By using the relation $\dmul (\uh^\dagger \uh ) = 0$, one also obtains
\beq
(\dmul + i \uv_\mu ) N \rightarrow \uh [(\dmul + i \uv_\mu ) N] \ , 
\qquad\qquad
\uU \rightarrow \uL \uU \uR^\dagger \ ,
  \label{eq:firstcov}
\eeq
which means that the indicated derivative of the nucleon field transforms
{\em covariantly} [that is, the same way as the nucleon field in 
Eq.~\eqref{eq:Njdw}], and the pion matrix $\uU$ transforms {\em globally}.
Thus the remaining parts of the lagrangian $\lagrang$ remain unchanged under
the transformation (except for the symmetry-violating
term proportional to $\mpisq$).

This new realization of the chiral symmetry has the following important
properties:
\begin{itemize}
\item
Parity is conserved.
\item
The fermions (baryons) appear in isospin multiplets.
\item
A baryon mass term is allowed in the lagrangian.
\item
The Goldstone bosons (pions) arising from the spontaneous chiral symmetry
breaking enter through the chiral matrices $\uU$ and $\uxi$.
\item
The pions are coupled to the nucleons with derivative couplings
[Eqs.~\eqref{eq:lfnew} and \eqref{eq:avdef}]; 
hence one reproduces all the soft-pion results implied by chiral symmetry.
\item
The isospin transformation $\uh (x)$, which is an element of the 
{\em unbroken\/}
$SU(2)_{\rm V}$ subgroup of the full $SU(2)_{\rm L} \times SU(2)_{\rm R}$
group, is {\em local}: it depends on $x$ because the pion fields contained
in $\uxi$ depend on $x$.
\item
The scalar field $\sigma$ (which still appears in the theory) and vector
field $\Vmu$ are chiral scalars.
They can therefore be removed without destroying the invariance of the
lagrangian, or alternatively, additional chiral scalars can be included.
\item
One can rewrite the preceding results in conventional form by identifying
\beq
s_0 = \fpi \ .
\eeq
\end{itemize}
\noindent
In the elegant papers \cite{Ca69,Co69}, it is shown that {\em any\/}
nonlinear realization of ${\rm SU}(2)_L \times {\rm SU}(2)_R$ with these
properties can be brought into the form of Eqs.~\eqref{eq:xijdw} and
\eqref{eq:Njdw}.



\subsection{A Nonlinear Chiral Model with Vector-Meson Dominance}
\label{sec:VMD}

We now show how the ideas of EFT can be combined with the hadronic
phenomenology discussed in Sections~\ref{sec:qhd}--\ref{sec:few} by
constructing a nonrenormalizable, effective lagrangian that realizes chiral
symmetry in a nonlinear fashion \cite{Fu95,Fu96a}.
Vector mesons are included in a manner consistent with both
chiral symmetry and vector-meson
dominance (VMD), and a light scalar field is introduced as before to
simulate the exchange of two correlated pions between nucleons.
Although the dynamics described by these non-Goldstone bosons could 
also be generated through pion-loop diagrams, the advantage is
that we can now avoid (at least initially) the evaluation of complicated
loop integrals in studying the many-body problem.
Moreover, the low-energy electromagnetic structure of the nucleon is
described {\em within the theory\/} using VMD, so that {\em ad hoc\/} form
factors are not needed, as in the calculations of Section~\ref{sec:finite}.

The general framework for  nonlinear realizations of chiral symmetry is
stated very compactly in the original work of Callan, Coleman, Wess, and
Zumino (CCWZ) \cite{Ca69,We96}, and here
we paraphrase their discussion.
(The results of the previous subsection illustrate a specific example
of these ideas.)
We assume that $G$ is a compact, connected, semisimple Lie group that has
a continuous subgroup $H$.
We denote by $V_i$ and $A_\ell$ a complete, orthonormal set of generators
of $G$, such that $V_i$ are the generators of $H$.
In some neighborhood of the identity of $G$, every group element 
$g \in G$ can be uniquely decomposed as
\beq
 g = \e^{\mkern2mu\alpha \cdot A}\, \e^{\mkern2mu\beta \cdot V}
    \ , \label{eq:gofG}
\eeq
where $\alpha \cdot A = \sum_\ell \alpha_\ell A_\ell$,
$\beta \cdot V = \sum_i \beta_i V_i$, and $\alpha_\ell$ and $\beta_i$
are real constants.

A nonlinear realization of $G$ that becomes a linear representation when
$g \in H$ is given on the local field variables $(\omega , \psi )$ by
the transformation
\beq
 (\omega , \psi ) \longrightarrow g(\omega , \psi ) = (\omega' , \psi' ) \ ,
\eeq
where
\beq
 g \e^{\omega \cdot A} = \e^{\omega' \cdot A}\, \e^{u' \cdot V} \ ,
   \quad \psi' = D(\mkern-2mu\e^{u' \cdot V})\psi \ ,
            \label{eq:CCWZkey}
\eeq
$\omega' = \omega' (\omega ; g)$, and $u' = u' (\omega ; g)$.
Here $\omega$ represents the Goldstone boson fields, $\psi$ denotes
the other fields in the theory, and $\omega'$ is generally a nonlinear
function of $\omega$.
$D(h)$, with $h \in H$, is a linear, 
unitary representation of the unbroken subgroup $H$,
which is assumed to be written in fully reduced form.
In \cite{Co69} it is proved that any nonlinear realization of $G$ that is
linear on $H$ can be brought into the preceding form through a suitable
redefinition of field variables.

Unlike a {\em linear\/} realization of the symmetry, in which all particles
must be assigned to a representation of the full group $G$, the non-Goldstone
fields $\psi$ in a {\em nonlinear\/} realization are in representations
of the unbroken subgroup $H$ only, just as in the example in the previous
subsection (where the unbroken group is the $SU(2)_{\rm V}$ of isospin).
Information about the structure of the full group $G$ is encoded in the
Goldstone fields $\omega$ and the local transformation 
$D(\mkern-2mu\e^{u' \cdot V})$, with $\e^{u' \cdot V} \in H$.
Thus a nonlinear realization eliminates the necessity of assigning particles
to chiral multiplets, as in the linear $\sigma$ model and its
extensions \cite{Se92b}.
These assignments are additional dynamical input that go beyond the 
assumption of group invariance alone \cite{Co69}.

We now proceed to construct our model following the strategy outlined at
the end of Section~\ref{sec:eftintro}.
We include fields for the nucleon, pion, rho meson, and omega meson, together
with an effective scalar-isoscalar field to simulate two-pion exchange;
the $\Delta$ will be omitted here for brevity, but its inclusion is discussed
in \cite{Ta96}.
The Goldstone pion fields $\pi^a(x)$, with $a=1,\, 2,\, {\rm and}\, 3\,$,
form an isovector, which can be considered as the phase of a chiral rotation
matrix:
\beq
 \uxi (x) \equiv \exp (i\upi(x)/ \fpi)  \ , \quad
 \upi (x) \equiv {1\over 2}\, \vectau \veccdot \vecpi (x)\ .
      \label{eq:newxidef}
\eeq
Here the $\tau^a$ are Pauli matrices and $\fpi \approx 93\MeV$
is the pion decay constant.
The isospinor nucleon field is represented by a column matrix
\begin{equation}
\uN (x)=\left (\,
     \begin{array}{c} 
         p(x) \\ 
         n(x)
      \end{array} \right )\ ,
\end{equation}
where  $p(x)$ and $n(x)$ are the proton and neutron fields
respectively. 
The rho fields $\rho^a_\mu(x)$ also form an isovector, and we use the 
notation $\urho_\mu(x) \equiv {1\over 2}\,\vectau \veccdot \vecrho_\mu (x)$.

A nonlinear realization of the chiral group
$SU(2)_{\rm L}\times SU(2)_{\rm R}$ is now defined such that
for arbitrary {\em global\/} matrices 
$\uL \in SU(2)_{\rm L}$ and $\uR \in SU(2)_{\rm R}$, there is a mapping
\beq
\uL\otimes \uR:\ \ \ (\uxi, \urho_\mu, \uN)\longrightarrow 
        (\uxi', \urho'_\mu, \uN')   
          \ .     \label{eq:nonlr}
\eeq
Because of the parity operation ${\cal P}$, which produces the transformation
\beq
{\cal P}:\qquad L \longleftrightarrow R\ , \quad
      \pi^a (\coords ) \longrightarrow -\pi^a (\pcoords)\ , \quad
      \uxi (\coords ) \longrightarrow \uxi^\dagger (\pcoords)\ ,
        \label{eq:piparity}
\eeq
the chiral mapping \eqref{eq:CCWZkey} can be written as [compare
Eqs.~\eqref{eq:xijdw} and \eqref{eq:Njdw}]
\begin{eqnarray}
\uxi'(x) &=& \uL \uxi(x) \uh^{\dagger}(x) = \uh (x) \uxi(x) \uR^{\dagger}
               \ , \label{eq:Xitrans} \\[4pt]
\urho'_\mu(x) &=& \uh (x) \urho_\mu (x) \uh^{\dagger}(x)
               \ , \label{eq:Rhotrans} \\[4pt]
 \uN'(x) &=& \uh (x) \uN(x)  \ .       \label{eq:Ntrans}
\end{eqnarray}
The second equality in Eq.~(\ref{eq:Xitrans}) defines
$\uh (x)$ as a function of $\uL$, $\uR$, and the local pion fields:
$
\uh (x)=\uh (\uL,\uR,\vecpi (x)).
$
It follows from Eq.~\eqref{eq:Xitrans} that $\uh (x)$ is invariant under
the parity operation \eqref{eq:piparity}, that is,
\beq
\uh (x) \in SU(2)_{\rm V} \ ,
\eeq
with $SU(2)_{\rm V}$ the unbroken vector subgroup of 
$SU(2)_{\rm L} \times SU(2)_{\rm R}$.\footnote{We can express 
$\uh$ in terms of the matrices $\uL$, $\uR$, and $\uU$ 
[see Eqs.~\protect\eqref{eq:Udef} and \protect\eqref{eq:Utrans}] as
$\uh(x) = \sqrt{{\uU'}^{\dagger}(x)}\,
\uL\sqrt{\vphantom{{U'}^{\dagger}}\uU(x)}
= \sqrt{\uR \uU^{\dagger}(x) \uL^{\dagger}}\,
\uL\,\sqrt{\vphantom{U^{\dagger}}\uU(x)}$.  Given
the decomposition \protect\cite{Co69}
$\uL=\exp(i\bbox{\alpha\cdot\tau})\exp(i\bbox{\beta\cdot\tau})$,
$\uR=\exp(-i\bbox{\alpha\cdot\tau})\exp(i\bbox{\beta\cdot\tau})$,
and  $\uh=\exp(i\bbox{\gamma\cdot\tau})$,
with $\bbox{\alpha}$, $\bbox{\beta}$, and $\bbox{\gamma}$ real,
the infinitesimal expansion of $\uh$ is determined by
$\bbox{\gamma} = \bbox{\beta} - (\bbox{\alpha}\times\bbox{\pi})/2\fpi 
  + O(\bbox{\alpha}^2, \bbox{\beta}^2, \bbox{\pi}^2)$ \cite{Me93a,Be95}.}
Equations~(\ref{eq:Rhotrans}) and (\ref{eq:Ntrans}) 
ensure that the rho and nucleon fields transform linearly under
$SU(2)_{\rm V}$ in accordance with their isospins. 
Note that the matrix $\uh (x)$  becomes constant only when 
$\uL = \uR$, in which case $g \in H = SU(2)_{\rm V}$ and
$\uh=\uL=\uR$.
The isoscalar fields $\Vmu (x)$ and $\phi (x)$ are chiral scalars and are
unaffected by both chiral and isospin transformations.

For discussing purely pionic interactions, it is convenient to define the
matrix [compare Eqs.~\eqref{eq:firstU} and \eqref{eq:clever}]
\beq
\uU(x) \equiv \uxi^2 (x) = \exp (2i \upi (x)/ \fpi ) \ , \label{eq:Udef}
\eeq
since the transformation law \eqref{eq:Xitrans} then implies
\beq
\uU(x) \longrightarrow \uU' (x) =
  \uL \uU (x) \uR^\dagger \ , \label{eq:Utrans}
\eeq
so that $\uU (x)$ {\em always transforms globally} 
[see Eq.~\eqref{eq:firstcov}].
Thus derivatives of $\uU (x)$ transform the same way as $\uU (x)$, and
chirally invariant interactions involving pions alone can be constructed from 
products of $\uU (x)$, $\uU^\dagger(x)$, and their derivatives.
As is well known, these terms can be organized according to the number of
derivatives, with the lowest-order term \cite{Wa95}
\beq
{\cal L}_2 = {1\over 4}\, \fpi^2 \tr\, (\dmul \uU \dmu \uU^\dagger ) 
          = \tr\, (\dmul \upi \dmu \upi )
     + {1\over 3\fpi^2}\, \tr\, ([\upi , \dmul \upi ]^2 ) + \cdots\ .
       \label{eq:Ltwo}
\eeq
${\cal L}_2$ determines all multipion scattering amplitudes to second order
in external momenta in terms of the single constant $\fpi$.
Terms with more derivatives $({\cal L}_4 , {\cal L}_6 , \ldots)$ can be used
to describe pion dynamics within the framework of ChPT \cite{Ga84}.
We will not need more than ${\cal L}_2$ in the model studied here.

For describing the interactions of pions with other particles, $\uU (x)$ is
not convenient, because other fields transform with the local $\uh (x)$
of the unbroken isovector subgroup $SU(2)_{\rm V}$.
It follows from the transformation laws given earlier, that interaction
terms that are invariant under {\em local\/} isospin rotations will
be invariant under {\em global\/} transformations of the full group 
$SU(2)_{\rm L} \times SU(2)_{\rm R}$.
Thus, to form chirally invariant interactions involving pions and other
fields, we need functions of the pion field that transform with $\uh (x)$
only.

The desired functions involving one derivative of the pion field can
be written as [compare Eq.~\eqref{eq:avdef}]
\begin{eqnarray}
\ua_{\mu}     & \equiv & -{i \over 2}(\uxi^{\dagger} \partial_{\mu} \uxi -
    \uxi \partial_{\mu}\uxi^{\dagger} )
          = \ua_{\mu}^\dagger \ ,\label{eq:adef}\\[4pt]
\uv_{\mu}     & \equiv & 
           -{i \over 2}(\uxi^{\dagger} \partial_{\mu} \uxi +
    \uxi \partial_{\mu}\uxi^{\dagger} )
         = \uv_{\mu} ^\dagger
            \ , \label{eq:vdef}
\end{eqnarray}
where the hermiticity follows from $\dmul (\uxi^\dagger \uxi) = 0 
= \dmul (\uxi \uxi^\dagger )$.
Under parity transformations, we have
\beq
{\cal P}:\qquad \ua_{\mu} (\coords) \longrightarrow -\ua^{\mu} (\pcoords)\ ,
    \quad
      \uv_{\mu} (\coords ) \longrightarrow \uv^{\mu} (\pcoords ) \ ,
             \label{eq:avparity}
\eeq
so that $\ua_{\mu}$ is an axial vector and $\uv_{\mu}$ is a polar vector.
To leading order in derivatives, one finds
\beqa
 \ua_{\mu} &=& {1\over f_\pi}\partial_{\mu}\upi+ \cdots \ ,\\
 \uv_{\mu} &=& 
     -{i\over 2f_\pi^2}[\upi,\partial_{\mu}\upi] + \cdots \ .
\eeqa
Moreover, under a chiral transformation, Eq.~\eqref{eq:Xitrans} implies
\begin{eqnarray}
\ua_\mu   &\rightarrow &  \ua'_\mu =  \uh \ua_\mu \uh^{\dagger} 
               \ , \\[3pt]
\uv_\mu   &\rightarrow  & \uv'_\mu =  \uh \uv_\mu \uh^{\dagger}
                   -i\uh\partial_\mu \uh^{\dagger} 
  = \uh \uv_\mu \uh^{\dagger} + i (\partial_\mu \uh ) \uh^{\dagger} \ .
\end{eqnarray}
Thus $\ua_{\mu}$ transforms {\em homogeneously\/} under the local 
$SU(2)_{\rm V}$ group and can be interpreted as a covariant derivative of
the pion-field matrix $\uxi (x)$.
In contrast, the {\em inhomogeneous\/} transformation law for $\uv_{\mu}$ 
resembles that of a gauge field, so that $\uv_{\mu}$ allows us to construct
chirally covariant derivatives of the other fields.
For example, it is straightforward to verify that the covariant derivatives
\beq
D_{\mu} \uN \equiv (\dmul + i \uv_{\mu}) \uN \ , \quad
    D_\mu\urho_\nu \equiv \dmul\urho_\nu + i [\uv_\mu , \urho_\nu]
          \label{eq:covderivs}
\eeq
transform homogeneously with $\uh(x)$ under the full group:
\beq
(D_{\mu} \uN )' = \uh (D_{\mu} \uN ) \ , \quad
(D_\mu\urho_\nu )' = \uh (D_\mu\urho_\nu ) \uh^{\dagger} \ .
\eeq

The covariant derivative of $\urho_\mu$ can be used to construct
the covariant field tensor
\beq
\urho_{\mu\nu} = D_\mu\urho_\nu - D_\nu\urho_\mu 
   + i\grho [\urho_\mu , \urho_\nu ]
      \ .   \label{eq:rhomunu}
\eeq
The antisymmetric combination of derivatives implies that the timelike
components $\rho^a_0$ of the rho field have no conjugate momenta and are
thus determined by equations of constraint, as appropriate for a massive
vector field with three dynamical degrees of freedom.
The final term in Eq.~\eqref{eq:rhomunu}
has the usual form for a non-abelian vector field and enables the
$\vecrho$ meson to couple to a conserved isovector current \cite{Se86}.
We can also construct a covariant tensor for the pion field as
\beq
\uv_{\mu\nu} = \dmul \uv_{\nu} -\dnul \uv_{\mu} + i [\uv_{\mu}, \uv_{\nu}] 
       = - i [\ua_\mu, \ua_\nu]   \ , \label{eq:vmunu}
\eeq
which transforms homogeneously with $\uh$, and which
will allow us to produce an invariant $\rho\pi\pi$ coupling through an 
interaction of the form $\tr\, ( \urho_{\mu\nu} \uv^{\mu\nu} )$.

Before exhibiting the lagrangian for the model, we consider electromagnetic 
interactions.
As discussed in \cite{Fu96a}, these can be included straightforwardly
by defining appropriate charge operators for the particles and by modifying
the preceding covariant derivatives so that they remain covariant under
the local U(1) transformations of electromagnetism.
We will not present the intermediate steps here and simply show the
contributions to the model lagrangian below.
The electromagnetic interactions induce small violations of both the isospin
and chiral symmetries.

To write a general effective lagrangian, we need an organizational scheme
for the interaction terms. 
We organize the lagrangian in increasing powers of the fields and their
derivatives, as motivated by the principles discussed earlier. 
We assign to each interaction term an index
\begin{equation}
\nu = d+{n\over 2} + b  \ ,   \label{eq:order}
\end{equation}
where $d$  is the number  of derivatives,
$n$ is the number of nucleon fields, and $b$ is
the number of  non-Goldstone boson fields in the interaction term.
The first two terms in Eq.~(\ref{eq:order}) are suggested by 
Weinberg's work \cite{We90a}.
Derivatives on the nucleon fields are {\em not counted in\/} $d$ because they
will generally introduce powers of the nucleon mass $M$, which will not lead
to small expansion parameters.
The last term is a generalization that arises because 
a non-Goldstone boson couples to {\it two} nucleon fields.
Equation~(\ref{eq:order}) is also consistent with finite-density applications 
when the density is not too much higher than the nuclear-matter equilibrium
density, as we will see below.

The effective lagrangian for the full model can be written as
\beq
  {\cal L} = {\cal L}_{\rm N} + {\cal L}_{\rm M} + {\cal L}_{\rm EM}\ ,
     \label{eq:fullL}
\eeq
where each term will be truncated by considering the various values of
$\nu$, as defined above.
Only non-redundant terms will be exhibited, and we will return to consider
the redundancy problem after describing the lagrangian.

The part of the effective lagrangian involving nucleons can be 
written through order $\nu =3$ as
\beqa
{\cal L}_{\rm N}(x) &=&
         \uNbar \Big[ i\gammamu (\dmul +i\uv_{\mu}+i\grho\urho_\mu
             +i\gv\Vmul)
             +g_{\rm \scriptscriptstyle A}\gammamu\gammafivel \ua_{\mu}
              -M +\gs\phi \Big] \uN \nonumber \\[3pt]
   & & \quad
       -{\frho \grho \over 4M}\uNbar
        \urho_{\mu\nu}  \sigmamunu \uN
         -{\fv \gv \over 4M}\uNbar\Vmunul
                  \sigmamunu \uN
           + \cdots  \ , \label{eq:lnucl}
\eeqa
where $\Vmunu \equiv \dmu V^{\nu} - \dnu\Vmu$.
Note that both vector and tensor couplings are included for the $\rho$
and $\omega$ mesons, together with a Yukawa coupling for the effective
scalar field $\phi$; we explain the motivation for a simple Yukawa coupling
below.
The ellipsis represents terms involving $\pi N$ interactions that are
not needed in the following discussion, as well as terms
with derivatives on the nucleon field, which will be considered below.
Four-nucleon contact terms are not included, since they can be represented
by appropriate powers of other meson fields \cite{Fu96a},
as discussed shortly.
Additional terms with $\nu = 4$ are either redundant or tiny.

We emphasize that because the nucleon field obeys the transformation law
\eqref{eq:Ntrans}, the mass term in Eq.~\eqref{eq:lnucl} is chirally
invariant.
Thus the nature of the spontaneous chiral symmetry breaking is 
{\em outside\/} the realm of our effective theory, unlike a linear 
realization, where the symmetry breaking is implemented by the fields in
the model [see Eq.~\eqref{eq:mdefs}].
In a linear model, the mechanism that generates the free-space mass $M$
is also responsible for shifting $M \rightarrow \Mstar$ at finite density,
which leads to the problems discussed in Section~\ref{sec:chiral}.
A nonlinear realization of the symmetry allows these two aspects of the
nucleon mass to be treated independently.

The mesonic part of the lagrangian is also organized in powers of $\nu$.
Keeping terms up to order $\nu =4$, we find\footnote{%
Note that the leading-order coupling of the scalar-isoscalar field to pions 
has the form $\phi \tr\,(\dmul \uU\dmu \uU^{\dagger})$.}
\begin{eqnarray}
{\cal L}_{\rm M}(x) &=&
                   {1\over 4}\, \fpi^2 \tr\,(\dmul \uU\dmu \uU^{\dagger})
      + {1\over 4}\, \mpisq \fpi^2 \tr\, (\uU + \uU^\dagger-2)
           + {1\over 2}\Big(
                 1  + \alpha_1 {\gs\phi\over M}
                \Big) \dmul \phi \dmu \phi
                \nonumber  \\[3pt]
       &  &  \null     
       -{1\over 2}\tr\, (\urho_{\mu\nu}\urho^{\mu\nu})
                   -{1 \over 4}\Big(
                 1  + \alpha_2 {\gs\phi\over M}
                \Big) \Vmunul\Vmunu
           -g_{\rho \pi\pi}{2\fpi^2 \over \mrhosq}
              \tr\,(\urho_{\mu\nu} \uv^{\mu\nu})
                          \nonumber  \\[3pt]
        &  &  \null  + {1\over 2}\bigg (
      1+ \eta_1 {\gs\phi\over M}  
        + {\eta_2\over 2} {\gssq\phi^2\over M^2} \bigg )
                   \mvsq \Vmul\Vmu
          +{1\over 4!}\zeta_0
             \gvsq (\Vmul\Vmu)^2
       \nonumber \\[3pt]
         &  &       \null    +
            \bigg ( 1+ \eta_\rho {\gs\phi\over M} \bigg )
             \mrhosq \tr\,(\urho_{\mu} \urho^\mu)
           - \mssq\phi^2\bigg ({1\over 2}+{\kappa_3\over 3!}\,
            {\gs\phi \over M} + {\kappa_4 \over 4!}\,
           {\gssq\phi^2\over  M^2}\bigg )
                \ ,\label{eq:NLag}
\end{eqnarray}
where we have included a small chiral-symmetry-violating term involving 
$\mpisq$.
Apart from conventional definitions of some couplings ($\gs$ and $\gv$)
and the masses, the parameters are defined so that they are of
order unity according to the naive dimensional analysis discussed earlier.
This hypothesis will be tested by fitting the parameters to nuclear 
properties.
Moreover, since the expectation value of the $\rho$ field is typically an 
order  of magnitude smaller than that of the $\omega$ field, we have retained
nonlinear $\rho$ couplings only through order $\nu = 3$.
Note that the $\alpha_1$ and $\alpha_2$ terms involve derivatives of the
meson fields and have $\nu = 5$, but these give contributions to the nuclear
surface energy that are numerically of the same
magnitude as the quartic scalar term, so we
have retained them.
Thus numerical factors such as $1/n!$, which are cancelled in scattering
amplitudes, are relevant in deciding the importance of contributions to
the energy.

Higher-order self-couplings and derivatives involving meson fields
alone [\eg $\phi^5$, or $(\Vmul\Vmu)^3$, or $(\dmul\phi\dmu\phi )^2$]
should be numerically small unless their coefficients are ``unnaturally'' 
large.
We show below that the parameters that have been retained indeed
exhibit naturalness, so that the omission of these terms is justified at
the level of accuracy we can expect in comparisons with observables in
finite nuclei.

Finally, the electromagnetic interactions are described by
\begin{eqnarray}
{\cal L}_{\rm EM}(x) &=& -{1\over 4} \Fmunu F_{\mu\nu}
         - e \uNbar \gammamu  {1\over 2}(1+\tau_3) \uN \Amul
                \nonumber \\[3pt]
    & & \quad
        -{e \over 4M}F_{\mu\nu}
         \uNbar \ulambda'   \sigmamunu \uN
           -{e\over 2M^2}
           \uNbar\gammamul(\beta^{(0)}+\beta^{(1)}\tau_3) \uN
               \partial_\nu F^{\mu\nu} \nonumber\\[3pt]
       &  &  \quad     
              -2e\fpi^2\Amu \tr\,(\uv_{\mu}\tau_3)
              -{e \over 2g_{\gamma}}F_{\mu\nu}\Big[
           \tr\,(\tau_3\urho^{\mu\nu})
            +{1\over 3}\,\Vmunu \Big]  
           + \cdots  \ , \label{eq:lem}
\end{eqnarray}
where $\Amu$ is the electromagnetic potential and $\Fmunu$ is now the usual
field tensor.
The lagrangian ${\cal L}_{\rm EM}$ is invariant under the $U(1)$ group
of electromagnetism, and the resulting current [see Eq.~\eqref{eq:EMcrnt},
below] is conserved, at least to $O(e)$.\footnote{%
The fully U(1)-invariant lagrangian is discussed in \protect\cite{Fu96a}.}
The composite structure of the nucleon is included here through an
anomalous moment $\ulambda'$ [see Eq.~\eqref{eq:2ba}, below]
and through terms that will generate a
$q^2$ dependence in both the isoscalar ($\beta^{(0)})$ and isovector
($\beta^{(1)}$) electromagnetic form factors.
Moreover, the coupling between the massive vector mesons and the photon
generate contributions to the nucleon form factor in accord with
vector-meson dominance.
The end result is momentum dependence that resembles the empirical
``dipole'' form.
Similar contributions will arise in the pion form factor due to the
$\rho\pi\pi$ coupling in Eq.~\eqref{eq:NLag}.
We will return to the electromagnetic structure of the nucleon
in this model in Section~\ref{sec:rmft}.

Turning now to ``redundant'' terms that have been omitted in the preceding
equations, we emphasize that there is considerable freedom
in the choice of generalized coordinates (fields) for the lagrangian.
It is known that a wide class of point transformations of the fields
do not change the on-shell scattering amplitudes 
\cite{Co69}.\footnote{We assume that this is also true for finite-density
observables, but we know of no proof.}
Thus the relevant question is which choice of coordinates leads to the
most practical and accurate truncation scheme, and this is currently
under active investigation.

Some redundant terms that we have omitted through our choice of field
variables are:
\begin{enumerate}
\item 
Contact terms involving nucleon fields beyond bilinear order
[\eg $(\uNbar \uN)(\uNbar \uN)$ 
or $(\uNbar\gammamu \uN)(\uNbar\gammamul \uN)$].
We observe that if one constructs the hamiltonian from the lagrangian
given above and then eliminates the meson fields using their equations of
motion, contact terms involving products of nucleon bilinears will arise.
Conversely, contact terms included originally in the hamiltonian
can be eliminated in favor of the scalar and vector fields, if we allow
products of fields to all orders.
The issue then becomes one of efficiency, since one will always have to
truncate in practice.
Although both ways of representing these nonlinear interactions may turn out
to be practical \cite{Bo77,Bo91,Ni92,Fu96}, the fits to nuclei presented 
below show that nonlinear meson interactions lead to an efficient and 
natural truncation.
\item 
More complicated meson--nucleon couplings [\eg $\gs (\phi )
\uNbar\uN\phi$]; these are often motivated by the claim that they
are necessary to incorporate the compositeness of the nucleon.
Here we rely on the freedom to redefine the fields to rewrite
complicated couplings in simple Yukawa form: rather than work with
$\gs (\phi )\uNbar\uN\phi$, where $\gs (\phi ) = \gs (1 + c_1 \phi
+ c_2 \phi^2 + \cdots )$,
we could define a new scalar field
$\widetilde\phi$ by ${\widetilde g}_{\rm s}
{\widetilde\phi} \equiv \gs (\phi )\phi$.
Inversion of this relation and substitution into the lagrangian would
result in additional nonlinear interactions in powers of $\widetilde\phi$ 
that have the same form as those that have already been included.
Note that this procedure can actually be used on all possible 
scalar-isoscalar field combinations, so the following couplings are all
redundant:
\[
             \uNbar  \uN\phi^2\, ,\quad \uNbar  \uN\phi^3
           \, ,\quad\uNbar  \uN\partial^2\phi
            \, ,\quad\uNbar  \uN\Vmul\Vmu \ .
\]
A similar observation holds for the vector-isoscalar and vector-isovector
couplings.
This discussion illustrates the important point that nucleon compositeness
can be incorporated through nonlinear meson interactions.
\item 
Couplings involving higher derivatives of the nucleon field [\eg
$(\uNbar \partial^2 \uN)\phi$, $\uNbar i \Vmu \dmul \uN$,
or $(\uNbar\dmul \uN)(\uNbar\dmu \uN)$].\footnote{%
Higher-derivative terms should of course be written in terms of
{\em covariant\/} derivatives [see Eq.~\protect\eqref{eq:covderivs}].
The contributions from the meson fields to these terms have already been
classified; here we focus on the gradient pieces.}
These are the most problematic, since derivatives acting on
the nucleon field produce factors of the nucleon energies $E_i$,
and since $E_i / M \approx 1$, this would spoil the expansion and
truncation procedure outlined above.
Fortunately, through partial integration, redefinition of the baryon field,
and the use of the equations of motion in the construction of the 
hamiltonian, these terms can be recast in the form of the terms we have
retained or can be shown to produce terms that should give only small
contributions.
Further discussion is contained in \cite{Ba88,Ge91,Fu96a} and in
\S7.7 of \cite{We95}.
\end{enumerate}

The mean-field equations and energy density resulting from the lagrangian 
\eqref{eq:fullL} can be derived straightforwardly.
For symmetric nuclear matter ($\gamma = 4$), one finds through order
$\nu = 4$
\newpage
\begin{eqnarray}
     {\cal E}[\Phi, W; \rhoB ]
        &=& 
           W\rhoB 
     + {4\over (2\pi)^3}\! \int_0^{\kfermi}
       {\kern-.1em}{\rm d}^3{\kern-.1em}{k} \,
           \sqrt{{\bf k}^2+\Mstarsq}
       + {1\over \gssq}\Big({1\over 2}+
               {\kappa_3\over 3!}{\Phi\over M}
              +{\kappa_4\over 4!}{\Phi^2\over M^2}\Big)\mssq\Phi^2
        \nonumber\\[4pt]
     & & \quad
        -{1\over 2\gvsq}
         \Big(1+\eta_1{\Phi\over M}+{\eta_2\over 2}{\Phi^2\over M^2}\Big)
                   \mvsq W^2
      - {1\over 4! \gvsq} \zeta_0 W^4 \ , \label{eq:vmdnm}
\end{eqnarray}
where $\Phi = \gs\phizero$ and $W = \gv \Vzero$ are the scaled fields
defined earlier.
One can also compute the bulk symmetry-energy coefficient \cite{Se86}
\begin{equation}
  a_4 = {\grhosq \over 12\pi^2 {\mrho^*}^2}\,\kfermi^3
          + {1 \over 6} {\kfermi^2
          \over \sqrt{\kfermi^2+\Mstarsq}} \ , \label{eq:vmdsym}
\end{equation}
where the effective rho mass $\mrho^*$ is defined by 
\begin{equation} 
{\mrho^*}^2 \equiv \mrhosq (1 + \eta_\rho \Phi/M)  \ . \label{eq:mrhostar}
\end{equation}
A comparison with Eq.~\eqref{eq:endens} shows that the nuclear matter energy
has been generalized to include additional nonlinearities that are not
allowed in the renormalizable model QHD--I.
The fields $\Phi$ and $W$ are again determined by extremization.

The Dirac--Hartree equations for finite nuclei can also be derived using
the procedures in Section~\ref{sec:finite}.
The resulting equations are lengthy and will not be reproduced here; the
interested reader is referred to \cite{Fu96a} for details.
One important result is that due to the additional nonrenormalizable 
interactions between the nucleon and the electromagnetic field, and also due
to vector-meson dominance, the computed nuclear charge density 
automatically contains the effects of nucleon structure, and it is 
unnecessary to introduce an {\em ad hoc\/} form factor.

It is clear from the discussion in Section~\ref{sec:qhd} that the present
model has more than enough parameters to give an accurate reproduction
of nuclear properties.
The more important question is whether the parameters fitted to nuclei
are natural.
In \cite{Fu96a}, the parameters were determined by calculating a set of
observables \{$X_{\rm th}^{(i)}$\} for each nucleus
and by adjusting the parameters to minimize the generalized $\chi^2$
defined by \cite{Ni92}
\begin{equation}
\chi^2 = \sum_{i} \sum_{X}
               \bigg[{ X_{\rm exp}^{(i)}-X_{\rm th}^{(i)}
                      \over W_{X}^{(i)} X_{\rm exp}^{(i)}} 
                \bigg]^2 \ ,  \label{eq:chisq}
\end{equation}
where $i$ runs over the set of nuclei, $X$ runs over the set of
observables, the subscript ``exp'' indicates the experimental value
of the observable, and $W_{X}^{(i)}$ are the relative weights.
The weights are chosen to be the relative accuracy expected for the
given observable in a good fit.\footnote{In practice, a reasonable range
of weights was tested, and the qualitative conclusions discussed below
were always reproduced.
Some of the considerations relevant in choosing the weights are discussed
in Section~\protect\ref{sec:rmft}.}
The nuclei chosen were $^{16}$O, $^{40}$Ca, $^{48}$Ca, 
$^{88}$Sr, and $^{208}$Pb. 

A total of 29 observables and their relative weights were taken as follows:
\begin{itemize}
\item The binding energies per nucleon $\epsilon /B$, with a relative
       weight of 0.15\%
\item The rms charge radii
        $\langle r^2 \rangle^{1\over 2}_{\rm chg}$, with
         a relative  weight of  $0.2\%$
\item The d.m.s. radii $R_{\rm dms}$,
         with a relative weight of  $0.15\%$
\item The spin-orbit splittings $\Delta E_{\rm SO}$ of the
      least-bound protons and neutrons, with
         a relative  weight of $5\%$ for $^{16}$O, $15\%$ for
     $^{208}$Pb,
         $25\%$ for $^{40}$Ca and $^{48}$Ca, and
         $50\%$ for $^{88}$Sr
\item The proton energy  $E_{\rm p}(1h_{9/2})$ and the
          proton level splitting
          $E_{\rm p}(2d_{3/2}) - E_{\rm p}(1h_{11/2})$
          in $^{208}$Pb, with relative
          weights of  $5\%$ and $25\%$, respectively
\item The surface-energy and symmetry-energy
    deviation coefficients $\delta a_2$ and $\delta a_4$, each with
        a  weight of $0.08$.
\end{itemize}
The so-called diffraction-minimum-sharp (d.m.s.) radius of a nucleus is
defined to be\cite{Fr82}
\begin{equation}
 R_{\rm dms}\equiv 4.493/Q_0^{(1)}  \ ,
\end{equation}
where $Q_0^{(1)}$ is the 
three-momentum transfer at the first zero of the nuclear 
charge form factor $F(Q) \equiv F_{\rm chg}({\bf q})$ with 
$Q=|\bf q|$.
The surface-energy and symmetry-energy
deviation coefficients $\delta a_2$ and $\delta a_4$
are {\em defined\/} by fitting the difference between experimental
and calculated binding energies 
$\delta\epsilon_i \equiv (\epsilon_i)_{\rm exp} - (\epsilon_i)_{\rm th}$
according to\footnote{Note that phenomenological surface-energy and 
symmetry-energy coefficients are {\em not\/} used, so there is no direct
input from nuclear matter to the fitting procedure.}
\begin{equation}
    \delta \epsilon_i = \delta a_1\, A_i - \delta a_2\, A_i^{2/3}
         - \delta a_4\, (N_i-Z_i)^2/A_i  \ .  \label{eq:SEMF}
\end{equation}
Here $N_i$ and $Z_i$ are the number of neutrons and protons in the
$i^{\rm th}$ nucleus and $A_i = N_i + Z_i$.
(An exact fit to the energies would have $\delta a_1 =
\delta a_2 = \delta a_4 = 0$.)
The deviations $\delta a_2$ and $\delta a_4$ are included as separate
terms in Eq.~(\ref{eq:chisq}) in the form $[\delta a_i/W_{\delta a_i}]^2$, 
with $W_{\delta a_i}=0.08$.
The motivation for this choice of observables is discussed more fully in
\cite{Fu96a}.

\def\mc#1{\multicolumn{1}{c}{$\quad #1$}}
\def\zz{\phantom{0}}

\begin{table}[t]
\renewcommand{\baselinestretch}{1.0}
\caption{Parameter sets from fits to finite nuclei, as described in the text.
Note that sets W1 and Q1 include the same interaction terms as sets L2
and NLC in Table~\protect\ref{tab:params}.
}
\vspace{.1in}
\begin{tabular}[t]{ccrrrrrr}
         & $\nu$ & \mc{W1} & \mc{C1} & \mc{Q1} & \mc{Q2} 
& \mc{G1} & \mc{G2} \\
   \hline
$m_{\rm s}/M$    & 2 
         & 0.60305 & 0.53874 & 0.53735 & 0.54268 
& 0.53963 & 0.55410   \\     
$g_{\rm s}/4\pi$ & 2 
        & 0.93797 & 0.77756 & 0.81024 & 0.78661 & 0.78532 & 0.83522   \\
$\gv/4\pi$   & 2 
        & 1.13652 & 0.98486 & 1.02125 & 0.97202 & 0.96512 & 1.01560   \\
$g_\rho/4\pi$   & 2 
         & 0.77787 & 0.65053 & 0.70261 & 0.68096 & 0.69844 & 0.75467   \\
$\eta_1$        & 3 
         &         & 0.29577 &         &         & 0.07060    & 0.64992 \\
$\eta_2$         & 4 
        &         &         &         &         & $-$0.96161 & 0.10975   \\
$\kappa_3$       & 3 
  &         & 1.6698\zz  & 1.6582\zz  & 1.7424\zz  & 2.2067\zz  
& 3.2467\zz \\     
$\kappa_4$      & 4 
  &         &     & $-$6.6045\zz & $-$8.4836\zz & $-$10.090\zz\zz  
& 0.63152  \\     
$\zeta_0$       & 4 
  &         &         &         & $-$1.7750\zz & 3.5249\zz     
& 2.6416\zz    \\     
$\eta_\rho$      & 4 
        &         &         &         &        & $-$0.2722\zz  
& 0.3901\zz    \\     
$\alpha_1$      & 5 
        &         &         &         &        & 1.8549\zz     
& 1.7234\zz    \\     
$\alpha_2$       & 5
        &         &         &         &        & 1.7880\zz     
& $-$1.5798\zz \\     
$\fv /4$     &  3
         &         &         &         &        & 0.1079\zz     
& 0.1734\zz    \\
 \hline     
$f_{\rho}/4$    &  3
         & 0.9332\zz & 1.1159\zz & 1.0332\zz & 1.0660\zz & 1.0393\zz     
& 0.9619\zz    \\     
$\beta_{\rm s}$ &  4
         & $-$0.38482 & $-$0.01915 & $-$0.10689 & 0.01181 
& 0.02844 & $-$0.09328 \\     
$\beta_{\rm v}$ &  4
         & $-$0.54618 & $-$0.07120 & $-$0.26545 & $-$0.18470 
& $-$0.24992    & $-$0.45964     
\end{tabular}
\label{tab:Gparams}
\end{table}

The nucleon, $\omega$, and $\rho$ masses are taken to have
their experimental values:
$M=939\,$MeV, $\mv = 782\,$MeV, and $m_{\rho}=770\,$MeV.
(Including the heavy meson masses as free parameters produces only minor
changes in the fits.)
The anomalous magnetic moments of the nucleon are fixed at
$\lambda'_{\rm p}=1.793$ and $\lambda'_{\rm n}=-1.913$, and 
$g_\gamma=5.01$ is chosen to reproduce the experimental partial 
width $\Gamma (\rho^0 \rightarrow e^+ e^- ) = 6.8\,{\rm keV}$.
The empirical free-space charge radii of the nucleon are used
to fix $\beta_{\rm s}$, $\beta_{\rm v}$, and $f_\rho$ by solving
Eqs.~(\ref{eq:size1}) and (\ref{eq:size2}), below. 
The remaining thirteen parameters $g_{\rm s}$, $\gv$,
$g_{\rho}$, $\eta_1$, $\eta_2$, $\eta_\rho$, $\kappa_3$, $\kappa_4$,
$\zeta_0$, $m_{\rm s}$, $\fv$, $\alpha_1$, and $\alpha_2$
for the $\nu=4$ parametrization
are then obtained by optimization of the generalized $\chi^2$.

\begin{figure}[t]
 \setlength{\epsfxsize}{5.0in}
 \centerline{\epsffile{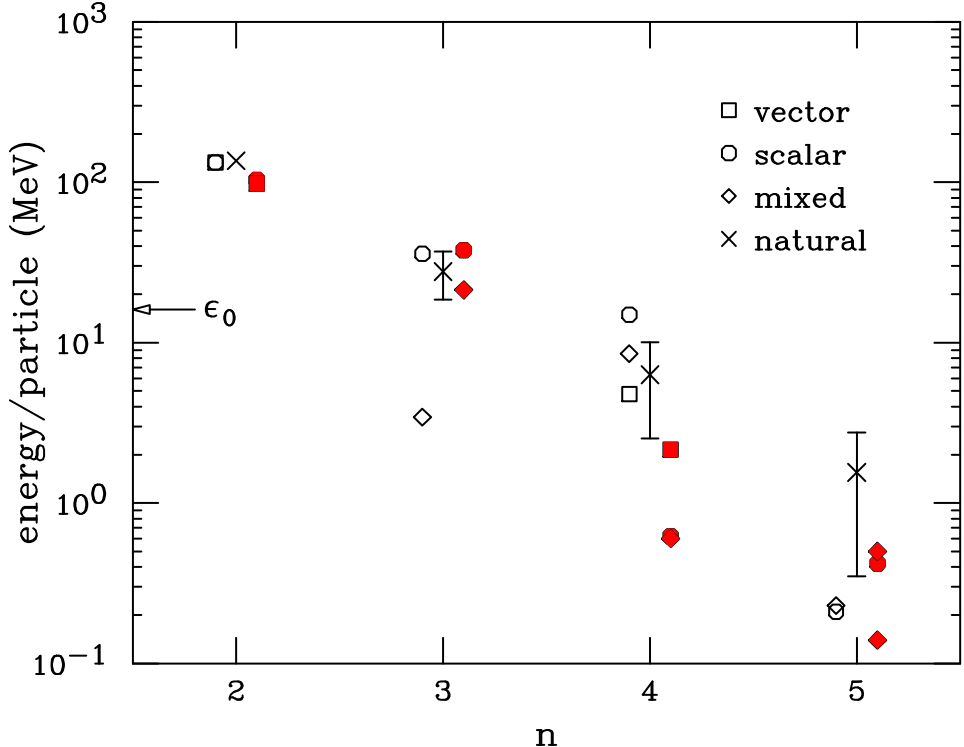}}
  \vspace{13pt}\renewcommand{\baselinestretch}{1.0}
    \caption{\protect\label{fig:nbody}%
Contributions to the energy per particle in nuclear matter for parameter
sets G1 and G2 from the $n^{\rm th}$-order terms
of the form $\Phi^\ell W^m$, where $n = \ell + m$.  
The boxes are terms with $\ell =0$, the circles are terms with $m=0$, and
absolute values are shown.
Results from set G1 are open and those from G2 are filled. 
The crosses are estimates based on Eq.~(\ref{eq:NDAgen}).
The arrow indicates the total binding energy $\epsilon_0 =16.1~\mbox{MeV}$.
}
\end{figure}

In Table~\ref{tab:Gparams}, we show two parameter sets (G1 and G2)
obtained from fits with roughly equal accuracy when all terms through
order $\nu = 4$ are retained.
The parameters have been displayed in such a way that they should all be
of order unity according to NDA and the naturalness assumption.
This is seen to be the case.\footnote{Natural parameters have also
been obtained in a ``point-coupling'' model that describes the NN
interaction through contact terms \protect\cite{Ni92,Fr96}.}
Most importantly, it is found that the accuracy of the fit and the
contributions to the nuclear matter energy/particle are not driven by
the last terms retained, as illustrated in Fig.~\ref{fig:nbody}.
This result was further checked \cite{Fu96a}
by including the $\nu = 5$ interactions
\begin{equation}
   {\cal L}_5 = 
                       - {1\over 5!}\kappa_5
                {\gs^3\phi^3\over M^3} \mssq\phi^2
                         + {1\over 3!}\eta_3 
                {\gs^3\phi^3\over M^3}\cdot
               {1\over 2}\, \mvsq\Vmul\Vmu
                         + {1\over 4!}\zeta_1
                   {\gs\phi\over M}
             \gvsq (\Vmul\Vmu)^2 \ ,
                      \label{eq:fifth}
\end{equation}
which do not improve the fits to the data; these contributions
are essentially negligible unless the coefficients are unnaturally large,
and no indication for such large parameters was found.
{\em Thus we conclude that NDA and the naturalness assumption are valid
when applied to finite nuclei, and that the truncation procedure defined
above is practical, at least for moderate densities.}
Moreover, although the parameters were obtained from a fit to a specific
set of nuclei, one can now extrapolate to study other features of nuclear
structure, such as nuclear deformations, isotope shifts in charge radii,
etc.

Also shown in Table~\ref{tab:Gparams} are the results of fits with fewer
parameters, which were obtained to verify that the best possible accuracy
for this set of input data has been achieved by keeping terms through
order $\nu =4$.
Roughly speaking, set W1 had $\chi^2 \approx 1700$, set C1 had
$\chi^2 \approx 400$, sets Q1 and Q2 had $\chi^2 \approx 100$, while
sets G1 and G2 achieved $\chi^2 \approx 50$.
Including the $\nu = 5$ parameters in Eq.~\eqref{eq:fifth} improved
$\chi^2$ only slightly (${<2}$ units).
Thus keeping terms through order $\nu = 4$ is essentially the best one
can do, and in fact, the parameters are already underdetermined at this
level, as is evident by the differences between sets G1 and G2.

\begin{table}[bt]
\caption{Nuclear matter equilibrium properties for sets
from Table~\ref{tab:Gparams} and for the point-coupling
model of Ref.~\protect\cite{Ni92} (set PC).
Values are given for the binding energy per nucleon (in MeV),
the Fermi momentum $\kfermi$ (in fm$^{-1}$), the
compression modulus $K$ (in MeV), the bulk symmetry energy coefficient
$a_4$ (in MeV), $\Mstar /M$,
and $\gv \Vzero$ (in MeV) at equilibrium.
}
\smallskip
\begin{tabular}[tbh]{cccccccc}
 Set & $E/B-M$ & $\kfermi$
            & $K$ & $a_4$ & $\Mstar M$ & $\gv \Vzero$  \\
 \hline
 W1   & $-16.46$ & 1.279 & 569 & 40.9 & 0.532 & 363 \\
 C1   & $-16.19$ & 1.293 & 304 & 32.0 & 0.657 & 255 \\
 Q1   & $-16.10$ & 1.299 & 242 & 36.4 & 0.597 & 306 \\
 Q2   & $-16.13$ & 1.303 & 279 & 35.2 & 0.614 & 292 \\
 G1   & $-16.14$ & 1.314 & 215 & 38.5 & 0.634 & 274 \\
 G2   & $-16.07$ & 1.315 & 215 & 36.4 & 0.664 & 248 \\
 PC   & $-16.13$ & 1.299 & 264 & 37.0 & 0.575 & 322 \\
\end{tabular}
\label{tab:vmdnucmat}
\end{table}

Nuclear matter properties {\em predicted} by these parameter sets are
given in Table~\ref{tab:vmdnucmat}.
Observe that sets G1 and G2 yield similar results in spite of the 
differences in the parameters, implying that the nuclear matter
properties are better determined than the parameters themselves.
Note also that all sets including parameters through orders $\nu = 3$
or $\nu = 4$ predict $\Mstar \approx 0.6\,M$ at equilibrium, in agreement
with our discussion in Section~\ref{sec:qhd}.
Although this result was obtained when spin-orbit information was
included as input, fits {\em without\/} such information leads
to similar values for $\Mstar$ \cite{Fu96a}, as was first shown in
\cite{Re86}.

\subsection{The Quantum Vacuum in QHD-I}
\label{sec:RHA}

In any consistent relativistic field theory, one must ultimately consider
loop diagrams.
These contributions are an integral part of a fully relativistic
description of nuclear structure, and as described in Section~\ref{sec:qhd},
it is impossible to construct a meaningful nuclear response or consistent 
nuclear currents without including the negative-energy  baryon states.  
Although the MFT ground state is causal and consistent with Lorentz
covariance and thermodynamics by itself, it is natural to ask about the role
of contributions from the filled Dirac sea.
This is one of the motivations for constructing QHD--I as a renormalizable
theory \cite{Wa74,Se86}.
Our goal in this subsection is to determine if the simplest evaluation of
these effects in QHD--I produces results that are consistent with NDA
and naturalness.

In Section~\ref{sec:qhd}, we studied the consequences of the mean-field 
hamiltonian of Eq.~\eqref{eq:hammft} and its generalization to finite nuclei.
Let us now return to infinite nuclear matter and include the
contribution from $\delta H$ in Eq.~\eqref{eq:deltah}.
The inclusion of this term defines the so-called relativistic Hartree 
approximation or RHA.  
(This is also often called the one-baryon-loop approximation.) 
 
An inspection of $\delta H$ reveals that, even with the indicated vacuum 
subtraction, the sum still diverges.  
Since QHD--I is a renormalizable model, however, the sum can be rendered
finite by including counterterms in the lagrangian \eqref{eq:lagrang}.  
These counterterms also appear in the
hamiltonian,  and they can be grouped with $\delta H$, resulting
in a correction to the energy density of the form %
\beq
    \Delta\edens (\Mstar)
       = -{1\over V} \sumkl \Big[ ({\kvec}^2 + \Mstarsq )^{1/2} -
                         ({\kvec}^2 + M^2)^{1/2} \Big]
      - \sum_{n=1}^4 {\alpha_n \over n!}  \phizero^n
                 \ .   \label{eq:newdeltah}
\eeq
The counterterms enter as a quartic polynomial in $\phizero$,
and the (infinite) coefficients $\alpha_n$ are determined by
specifying appropriate renormalization conditions on the energy.
Following \cite{Ch77} and \cite{Se86,Wa95}, we will choose the
counterterms to cancel the first four powers of $\phizero$
appearing in the expansion of the infinite sum.
This is equivalent to defining the renormalized parameters $\kappa$ and
$\lambda$ to be zero.
Although this procedure is not unique (and is also unnatural), it
minimizes the contributions from this vacuum
correction, and it is easy to verify that only the first four
terms in this expansion produce divergent results.  
The divergences can be defined by converting the sum to an integral
and then by regularizing dimensionally \cite{Ch77,Se86,Wa95}.
 
After removing the divergences with the counterterms, the remaining terms
are finite, and one finds (for spin-isospin degeneracy $\gamma = 4$)
\beqa
    \Delta \edens (\Mstar) &=&
      {}-{1\over 4\pi^2} \Big\{ \Mstar{}^4 \ln (\Mstar/M) + M^3
           (M - \Mstar) - {7\over 2}  M^2 (M - \Mstar)^2
              \nonumber \\[2pt]
      & &\qquad\qquad\qquad
           {}+ {13\over 3} 
           M (M - \Mstar)^3 - {25\over 12}  (M- \Mstar)^4 \Big\}
                     \\[5pt]  \label{eq:oldvf}
     &=& {M^4\over 4 \pi^2} \Big\{ {\Phi^5\over 5 M^5} 
             +{\Phi^6\over 30 M^6} + {\Phi^7\over 105 M^7} + \cdots
             + {4!(n-5)!\over n!}\, {\Phi^n \over M^n} + \cdots\Big\} \ ,
                    \label{eq:newvf}            
\eeqa
where $\Mstar \equiv M\!-\!\gs\phizero \equiv M\!-\!\Phi$.
$\Delta\edens$ is the finite shift in the baryon zero-point
energy that occurs at finite density and is analogous to the
``Casimir energy'' that arises in quantum electrodynamics.  
Just as in the MFT, $\Mstar$ is determined at each $\rhoB$ by minimization, 
which produces the one-loop
(RHA) self-consistency condition [compare Eq.~\eqref{eq:mftsc}]
\beqa
     \Mstar &=&
           M - {\gssq \over \mssq}\,  \rhos
     + {\gssq\over\mssq} {1\over \pi^2}  \Big\{ \Mstar{}^3 \ln (\Mstar/M)
              - M^2(\Mstar - M)
          \nonumber \\[3pt]
     & &\qquad\qquad\hspace{1in}
       {}-{5\over 2}  M (\Mstar - M)^2 - {11\over 6}  (\Mstar
          - M)^3 \Big\} \ .  
              \label{eq:RHAsc}
\eeqa
Note that the solution to this equation contains all orders in the coupling
$\gs$.
 
To discuss the size of the one-loop vacuum correction, we apply the NDA.
Based on the scaling rules discussed above, a term of $O(\phizero^5)$
should be scaled as
\beq
	{M^2\over 5! \fpi^3}\, \phizero^5  \ , \label{eq:good}
\eeq
and if this contribution is natural, any residual overall constant should
be of order unity.
However, if we perform a similar scaling on the leading term in
Eq.~\eqref{eq:newvf}, we find
\beq
   {M^4\over 4 \pi^2}\, {\gs^5\phizero^5\over 5 M^5} \longrightarrow
    {4\over 5}\,{M^2 \over \fpi^3}\, \phizero^5
    = 96 \bigg( {M^2\over 5! \fpi^3}\, \phizero^5 \bigg) \ ,
        \label{eq:bad}
\eeq
where we used $4\pi\fpi \approx M$ and $\gs \approx M/\fpi$.
Thus the one-baryon-loop contribution to the vacuum energy in QHD--I is
roughly two orders of magnitude {\em larger\/} than naturalness requires.
It is not hard to show from Eq.~\eqref{eq:newvf} that all higher
powers of $\Phi$ contain essentially the same large overall factor.

Similar behavior occurs in the linear sigma model.
If we rewrite the coefficients in Eq.~\eqref{eq:swNLs} in terms of the
nonlinear parameters in Eq.~\eqref{eq:SP}, we find (in the chiral limit)
$\kappa_3 = -\kappa_4 = -3$, so that the nonlinear parameters are natural
at the mean-field level.
However, if one includes the one-baryon-loop vacuum corrections,
renormalized in a fashion that preserves the chiral 
symmetry \cite{Ma82,Fu93}, one finds unnatural corrections to the
cubic and quartic couplings: 
$\Delta\kappa_3 = 2 M^2 / \pi^2 \fpisq \approx 20$, 
$\Delta\kappa_4 = -8 M^2/ \pi^2 \fpisq \approx -80$.
The quintic and higher corrections are exactly the same as in 
Eq.~\eqref{eq:newvf}.
Thus the one-baryon-loop vacuum contributions again produce unnatural
coefficients.

\begin{figure}[t]
 \setlength{\epsfxsize}{5.0in}
 \centerline{\epsffile{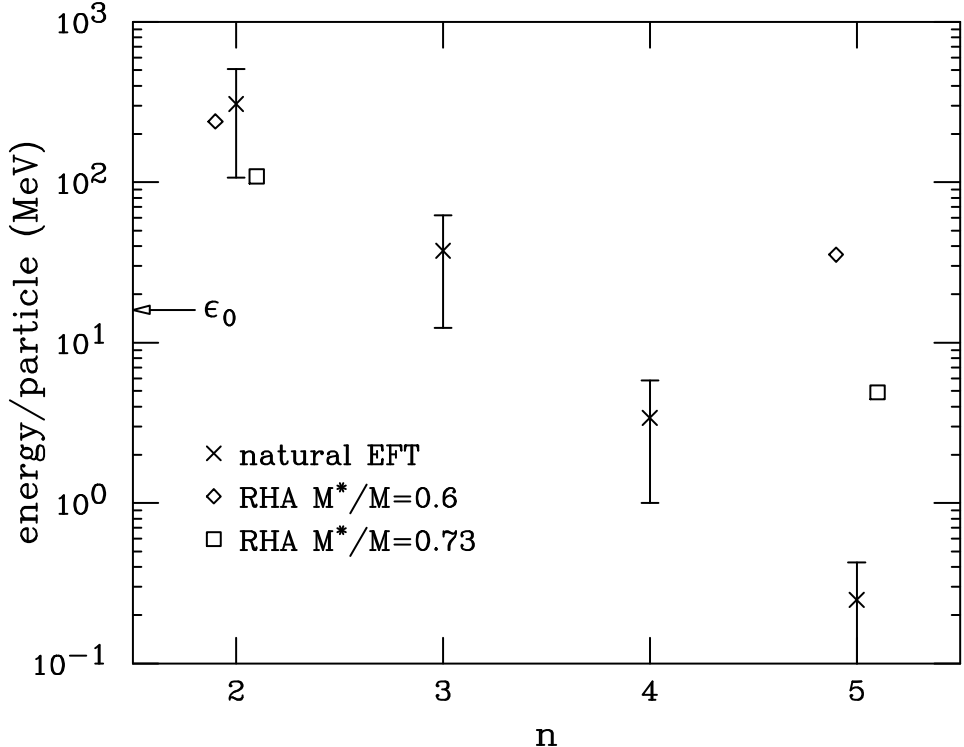}}
  \vspace{13pt}\renewcommand{\baselinestretch}{1.0}
    \caption{\protect\label{fig:rhabad}%
Contributions to the scalar potential
per particle in nuclear matter from the $n^{\rm th}$-order terms
of the form $\Phi^n$ for the RHA model.  
The crosses are estimates based on Eq.~(\ref{eq:NDAgen}).
The arrow indicates the total binding energy $\epsilon_0 =15.75~\mbox{MeV}$.}
\end{figure}

These unnatural coefficients generate correspondingly large corrections to
the MFT.
If we consider QHD--I and adjust the model parameters from their MFT values
to reproduce the desired nuclear matter properties in the RHA (equilibrium
at $\kfermi^0 = 1.30\infm$ with a binding energy of 15.75 MeV),
the baryon effective mass at equilibrium becomes $\Mstar / M \approx 0.73$.
This translates into a change in the scalar potential $\Phi$ from
$430\MeV$ in the MFT to $250\MeV$ in the RHA, which is a large effect.
In fact, the $O(\Phi^5 )$ term in the energy density forces
$\Phi$ to significantly lower values, and the contributions to the 
energy/nucleon from $\Delta\edens$ are much larger than what one would
expect from this term in a natural model, like the one discussed 
in the preceding subsection.
This is illustrated in Fig.~\ref{fig:rhabad}, where
the RHA $O(\Phi^5)$ contribution for $\Mstar /M = 0.6$ is
as large as a typical $O(\Phi^3)$ contribution in a natural model.

If one accepts the assumption of naturalness, the conclusion is that the
treatment of the quantum vacuum at the one-baryon-loop level in
the renormalizable model QHD--I is, at best, inadequate.\footnote{%
We emphasize, however, that naturalness is a strong assumption, just as
the assumption of renormalizability is.
In the latter, one sets all but a few of the model parameters to zero, while
in the former, one maintains that all parameters are roughly of the same
magnitude.}
Although higher-order corrections (some of which are discussed in
Section~\ref{sec:loops}) might reduce the size of the one-loop terms and
ultimately yield a natural size for the vacuum contributions, this can only
occur through sensitive cancellations against the one-loop terms.

In contrast, in an effective field theory, the presence of nonrenormalizable
couplings will generally cause the sum in Eq.~\eqref{eq:newdeltah} to have 
divergences {\em at all orders in $\phizero$}, and an infinite number of 
counterterms must be added to produce finite results.
(This will happen, for example, if $\Mstar = M - \Phi$ becomes a more 
complicated function of $\Phi$.)
An explicit calculation of the counterterms is unnecessary, however, 
since the end result is simply an infinite polynomial in
the scalar field, with finite, unknown, and presumably natural coefficients
arising from the underlying dynamics of the QCD vacuum.
To have any predictive power, one must rely either on the truncation scheme
discussed above, so that only a small, finite number of unknown coefficients
are relevant, or on some other dynamics to constrain the form of the
renormalized scalar potential.\footnote{%
For example, a simple model is used in \protect\cite{Fu95} to show how the
broken scale invariance of QCD leads to dynamical constraints on the scalar 
potential, and fits to the properties of finite nuclei also generate
coefficients that are natural.}
There is, of course, no guarantee that an effective hadronic theory
{\em should\/} be predictive, and until one can derive the hadronic theory
from the underlying QCD, the range of predictive power must be determined
by comparison with experiment.
Moreover, since the non-Goldstone bosons are always off mass shell in
nuclear structure calculations, it is likely that the ability to make
realistic predictions depends on the choice of generalized
coordinates (fields).

In summary, even in an effective hadronic field theory, one must include
loop contributions that contain negative-energy baryon wave functions,
since it is essential
to maintain the completeness of the Dirac basis, which plays a crucial role
in the field theory \cite{We95}.
The new ingredient, compared to renormalizable models, is that there is now
an {\em infinite\/} number of counterterms.
Since the divergent contributions must also respect the symmetries in the
lagrangian, the appropriate counterterms to remove the divergences will 
always exist, but final, finite results will generally depend on an infinite
number of unknown parameters.
Thus one must rely on truncation or on additional dynamical input to 
limit the parameters to a manageable number.
We will discuss these observations further in Section~\ref{sec:loops}.



\section{Relativistic Mean-Field Theory}
\label{sec:rmft}

The background for relativistic mean-field theory (RMFT) and its 
application to finite nuclei\footnote{We sometimes use the terminology
``Dirac--Hartree theory'' for the RMFT of finite nuclei.}
is developed in detail in \cite{Se86,Se92,Wa95}; we have summarized 
this material in Section~\ref{sec:qhd}.  
Important background references also include
[Bo77,Ho81,Re86,
Fu87,Ru88,Fu89c,Re89,Ga90,Fu91a].
The main conclusion from this material is that by using
local relativistic Hartree equations for the baryon and
meson fields $(\psi, \sigma, \omega)$, with linear couplings
and a minimal set of parameters fitted to the properties of nuclear
matter, {\em one derives the nuclear shell model}.  

Moreover, by including phenomenological nonlinear meson
couplings of the form $(\kappa/3!)\phi^3 + (\lambda/4!)\phi^4$,
one can extend this model, fit the nuclear compression modulus and baryon 
effective mass \cite{Bo91,Fu96}, and obtain an accurate description of 
nuclear deformations in light 
nuclei \cite{Bo77,Bo84,Re86,Fu87,Ru88,Fi89,Se92,Va92}.

\subsection{Density Functional Theory}
\label{sec:DFT}

To understand the success of the RMFT and to put the
calculations discussed in this section in context, it us useful to consider
some elements of density functional theory (DFT).
A discussion of DFT applied to nonrelativistic systems can be found, for
example, in \cite{Dr90}.
Speicher, Dreizler, and Engel \cite{Sp92} extend the framework to study
the relativistic many-body problem.
These authors outline the density functional approach to the 
strong-interaction model of QHD and apply the Hohenberg--Kohn theorem, 
widely used in {\em ab initio\/} calculations of the structure of solids, 
to this situation.
The nonrelativistic version of this theorem can be stated as follows:
The ground-state expectation value of any observable is a {\em unique\/}
functional
of the exact ground-state density; moreover, if the expectation value
of the hamiltonian is considered as a {\em functional\/} of the density, the
exact ground-state density can be determined by minimizing the energy
functional.

Speicher, Dreizler, and Engel derive an approximate energy functional by
using a gradient expansion of the noninteracting kinetic energy to 
order $\hbar^2$.
The energy functional includes the effects of four-vector 
meson exchange and of vacuum contributions, and the variational equations
of the corresponding extended Thomas--Fermi model are discussed.  
A great deal of work has been done on the extended Thomas--Fermi model, 
where it is relatively straightforward to generalize the results of infinite 
nuclear matter calculations to finite systems.
The extended Thomas-Fermi approach is discussed in detail in
\cite{Ce92b,Mu92,Vo92b,Ce93a,Ce93b,Sp93,Ce94,Ha94,Vo94a,Vo94b,Sc95,Sc95a}.

In more general terms, the central object in a DFT formulation of the 
relativistic nuclear many-body problem is an energy functional of scalar 
and vector densities (or more precisely, vector four-currents).
Minimization of the functional gives rise to variational equations that
determine the ground-state densities.
By introducing a complete set of Dirac wave functions, one can recast these
variational equations as Dirac equations 
for occupied orbitals; the single-particle hamiltonian contains
{\it local\/} scalar and vector potentials, 
not only in the Hartree approximation, but in the general case as well.%
\footnote{Note that the Dirac eigenvalues do not correspond precisely
to physical energy levels in the general case \protect\cite{Dr90,Sp92}.}
Rather than work solely with the Dirac wave functions and the resulting
densities (as in \cite{Ni92,Sc95,Fr96}),
one can introduce auxiliary fields corresponding to the local potentials,
so that the energy functional depends also on classical meson fields.
The resulting DFT formulation produces field equations that resemble those
in a Dirac--Hartree calculation, but correlation effects can be included, 
{\it if\/} the proper functional can be found.

The procedure described above
is analogous to the well-known Kohn--Sham \cite{Ko65}
approach in DFT, which is based on the following theorem \cite{Dr90}
(generalized here to relativistic systems):
\begin{quote}
The exact ground-state scalar and vector densities, energy, and chemical
potential for the fully interacting many-fermion system can be reproduced by
a collection of (quasi)fermions moving in appropriately defined local, 
classical fields.
\end{quote}
In the QHD case, the local scalar and vector fields play the role of 
(relativistic) Kohn--Sham potentials, and by introducing nonlinear couplings 
between these fields, one can implicitly include additional density
dependence in the single-particle potentials, as well as the composite
nature of the nucleon \cite{Fu96a}.
Thus, even though the Dirac nucleons in an RMFT calculation move in local,
classical potentials, this does not preclude an {\em exact\/} description of
the observables mentioned in the theorem.

The exact energy functional has kinetic-energy and Hartree parts 
(which are combined in the relativistic formulation) plus 
an ``exchange-correlation'' functional, which is a
nonlocal, nonanalytic functional of the densities that contains all the
other many-body and relativistic effects \cite{Mu96}.
Rather than try to {\em construct\/} the latter functional from the
lagrangian using explicit many-body techniques \cite{Se86,Se92,Sc95}, the
basic idea behind the RMFT
approach is to {\em approximate\/} the functional 
using an expansion in classical meson fields and their derivatives, based 
on the observation that the ratios of these quantities to the nucleon mass
are small, at least up to moderate density.\footnote{Since the meson
fields are roughly proportional to the nuclear density, and since the
spatial variations in nuclei are determined by the momentum distributions
of the valence-nucleon wave functions, this organizational scheme is 
essentially an expansion in $\kfermi /M$, for $\kfermi$ corresponding
to ordinary nuclear densities.
Here the nucleon mass $M$ is the generic large mass scale characterizing
physics beyond the Goldstone bosons.}
The parameters introduced in the expansion can be fitted to experiment, and 
if we have a systematic way to truncate the expansion, the framework is 
predictive.

Thus a conventional RMFT energy functional fitted directly to nuclear 
properties, if allowed to be sufficiently general, will automatically 
incorporate effects beyond the Hartree approximation, such as those due to
short-range correlations.
Future work can then be focused on the explicit inclusion of higher-order
many-body effects (as discussed later in this work), to examine the accuracy 
and limitations of the relativistic Kohn--Sham approach.

Why should we expect an approximate, mean-field functional to work well?
We observe that while the mean scalar and vector potentials $\Phi$ and $W$ 
are small compared to the nucleon mass, they are large on nuclear energy 
scales\cite{Bo91,Fu95}.
Moreover, as is illustrated in Dirac--Brueckner--Hartree--Fock (DBHF)
calculations \cite{Ho87,Te87,Ma89},
the scalar and vector potentials (or self-energies) are nearly 
state independent and are nearly equal to those obtained in the Hartree 
approximation.
Thus the Hartree contributions to the energy functional should dominate,
and an expansion of the exchange-correlation functional in terms of mean
fields should be reasonable.
This ``Hartree dominance'' also implies that it should be a good 
approximation to associate the single-particle Dirac eigenvalues with the 
empirical nuclear energy levels, at least for states near the Fermi 
surface \cite{Dr90}.

We also observe that the nuclear properties of interest include: 1)~nuclear 
shape properties, such as charge radii and charge densities, 2)~nuclear 
binding-energy systematics, and 3)~single-particle properties such as level 
spacings and orderings, which reflect spin-orbit splittings and shell 
structure.
Since the Kohn--Sham approach is formulated to reproduce
exactly the ground-state energy and density, and the Hartree contributions
are expected to dominate the Dirac single-particle potentials,
these observables are indeed the ones for which meaningful comparisons
with experiment should be possible.
Nevertheless, because the Dirac eigenvalues do not correspond
precisely to observed single-particle energies (except exactly at the Fermi
surface), we should not expect to reproduce spin-orbit splittings at
the same level of accuracy as rms charge radii and total binding 
energies.\footnote{%
These arguments are relevant to the choice of weights for the fitting
procedure described in Section~\protect\ref{sec:VMD}.}

As discussed in Section~\ref{sec:VMD}, an RMFT energy functional of 
the form in Eq.~\eqref{eq:Efunc}, extended to include meson 
self-interactions as in Eq.~\eqref{eq:vmdnm}, successfully reproduces these 
nuclear observables with parameters of natural size \cite{Fu96a}.
This justifies a truncation of the energy functional at the first few
powers of the fields and their derivatives, as is evident from
Fig.~\ref{fig:nbody}.
Moreover, the full complement of parameters is underdetermined, so keeping
only a subset does not preclude the possibility of
a realistic fit to nuclei.
Both the early RMFT calculations mentioned above and the newer calculations
discussed below should be interpreted within the context of this Kohn--Sham
approach to DFT, since they typically involve truncation at low powers of
the fields and include only a subset of the possible parameters.
We also emphasize that the Dirac--Hartree approach to nuclei is not really
a Hartree approximation in a ``strict'' sense, in which one would determine
the parameters in the lagrangian from other sources (NN scattering, for
example) and then solve the mean-field equations for nuclei with the 
{\em same\/} parameters.
The DFT interpretation implies that the model parameters fitted to nuclei
implicitly contain effects of both short-distance physics and many-body
corrections.

\subsection{Nuclear Structure}

Recent applications of these concepts exist for a wide variety of
nuclei.
Properties of light nuclei with neutron halos
are examined in \cite{Ta92,Zh94}.  The single-particle structure
of odd-$A$ nuclei is discussed in \cite{Fu89c,Pe94,Wa94,Ne95}.
``Islands of inversion'' in neutron-rich Ne, Na, and Mg nuclei are
studied in \cite{Pa91}.  Exotic nuclei near $Z=34$ and the
proton drip line, which play a role in the nuclear $r$-process and
in astrophysics, are studied in \cite{Sh93b,Ga94a}.  The shapes of
nuclei with $N=Z$ and $20\leq A\leq 48$ are discussed in \cite{Pa93b,Ma96a},
those of superdeformed Hg isotopes, in \cite{Pa94a}, and superdeformation
for $140\leq A\leq 150$, in \cite{Af96}.
The shapes of neutron-deficient Pt isotopes are discussed in \cite{Sh92a};
similar discussions exist for Pt, Hg, and Pb in \cite{Yo94}, for
Sr and Zr in \cite{Ma92a}, for Sr and Zn isotopes near the proton
drip line in \cite{Ma92b}, for Sn in \cite{En93a,Ho94a}, for Ho
in \cite{Pa93}, and for rare-earth nuclei in \cite{La96}.
Hexadecapole moments of Yb isotopes are examined in \cite{Pa95}.  
Exotic Ba isotopes are studied in
\cite{Sh93c}, and superheavy nuclei are studied in
\cite{Bo93a,La96a}.
Fission barriers in heavy nuclei are discussed in \cite{Ru95}.

Light nuclei are also examined in \cite{Pa93a}, and the effects of
pairing are included in \cite{Pa93c}.
Kinks in the isotope shifts of charge radii near $Z=40$ are
examined in \cite{La95} and for the Pb isotopes, in \cite{Sh93a}.
They appear to come out quite naturally in the RMFT.  

The general conclusion of this body of work is that the 
relativistic mean-field theory provides an economical means of
describing much of the structure of observed nuclei and a
relatively reliable way to extrapolate to new regions of nuclear structure.

The charged mesons $(\vecpi,\vecrho)$ first enter these calculations at
the relativistic Hartree--Fock (HF) level, where the exchange
interaction is included.  Systematic studies at this level are
contained in \cite{Vo92a,Sh93,Zh93,Su93,Bo94,Ha94,Sc95a}.
In nonrelativistic Hartree--Fock calculations with Skyrme
interactions, one makes a phenomenological fit to nuclei using a 
contact NN interaction containing various powers of the density.
(A computer code to carry out such calculations now exists in the literature
\cite{Re91}.)  
Comparisons between the RMFT and Skyrme
calculations of various nuclear properties are presented in
\cite{Ma92b,Sh92,Su93,Pe94,Sh94}.

The effects of retardation and of medium modifications to the Dirac wave
functions are examined in more detail in \cite{Mi92,Zh91,Za92,Zh92}.
The relationship to Landau Fermi-liquid theory is studied 
in \cite{Ue92,Ta93}.
Extensions to include other types of phenomenological nonlinear
self-couplings of the meson and baryon fields are discussed in
\cite{Gr92,Ni92}.

As noted above, an important goal is to relate calculations of nuclear
properties more directly to parameters determined from NN observables by
explicitly computing many-body contributions.
Because the relevant equations must be solved self-consistently, this is
a very difficult problem for finite systems, and an important advance was
made by Gmuca \cite{Gm92,Gm92a}.  
Here the DBHF
calculations of nuclear matter \cite{Br90a,De91} are {\em para\-me\-trized} 
by fitting the RMFT with scalar
and vector nonlinear self-interactions to the DBHF results for the 
energy/nucleon {\em and\/} the self-energies $\Phi$ and $W$ over a range of
densities.
(The nonlinear parameters used correspond to $\kappa_3$, $\kappa_4$, and
$\zeta_0$ in Eq.~\eqref{eq:vmdnm}.
The important advance is that it is more efficient to fit the nucleon 
self-energies than the effective NN interaction (or $G$ matrix), as
in \cite{Br92,Bo94,Ha94}.)
The effective interactions thus obtained are used in RMFT studies of the
structure of \nucleus{O}{16} and \nucleus{Ca}{40} nuclei 
{\em without the introduction of additional free parameters}.  
The calculated binding energies, single-particle spectra, and
charge radii agree reasonably well (although not completely
satisfactorily) with experimental data and present an
improvement over the nonrelativistic Brueckner--Hartree--Fock approximation.
This approach provides a framework for relating the RMFT to the DBHF
calculations of nuclear matter in a quantitative manner, and thus,
ultimately, to the free NN interaction.  
The relationship between the RMFT calculations and 
DBHF studies of nuclear
matter is also explored in \cite{Zh92,Za92,Su94}.

Lenske, Fuchs, and Wolter have also made an important contribution
\cite{Le95,Fu95b}. Here a fully covariant, density-dependent hadronic
field theory is obtained in which nonlinear effects are described
through a functional dependence of meson--nucleon
vertices on the baryon field operators.  
Rearrangement self-energies arise in the baryon field equations 
from the variational derivatives of the vertices.  
Solutions are
studied in the Hartree limit and compared to the local-density approximation 
to DBHF theory.
Parametrizations of nonlinear effects in terms of the scalar density
or baryon (vector) density are discussed.
Hartree calculations for nuclei between \nucleus{O}{16} and 
\nucleus{Pb}{208} show that rearrangement corrections simultaneously improve
the description of binding energies,
root-mean-square radii, and density distributions.
This approach provides an alternative to the method used by Gmuca for
connecting nuclear observables to the NN interaction.

In Section~\ref{sec:sigma}, we noted that a linear realization of
chiral symmetry, with the usual form of symmetry breaking,
cannot produce successful nuclear phenomenology at the RMFT level.
This was the primary motivation for constructing a model with
a nonlinear realization of the symmetry in Section~\ref{sec:VMD},
which includes a light scalar to simulate the exchange of two
correlated pions between nucleons.
The failure of the linear models, however, lies more with the
form of the scalar potential responsible for spontaneous
symmetry breaking than with the linear realization of
the symmetry.

In \cite{He94a,Ca96}, nuclear matter and finite nuclei are 
studied in the RMFT with a chiral lagrangian that generalizes 
the linear $\sigma$ model and also accounts for the QCD trace anomaly.
A logarithmic meson potential that involves the $\sigma$ and $\vecpi$ fields
and also a heavy glueball field $\varphi$ is used to spontaneously break
both scale invariance and chiral symmetry.
The scale-invariant term that leads to an $\omega$
meson mass after spontaneous symmetry breaking is strongly 
favored to be of the form $\omega_{\mu} \omega^{\mu} \varphi^2$ 
by the bulk properties of nuclei; they also rather strongly 
constrain the other parameters.  
A reasonable description of the closed-shell nuclei oxygen, 
calcium, and lead can be achieved, and the results are
improved by including a quartic omega self-interaction term 
in the lagrangian.  
These results are consistent with the discussion at the
beginning of this section, since this linear chiral model
contains meson self-interactions corresponding to the
$\kappa_3$, $\kappa_4$, and $\zeta_0$ terms in
Eq.~\eqref{eq:vmdnm}.
(See also parameter set Q2 in Table~\ref{tab:Gparams}.)
The important point is that the use of a more general (\ie nonrenormalizable)
interaction potential provides more
freedom to adjust the self-interactions, and the number
of free parameters is sufficient to describe nuclei,
in contrast to the usual (renormalizable) sigma model, 
as discussed after Eq.~\eqref{eq:lswafter}.

In \cite{Pr94}, a single, adjustable
scalar-nonlinearity parameter is included in the energy, together with
the zero-point energy $\Delta\edens$ discussed in Section~\ref{sec:RHA}.
Although observed neutron star masses do not constrain this parameter,
it can be chosen to provide a reasonable description of bulk nuclear 
properties, when the scalar mass is roughly 600\,MeV.
In this case, however, the coefficients of the nonlinear $\Phi^3$ and
$\Phi^4$ terms are unnatural; in particular, the quartic coefficient is
unnaturally large and negative.
This is necessary to cancel the effects of the unnaturally large, positive
coefficients introduced by $\Delta\edens$ [see Eqs.~\eqref{eq:newvf} and
\eqref{eq:bad}].



\subsection{Electroweak Interactions in Nuclei}
\label{sec:ew}

\subsubsection{Electromagnetic currents in QHD}
In QHD--I there are no charged mesons.
One can, however, introduce an effective electromagnetic current, to be used 
in lowest order, that incorporates some of the nucleon's internal structure
\cite{Se86,Se92,Wa95}:
\begin{eqnarray}
     J_{\mu} & = & \bar{\psi} \gamma_{\mu} \uQ \psi
              + \frac{1}{2M}\, \partial^{\nu} \mkern-4mu\left( \bar{\psi}
                  \ulambda^{\prime} \sigma_{\mu \nu} \psi \right)
                             \ ,    \label{eq:jemone} \nonumber \\
      \uQ & = & \frac{1}{2} ( 1 + \tau_3 )\ , \qquad
      \ulambda^{\prime}  =  \lambda^{\prime}_{\rm p}\,
                \frac{1}{2} ( 1 + \tau_3)
                +\lambda^{\prime}_{\rm n}\, \frac{1}{2} ( 1 - \tau_3)
                                 \ .   \label{eq:2ba}
\end{eqnarray}
Here $\lambda^{\prime}_{\rm p} = 1.7928$ and $\lambda^{\prime}_{\rm n}
 = - 1.9131$.
This current is covariant and local, and it is conserved by virtue of the
QHD--I field equations.
It also contains the correct anomalous magnetic moment.
To include the spatial extent of the nucleon, one can introduce a single
{\em overall\/} form factor
\begin{eqnarray}
      f_{\rm s.n.}(q^2) = \left[ \frac{1}{  1 - q^2 / (855\MeV )^2 }
                                \right]^2 \ ,
                                       \label{eq:2bb}
\end{eqnarray}
where $q^\mu$ is the four-momentum transfer,
to be used in all matrix elements of the current \eqref{eq:jemone}.
This is equivalent to replacing the photon propagator $1/q^2$ with
the effective M{\o}ller potential $f_{\rm s.n.}(q^2)/q^2$.
This current can be used consistently with the RMFT solutions in QHD--I, such
as the Dirac--Hartree wave functions.  
Many applications exist \cite{Se86,Se92,Wa95}.

The use of a single function $f_{\rm s.n.}(q^2)$ assumes that the
momentum dependence of the charge and anomalous form factors is the same.
This assumption breaks down at large $q^2$, since it is actually the
Sachs form factors that scale similarly.
This observation can be implemented easily with a simple change to
$f_{\rm s.n.}(q^2)$ and $J_\mu$ \cite{Wa95}.

Alternatively, one can attempt to calculate the single-nucleon structure
by starting with the electromagnetic current in the renormalizable model
QHD--II, which contains charged $\vecrho$ and $\vecpi$ mesons, and then by
evaluating quantum loop diagrams.
This is discussed in \cite{Se86}.
The structure of the electromagnetic current in QHD--III \cite{Se92b}, a 
chirally invariant, renormalizable extension of QHD--II, is currently under 
investigation \cite{Pr96}.

In the context of effective field theory, the composite structure of the
particles is described with increasing detail by including more and more
{\em nonrenormalizable\/} interactions in a derivative expansion.
For example, the electromagnetic current obtained from Eq.~\eqref{eq:lem}
by taking $\delta\lagrang /\delta (eA_\mu)$ is, after some partial 
integration,
\begin{eqnarray}
 J^\mu &=& {1\over 2} \!\uNbar (1+\tau_3)\gammamu \uN
       +{1\over 2M}\,\dnul(\uNbar\ulambda'\sigmamunu \uN )
         -{1\over 2M^2}\,\partial^2[\uNbar
           (\beta^{(0)}+\beta^{(1)}\tau_3)\gammamu \uN ] 
          \nonumber  \\[4pt]
     & &    
         +{1\over 2M^2}\,\partial^\mu \partial^\nu [\uNbar
           \beta^{(1)}\tau_3\gamma_{\nu} \uN ] 
        +  {1\over g_\gamma}(\dnul \rho_3^{\mu\nu}
             +{1\over 3} \dnul\Vmunu )
         +2 \fpi^2\, {\rm tr}(\uv^\mu \tau_3) \ . \label{eq:EMcrnt}
\end{eqnarray}
Note that the photon can couple to the nucleon either directly or through 
the exchange of neutral vector mesons (rho or omega). 

We can determine the tree-level electromagnetic form factors of the
nucleon from the current (\ref{eq:EMcrnt}) and the 
lagrangian (\ref{eq:lnucl}). 
For spacelike momentum transfers $Q^2=-q^2$, the isoscalar and isovector 
charge form factors are
\begin{eqnarray}
F_1^{(0)}(Q^2) &=& {1\over 2} - {\beta^{(0)}\over 2}\,{Q^2 \over M^2}
     -{\gv\over 3g_\gamma}\,
       {Q^2 \over Q^2+\mvsq}+\cdots \ ,     \\[3pt]
F_1^{(1)}(Q^2) &=& {1\over 2} - {\beta^{(1)}\over 2}\,{Q^2 \over M^2}
     -{\grho\over 2g_\gamma}\,
       {Q^2 \over Q^2+\mrhosq}+\cdots  \ ,
\end{eqnarray}
and the anomalous form factors are
\begin{eqnarray}
F_2^{(0)}(Q^2) &=& {\lambda^{\prime}_p+\lambda^{\prime}_n\over 2}
     -{\fv \gv\over 3g_\gamma}\,
       {Q^2 \over Q^2+\mvsq}+\cdots \ ,   \\[3pt]
F_2^{(1)}(Q^2) &=& {\lambda^{\prime}_p-\lambda^{\prime}_n\over 2} 
     -{\frho \grho\over 2g_\gamma}\,
       {Q^2 \over Q^2+\mrhosq}+\cdots \ .
\end{eqnarray}
The corresponding mean-square charge radii are
\begin{eqnarray}
\langle r^2 \rangle^{(0)}_1 &=& 
      6\biggl( {\beta^{(0)}\over M^2}
                 +{2\gv \over 3g_\gamma \mvsq}\biggr)  \ , \qquad
\label{eq:size1} 
\langle r^2 \rangle^{(1)}_1 =
      6\biggl( {\beta^{(1)}\over M^2}
                 +{\grho \over g_\gamma \mrhosq}\biggr)\ , \\[3pt]
\langle r^2 \rangle^{(0)}_2 &=&
      {4\over \lambda^{\prime}_p+\lambda^{\prime}_n}\,
      {\fv \gv \over g_\gamma \mvsq} \ , \qquad
 \langle r^2 \rangle^{(1)}_2 =
      {6\over \lambda^{\prime}_p-\lambda^{\prime}_n}\,
      {\frho \grho \over g_\gamma \mrhosq} \ . \label{eq:size2}
\end{eqnarray}

The form factors have a contribution
from vector dominance and a correction from the intrinsic structure
of order $Q^2$, that is, to second order in a derivative expansion.
This correction is adequate for most applications to
nuclear structure; thus, when the model of Section~\ref{sec:VMD} is
applied to finite nuclei, an {\em ad hoc\/} form factor need not be
introduced.\footnote{A form factor for the pion also arises directly 
from vector-meson dominance \protect\cite{Fu96a}.}
In practice, as the values of $\gv$ and $\grho$ are varied to fit
nuclear properties, $\beta^{(0)}$, $\beta^{(1)}$, and $\frho$ are
chosen to reproduce the empirical isoscalar and isovector charge radii
and isovector anomalous radius of the nucleon.
(The parameter $\fv$ is determined from nuclear properties, since the
nucleon's isoscalar anomalous radius is poorly known.)

We note, however, that although the derivative expansion for the current is
adequate for most RMFT calculations, applications involving large
energy-momentum transfers will require additional terms of higher order in
$Q^2$.
The utility of this expansion for $Q^2 \gtrsim M^2$ is an open question that
is currently under active investigation.

\subsubsection{Weak currents in QHD}
For the effective weak currents, one can proceed analogously.
The effective weak current to be used in lowest order in QHD--I is given 
as the sum of a polar-vector and an axial-vector part by \cite{Se86,Wa95}
\beq
  {\cal{J}}_{\mu}^{\pm} = J_{\mu}^\pm + J_{ \mu 5}^\pm \ ,
\eeq
where the charge-changing, polar-vector current is defined by
\begin{equation}
     J_{\mu}^{\pm}  =  \bar{\psi} \gammamul \tau_{\pm} \psi
              + \frac{(\lambda^{\prime}_p - \lambda^{\prime}_n)}{2M} 
         \, \dnu\! \left( \bar{\psi}
                  \tau_{\pm} \sigmamunul \psi 
                             \right)\ ,
                                   \label{eq:2bd}
\end{equation}
with $\tau_{\pm}  =  \frac{1}{2} ( \tau_1 \pm i \tau_2 )$.
Similarly, the charge-changing, axial-vector current is given by

\beq
 J_{\mu 5}^{\pm}  =  F_A(0) \left( g_{\mu}^{\phantom{\mu}\nu} - 
          \frac{1}{m_{\pi}^2 + \partial^2} 
          \partial_{\mu} \partial^{\nu} \right)
   \bar{\psi} \gamma_{\nu} \gamma_5 \tau_{\pm} \psi \ .
                                   \label{eq:2be}
\eeq

The weak current so defined is covariant, it satisfies 
PCAC (partial conservation of the axial-vector current), it contains
a nonlocality implied by pion-pole dominance, and it gives the correct
result for semileptonic weak interactions on a free nucleon.

The effective weak {\em neutral\/} current for QHD--I is defined to have
the symmetry properties of the standard model \cite{Se86,Wa95}
\beq
  {\cal{J}}_{\mu}^{\mkern2mu0} =   
   J^{\mkern2mu0}_\mu + J^{\mkern2mu0}_{\mu 5} - 2 \sin^2{ 
      \theta_{\mkern2mu\rm W}} J_{\mu} \ ,
                                     \label{eq:2bf}
\eeq
where $J^0_\mu$ and $J^0_{\mu 5}$ are obtained from Eqs.~\eqref{eq:2bd}
and \eqref{eq:2be} by the replacement $\tau_\pm \rightarrow \frac{1}{2}
\tau_3$, and $J_\mu$ in the final term is the electromagnetic current.
Several applications of these currents exist \cite{Se86,Wa95}.
The weak current in the effective theory of Section~\ref{sec:VMD} remains
to be studied.

\subsubsection{Electroweak exchange currents}
Within the framework of a consistent hadronic field theory, one can 
also calculate two-body exchange currents. 
Early work on electromagnetic exchange currents is summarized and 
referenced in \cite{Se86}, where explicit expressions are given for the 
long-range pion and pair currents.
By the conserved-vector-current theory (CVC), these
also give the exchange-current contributions to the weak vector current.
Exchange currents play a crucial role in the accurate reproduction of
the high-momentum-transfer behavior of electromagnetic observables for
light nuclei, as discussed in Section~\ref{sec:NN}.
  
One can also compute the exchange-current
contributions to the weak axial-vector current.  
Instead of strict current conservation, one must now respect chiral
symmetry and PCAC \cite{Wa95}.  
While the forms of the relativistic one-body and two-body
axial-vector currents are model dependent, a simple
and useful version arises from the chirally transformed
(nonlinear) sigma model discussed in Section~\ref{sec:nlsigma}, where a 
projection operator of the form in Eq.~\eqref{eq:2be} appears directly 
in the current \cite{An96}.
The long-range pion-exchange-current contribution to the isovector
axial charge operator in this approach is given by
\beq
{\bf J}_{05} ({\bf x}_1, {\bf x}_2, {\bf x}) =
- \frac{ F_A f^2 }{ 4 \pi}[ \vectau (1) \times \vectau (2) ] 
[ \delta ( {\bf x}_1 - {\bf x} )  \vecsig_2
+ \delta ( {\bf x}_2 - {\bf x} )  \vecsig_1 ] \cdot \hat{r} \;
( 1 + x_{\pi} ) \frac{\exp{( - x_{\pi} )}}{ x_{\pi}^2} \ ,
\label{eq:2bk}
\eeq
where $x_{\pi} = m_{\pi}|{\bf r}|$, with ${\bf r} \equiv 
{\bf x}_1 - {\bf x}_2$,  and $f/m_{\pi} = g_{\pi} / 2M$.  
The Goldberger--Treiman relation, which follows from PCAC, states that 
$ MF_A(0) = - \gpi \fpi$, where $\fpi \approx 93\MeV$ 
is the pion decay constant \cite{Wa95}.
Many applications of the above expression, including additional,
shorter-range, hadronic-exchange contributions, also appear
in the literature \cite{To95}.\footnote{Equation~\protect\eqref{eq:2bk},
based on pion-pole dominance of the relevant amplitudes, contains an
extra factor of $F_A^2$ relative to that used in \protect{\cite{To95}}.}

\subsubsection{Recent developments in electromagnetic interactions}  
The simplest and most informative electromagnetic process to study
is electron scattering $(e,e^{\prime})$.  
Relativistic analyses of scattering in the region of the quasielastic peak,
corresponding to single-nucleon knockout, are found in
\cite{De92,Ji92,Ei94,Ri94,Fr94,Pi95}. 
The systematics of the location and shape of the peak, from low energies to 
the high-energy data from SLAC on \nucleus{Fe}{56}, is examined in
\cite{De92,Fr94}. In the former work, the roles of $\Mstar$ and 
exchange currents at high $q^2$ are emphasized. 
In the latter, it is shown that one must include momentum dependence in
the relativistic self-energy $\Sigma (k)$ to understand the data;
the RMFT gives a constant $\Sigma$.

The failure of the $(e,e^{\prime})$ data to satisfy the Coulomb sum
rule throughout the periodic table has long been one of
the significant puzzles in nuclear physics \cite{Wa95}.  The
correct form of the relativistic sum rule, the effects of nuclear
binding, and the use of the off-shell current are analyzed in \cite{Fe94,Ko95}. 
An important recent contribution \cite{Jo95} indicates that the solution
to this problem may now be in hand. 
In this work, the world data on inclusive quasielastic electron scattering
have been used to separate the longitudinal and transverse response
functions of  $^{56}{\rm Fe}$ and $^{12}{\rm C}$.  
The resulting longitudinal response functions lack the ``quenching'' that
has been such a problem.  
There are still some inconsistencies in the total collection of world data,
however, and this is a subject that cries for a complete experiment with
all the kinematic flexibility of CEBAF for definitive resolution.
\begin{figure}[t]
 \setlength{\epsfxsize}{5.0in}
 \centerline{\epsffile{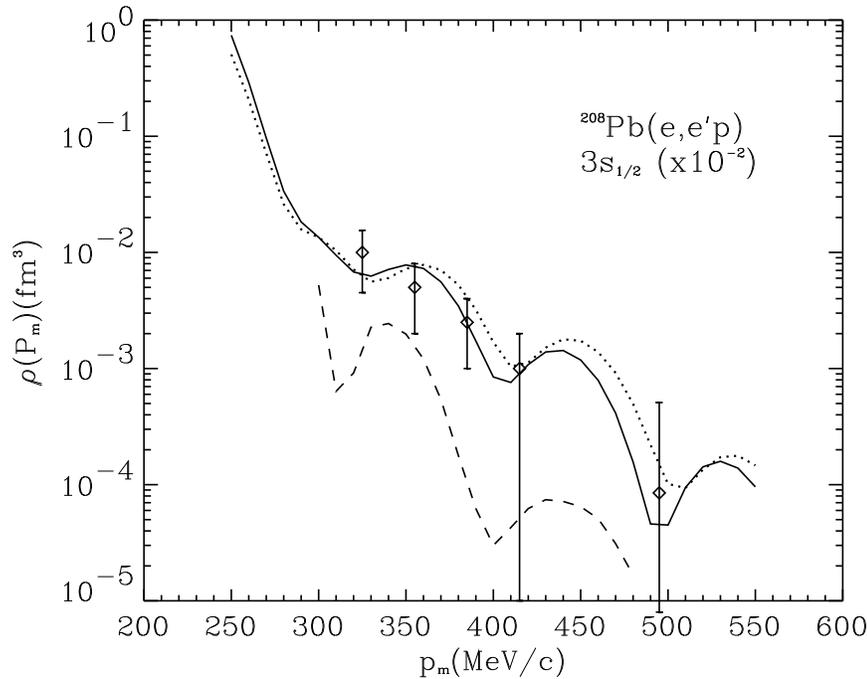}}
  \vspace{13pt}\renewcommand{\baselinestretch}{1.0}
\caption{\protect\label{fig:high}%
The ($e,e'p[3s_{1/2}]$) reduced cross section of ${^{208}}$Pb for high missing
momentum with incoming kinetic energy $E_{i}=487$ MeV
and the binding energy $E_{b}=10$ MeV. The solid and dotted lines  
are the results of Kim and Wright \protect\cite{Ki96},
who use an RMFT model with a spectroscopic factor  
fitted to low-missing-momentum results.
The solid line includes Coulomb distortion in an approximate way,
while the dotted line is the plane-wave result.
The dashed line is a nonrelativistic result by the Belgium
group \protect\cite{Va96}, which includes only the one-body current.
The data are from NIKHEF \protect\cite{Bo94b}. }
\end{figure}

With coincidence capabilities at extreme kinematics now available at Bates,
Mainz, NIKHEF, and CEBAF, one can use the reaction $(e, e^{\prime} X)$ to
probe, among other things, the high-momentum tail of the momentum
distribution in the nucleus.  
There are many effects that will contribute at high momentum transfer
and at high missing momentum, in $(e,e^{\prime}p)$ from a filled orbital.
For example, in $^{208}{\rm Pb} (e, e^{\prime} p)$ from the $ 3 s_{1/2}$ 
level, one must consider long- and short-range correlations, 
meson-exchange currents,
isobar currents, the spin-orbit interaction, other relativistic effects, etc.  
One of the prime motivations for such experiments is to disentangle and
study these effects.  
The RMFT analysis of such a process, where the initial wave function is the
solution to the Dirac--Hartree equations and the final wave function
is generated in the RIA optical potential, has the distinct advantage that
much of this physics, as in nucleon--nucleus scattering 
(Section~\ref{sec:RIA}),
is incorporated in the starting approximation.  
Figure \ref{fig:high}, for example, shows recent calculations of Kim and 
Wright \cite{Ki96} for this transition.  
This calculation, with no free parameters, accurately reproduces the existing
data. 
It is only by pushing to more extreme kinematics, with precision experiments,
and with many nuclear transitions, that one will be able to unambiguously
disentangle the role of the various additional contributions.

The subject of $(e,e^{\prime}X)$ is also discussed within the framework of
QHD in \cite{Pi92,Ji93,Ga94}.  
It is emphasized in the final paper that one must understand
relativistic hadron dynamics before drawing conclusions about new phenomena,
such as color transparency.

Relativistic analyses of the photonuclear
reactions $(\gamma,p)$ and $(\gamma, \pi^- p)$ are discussed in
\cite{Lo92,He94,Jo94}.
Other phenomena such as $(\sigma \omega \gamma)$ mixing, 
$(\rho \omega)$ mixing, charge-symmetry breaking, and the production
of $(e^+ e^-)$ pairs to probe the quark-gluon plasma are
discussed in \cite{Kr92,Li95a,Li95b}.

In a recent review article on {\em Electromagnetic Response Functions in
Quantum Hadrodynamics} \cite{We93}, elastic, inelastic, and quasielastic
electron scattering over a wide range of energy and momentum transfers
are discussed as a probe of the electromagnetic response of nuclei.
Electromagnetic response functions obtained with different QHD models of
nuclear structure and at different levels of approximation are compared
with data.  
It is shown that RPA correlations are important and have different effects
in different kinematical regimes.

A second recent review of {\em Nuclear Response in
Electromagnetic Interactions with Complex Nuclei} is given in
\cite{Bo93}.  Here the response of nucleons and complex nuclei
to an external electromagnetic probe at intermediate and high
energies is illustrated by considering both inclusive and
semi-inclusive electron scattering.  
Form factors for elastic scattering and structure functions for inelastic
scattering are derived, and several examples are discussed, including
polarization observables.

\subsubsection{Recent developments in electroweak interactions}
One can obtain information about the single-nucleon matrix elements of the 
weak current by studying quasielastic neutrino scattering $(\nu,
\nu^{\prime})$ from nuclei.  
The importance of the single-nucleon self-energies and of a realistic
(RPA) description of the nuclear excitation spectrum 
in the extraction of the axial-vector form factor of the nucleon
$F_A(q^2)$ is discussed in \cite{En93,Ki95}.

The single-nucleon matrix element of the weak axial-vector
current contains a one-pion-exchange contribution leading to the
induced pseudoscalar coupling $F_P(q^2)$, and the weak axial-vector 
exchange current also has a one-pion-exchange contribution 
\eqref{eq:2bk}; one might expect this longest-range contribution
to be modified in the nuclear medium.  A relativistic hadronic
field theory of the nucleus (QHD) allows one to
estimate these corrections.  A great deal of recent work has progressed
in this direction \cite{To92,Ga92,Ba93,Iz94,Pa94,Gi95}, much of it carried
out within the RMFT of nuclear structure.  The basic conclusions
are that weak axial-vector exchange currents can play an important role if 
one looks in the right place, for example, in first-forbidden $\beta$-decay, 
and that there can be significant nuclear-structure effects in the weak
axial-vector coupling to the nucleus.

In \cite{To92}, the ``enhancement factor'', defined as the ratio of
the axial-charge matrix element in first-forbidden
$\beta$-decay to its one-body value, is
calculated in a meson-exchange model for both light and heavy
nuclei.  Pion-exchange processes are computed for a
chirally symmetric, phenomenological lagrangian and compared with
results obtained in the soft-pion approximation [Eq.~\eqref{eq:2bk}].
Heavy-meson pair graphs are included, with
coupling constants determined from the Bonn NN potential \cite{Ma89}.
Nonrelativistic reductions are performed to both leading order and
next-to-leading order.  The results are sensitive to the
choice of the short-range correlation function used in
conjunction with harmonic oscillator wave functions.  A ratio of
transition matrix elements, however, is less sensitive to this choice.
The main result in this paper is the comparison of
a transition in a heavy nucleus with a transition in a light nucleus:
$r= \delta_{\rm mec}(A=208; 1g_{9/2} \rightarrow 0h_{9/2}) /
\delta_{\rm mec}(A=16; 1s_{1/2} \rightarrow 0p_{1/2}) =1.38$.
This result is smaller than the value of $1.58 \pm 0.09$ deduced
by Warburton \cite{Wa94a} in a fit between one-body shell-model 
matrix elements and experiment.

In \cite{Pa94}, shell-model matrix elements of the
axial-charge exchange-current operator are calculated through
next-to-leading order in heavy-fermion chiral perturbation theory.
It is found that the loop
corrections to the one-soft-pion-exchange contribution in Eq.~\eqref{eq:2bk}
are small (roughly 10\%) and have no significant
dependence on the nuclear mass number or on the valence-nucleon orbits. 

Radiative $\mu$-capture is studied in the RMFT in \cite{Fe92}.  

\subsubsection{Parity violation in $(\vec{e}, e^{\prime})$}
One of the most important areas of future research at CEBAF will be the
study of parity violation in electron scattering.
This arises from the interference between the amplitudes for photon
exchange and for $Z^0$ exchange, which involve couplings to the 
electromagnetic and weak neutral currents, respectively \cite{Wa95}.  
Feasibility has been demonstrated in pioneering experiments at Bates 
and Mainz. 

Two particularly interesting contributions are \cite{Al93,Ba94}.
In \cite{Al93}, the impact of pionic correlations and
meson-exchange currents in determining the (vector) response
functions for electroweak, quasielastic lepton scattering from
nuclei is discussed.  
The Fermi-gas model is used to maintain consistency in
treating forces and currents (gauge invariance) and to provide
a Lorentz-covariant framework.  
Results obtained in first-order
perturbation theory are compared with 
HF and RPA calculations and are found to provide quite successful
approximations for the quasielastic response functions.  
The role of pionic correlations is investigated in some detail, 
and meson-exchange currents are shown to provide a small, but 
non-negligible contribution to the vector response.

In the second paper, parity-violating quasielastic electron scattering
is studied within the context of the relativistic Fermi-gas model and
its extension to include pionic correlations and meson-exchange currents.  
The work builds on previous studies
using the same model; here, the part of the parity-violating
asymmetry that contains axial-vector hadronic currents is
developed in detail, and a link is
provided to the transverse vector-isovector response.
Various integrated observables are constructed from the
differential asymmetry, and the most favorable observables for studying
pionic correlations and the strangeness form factors of the nucleon are
determined.
Comparisons are also made with recent predictions based on the RMFT.

Nuclear-structure effects pertinent to the extraction of the
weak, neutral, axial-vector form factor of the nucleon, a quantity
of particular interest because of the potential role played by
$s \bar{s}$ quark-antiquark pairs, are emphasized in \cite{Ho93,Ho93a}.

Parity-violation in the structure of heavy nuclei is examined in the 
relativistic HF approximation in \cite{Ho94}.  
The role of nuclear structure in atomic parity-violating experiments
is discussed in the definitive work \cite{Po92a}.

\subsection{Strangeness in Nuclei}

The addition of strangeness adds another dimension to nuclear structure.  
Properties of hypernuclei with $S\!=\!-1$ are investigated in the 
RMFT in \cite{Co93,Gl93,Do95,Lo95}.  
The two primary systematic features of $\Lambda$ hypernuclei are the
relatively small depth of the central potential and the small spin-orbit
splittings.  
The scalar and vector couplings in a ($\sigma$, $\omega$)
model can be adjusted phenomenologically to describe the central
potential, but the spin-orbit splittings have been a more vexing problem. 

It was observed in \cite{Je90,Co91} that a tensor coupling of the
$\omega$ to the hyperon, with the appropriate sign, can indeed produce a 
small spin-orbit splitting.
The systematics of the interaction of hyperons with this additional tensor
coupling are investigated in \cite{Co92,Co94,Co95,Ma94,Ma95}.  
In particular, it is shown in \cite{Co94,Co95} that a relativistic optical
potential with a tensor coupling can describe the data.

If one views QHD as an effective hadronic field theory, 
such a coupling will exist in the lagrangian.
Even within a renormalizable framework, one has an induced coupling of 
this form.
While motivation for a tensor coupling appears to exist within a 
quark-model framework, the theoretical challenge is to quantitatively 
explain the difference in size between spin-orbit effects in the hyperon 
and nucleon sectors.

One interesting goal of the study of strangeness in nuclei is to
extrapolate to large $|S|$; the RMFT provides a
convenient basis for this extrapolation, and  
this problem is examined in \cite{Sc92,Do93,Ma93a,Sc93,Ga95}.  
A large class of bound, multi-strange objects is one distinct possibility
raised by these studies.

The $K^+$ has a relatively weak coupling to nucleons, and thus it
provides a useful hadronic probe for viewing the nuclear interior.  
$K^+$  scattering and production from nuclei are studied within the
framework of the RMFT in \cite{Ca92,La92,Ru92,Ko95a}.  
In \cite{Ko95b}, it is shown that $K^-$ quasielastic scattering from
nuclei can be explained within the framework of the local-density RPA
and isoscalar correlations contained in QHD--I.  
Subthreshold kaon production is discussed in \cite{Li95}, and
strong-interaction effects in $\Sigma^-$ atoms, in \cite{Ma95a}.  
Other properties of hypernuclei are examined in \cite{Lo94}.

On a more basic level, the properties of a hyperon in nuclear matter are
examined within the framework of QCD sum rules in \cite{Ji94,Ji95}.



\section{Matter Under Extreme Conditions}
\label{sec:matter}

\subsection{Relativistic Transport Theory}

One of the principal thrusts of nuclear physics has been, and
will continue to be, the use of relativistic heavy-ion
reactions to study the properties of nuclear matter under
extreme conditions of density, temperature, and flow. 
A great deal of work has been done within the framework of QHD
using relativistic transport theory to describe
heavy-ion collisions. For example, the foundations of
relativistic transport theory are discussed in
\cite{Ko87,Bl88a,Ko88,Li89a,La90,Ma92,Ma93,Mo94,Sc94} and further
developed in \cite{De91,De92a,Me93,Ni93,Mr94,Fu95a,Po95}.  
In this section, we start the discussion with a simple introduction to
transport theory.

The Vlasov--Uehling--Uhlenbeck (VUU) model describes the transport
of the distribution function in phase space.  It is a one-body
transport model that includes the effects of a long-range
mean field on the one-body dynamics, together with a short-range, two-body
collision term.  The mean field is calculated from the
distribution function, and when parametrized as a function of the
density, it probes the equation of state of the medium.
The dynamical description is classical, except that collisions are
treated stochastically, and a Pauli-blocking factor is included
for the final states in the collision term.  The approach can be
generalized to include inelastic channels. The basic ideas
behind various transport models are described in the review
article by Greiner and St\"{o}cker \cite{St86}, and a computer code to
carry out such calculations is now available \cite{Ha93a}. 
We review some of the basic concepts.

Consider the microcanonical ensemble that consists of a
collection of identical, randomly prepared, microscopic systems
(particles).  The ensemble can be characterized by the
distribution in phase space:
\begin{eqnarray}
  {\rm d}N & = & {\rm Number \; of \; members \; of \; the \; 
            ensemble \; in \; }
            \frac{{\rm d}^3p \, {\rm d}^3q}{(2 \pi)^3} 
                                                     \nonumber  \\
     & \equiv & f( {\bf p}, {\bf q}, t)\ \frac{{\rm d}^3p \,
                {\rm d}^3q}{(2 \pi)^3} \ .
                                                       \label{eq:7ba}
\end{eqnarray}
The {\em probability} to find a particle in this region of the
ensemble is the probability of picking such a member of the
ensemble at random; this is equal to ${\rm d}N/N$, where $N$ is the
total number of members in the ensemble, obtained by integrating
Eq.~\eqref{eq:7ba} over all phase space.  This probability
can be used to compute expectation values over the ensemble.

As a function of time, a particle moves from $({\bf p}_0, {\bf
q}_0)$ to $({\bf p}, {\bf q})$ in phase space.
Liouville's theorem (see, for example, \cite{Wa89}) states that 
with hamiltonian
dynamics, the volume in phase space is unchanged with time.
Since the number of particles ${\rm d}N$ in this volume is also
conserved, one concludes that the distribution function is
unchanged along a phase trajectory:
\begin{eqnarray}
  f[ {\bf p}(t), {\bf q}(t), t]  =
  f[ {\bf p}_0, {\bf q}_0, t_0] = {\rm constant}.
                                                  \label{eq:7bb}
\end{eqnarray}
Differentiation with respect to time yields
\begin{eqnarray}
  \frac{{\rm d}f}{{\rm d}t} = \frac{\partial f}{\partial t} + 
        \frac{\partial f}{\partial {\bf p}} \veccdot 
        \frac{{\rm d} {\bf p}}{{\rm d}t}
  +      \frac{\partial f}{\partial {\bf q}} \veccdot 
        \frac{{\rm d} {\bf q}}{{\rm d}t}
                = 0 \ .
                                                   \label{eq:7bc}
\end{eqnarray}
Now Hamilton's equations state
\beq
  \frac{{\rm d} {\bf q}}{{\rm d}t} = {\bf v} = 
          \frac{\partial H}{\partial {\bf p}} \ , \qquad
  \frac{{\rm d} {\bf p}}{{\rm d}t} = {\bf F} = 
       - \frac{\partial H}{\partial {\bf q}}
                               \ ,  \label{eq:7bd}
\eeq
and insertion into Eq.~\eqref{eq:7bc} gives
\begin{eqnarray}
  \frac{\partial f}{\partial t} =
              \frac{\partial H}{\partial {\bf q}} \veccdot        
      \frac{\partial f}{\partial {\bf p}} 
 -               \frac{\partial H}{\partial {\bf p}} \veccdot        
      \frac{\partial f}{\partial {\bf q}} 
       \equiv \left\{ H,f \right\}_{\rm PB}\ .
                                                      \label{eq:7be}
\end{eqnarray}
The last equality identifies the Poisson bracket.  In {\em
equilibrium}, there is no time dependence to the distribution
function, and this expression must vanish.  The solution to this
condition is that $f=f(H)$, and $f$ is a constant of the motion.

A short-range, two-body Boltzmann collision term is now included.  Go to
the canonical ensemble, which consists of identical, randomly prepared
assemblies (collections of systems).
Particles are now scattered into and out of the
region in phase space in Eq.~\eqref{eq:7ba}.  Assume that 
\begin{eqnarray}
 \frac{{\rm d}f}{{\rm d}t} = 
    \left( \frac{\partial f}{ \partial t } \right)_{\rm collisions}\ .
                                                     \label{eq:7bee}
\end{eqnarray}
Momentum is conserved in the collisions, so that ${\bf p}_1 +
{\bf p}_2 = {\bf p}_1^{\prime} + {\bf p}_2^{\prime}$. Detailed
balance then states that ${\rm Rate}(i \rightarrow f) = {\rm Rate}(f
\rightarrow i)$, or $v_{12} \sigma = v_{1^{\prime} 2^{\prime}}
\sigma^{\prime}$, where $v_{ij}$ is the relative velocity.\footnote{%
Note that the number of final states in momentum space ${\rm d}^3
n_f = {\rm d}^3 p_f /(2\pi )^3$ has now been explicitly removed from the
definition of $\sigma$.}
The number of transitions per unit time in the direction 
$ i \rightarrow f$ in the assembly is given by
\beq
  \left[ \frac{\rm number \; of \; transitions}{\rm time} 
          \right]_{ i \rightarrow f}
  = \left[ f({\bf p}_1, {\bf q} ,t)\, 
     \frac{{\rm d}^3 p_1}{(2 \pi)^3}\, v_{12} \right]
        \sigma 
    \left[ f({\bf p}_2, {\bf q} ,t) \,
       \frac{{\rm d}^3 p_2\, {\rm d}^3 q}{(2 \pi)^3} \right]
                                            \ .     \label{eq:7bf}
\eeq
The first factor is the incident flux, and the last factor is the
number of target particles, both in the appropriate momentum
interval.  The Boltzmann collision term at the point {\bf q} 
can thus be written as \cite{St86}
\begin{eqnarray}
  \left(\frac{ \partial f}{\partial t} \right)_{\rm collision}
   & = & \int {\rm d}^3 p_2 \,{\rm d}^3 p_1^{\prime} 
   \,{\rm d}^3 p_2 ^{\prime}
    \,\frac{1}{(2 \pi)^6}\, \delta^{(3)}
      ( {\bf p} + {\bf p}_2 - {\bf p}_1^{\prime} - {\bf p}_2^{\prime})
    \sigma v_{12} 
                                            \nonumber \\[4pt]
   &&  \quad\times 
      [ f({\bf p}^{\prime}_1, {\bf q} ,t)
      f({\bf p}^{\prime}_2, {\bf q} ,t) -  f({\bf p}, {\bf q} ,t)
          f({\bf p}_2, {\bf q} ,t)] \ .
                                              \label{eq:7bg}
\end{eqnarray}
The last term counts particles scattered {\em out\/} of this
region in phase space, and the first term counts those particles
scattered {\em in}.  This relation will be abbreviated as
\beq
  \left(\frac{ \partial f}{\partial t} \right)_{\rm coll}
    =  \int {\rm d}^3 p_2 \,{\rm d}^3 p_1^{\prime} \,
     {\rm d}^3 p_2 ^{\prime}\,
      \frac{1}{(2 \pi)^6} \,
     [ f^{\prime}_1 f^{\prime}_2 -  f f_2]  \sigma v_{12}
     \, \delta^{(3)}
      ( {\bf p} + {\bf p}_2 - {\bf p}_1^{\prime} - {\bf p}_2^{\prime})
                                  \ .   \label{eq:7bh}
\eeq
This is the result of classical transport theory; with
zero-range collisions, the angular distributions will be isotropic.

If the one-body hamiltonian has the form
\beq
  H = \frac{{\bf p}^2}{2m} + U({\bf q}) \ ,
                                             \label{eq:7bi}
\eeq
where $U$ is a mean field, then the transport equations take the
form
\beq
   \left(\frac{ \partial f}{\partial t} \right) 
     + {\bf v} \veccdot \nabla_{\bf q} f - \nabla_{\bf q} U 
                    \veccdot \nabla_{\bf p} f =
   \left(\frac{ \partial f}{\partial t} \right)_{\rm coll} \ .
                                               \label{eq:7bj}
\eeq
These are nonlinear, integro-differential equations for the
distribution function $f$.  In equilibrium, with no time dependence,
the collision term must vanish, which  implies $ f^{\prime}_1
f^{\prime}_2 =  f f_2 $, or $ f(E_1^{\prime})  f(E_2^{\prime}) =
f(E) f(E_2) $.  Since energy is conserved in the two-body
collision cross section, one has $E+E_2 = E_1^{\prime}
+E_2^{\prime}$.  If these relations are to hold for all $E$ and $E_2$,
then $f(E)f(E_2) = g(E+E_2)$.  Differentiation with respect to $E$
and $E_2$ in turn leads to $ f^{\prime}(E) / f(E) = {\rm
constant} \equiv - 1 / \kboltz T $, which yields the Boltzmann
distribution $f(H) = \exp{ (- H / \kboltz T ) }$ as the equilibrium
solution to these kinetic equations.

In {\em molecular dynamics} calculations [with no $U( {\bf q} )$], one
simply follows the classical equations of motion of
all the members of the canonical ensemble numerically.
In contrast, in the 
{\em Boltzmann--Uehling--Uhlenbeck\/} approach [still with no
$U( {\bf q}) )$],\footnote{This is sometimes called the
Nordheim--Uehling--Uhlenbeck approach.}
one includes a Pauli-blocking term for identical
fermions that prevents them from scattering into occupied states. 
An occupied state is assumed to consist of one
particle in the unit cell ${\rm d}^3 p \,{\rm d}^3 q /(2 \pi )^3$ 
in phase space.  Thus one makes the replacement
\beq
  [f_1^{\prime} f_2^{\prime} - f f_2] \rightarrow
  [f_1^{\prime} f_2^{\prime}(1-f)(1-f_2) 
   - f f_2 (1- f_1^{\prime})(1-f_2^{\prime})] \ .
                                             \label{eq:7bl}
\eeq
The equilibrium solutions to these kinetic equations are
the familiar Fermi distribution functions.

In the {\em VUU\/} approach, one also includes the additional
long-range mean field $ U({\bf q})$ produced by the local
particle density $ \rho$, which is obtained in turn by
integrating the distribution function over ${\rm d}^3 p/(2 \pi)^3$.
Parametrization of the $\rho$ dependence of $U$ then probes
various equations of state.  

What advantages does QHD have for studying these transport equations?

\begin{itemize}

\item It provides a covariant description of the nuclear
many-body system.

\item It thus provides a basis for a relativistic treatment of
the transport equations.

\item It allows for a consistent treatment of all hadronic channels.

\item In the RMFT, it provides an excellent first approximation to
the nuclear mean field, including an energy dependence in the equivalent
nonrelativistic optical potential \cite{Se86}.

\end{itemize}

A great deal of work on relativistic transport theory, mostly in
connection with relativistic heavy-ion reactions, has been done
in the past few years.
A covariant 
Boltzmann--Uehling--Uhlenbeck (BUU) approach, the basic ideas of which have
been presented above, is developed in \cite{Bl88a,La90,Ma92,Ma94a} and
applied in \cite{Bl89,Bl91,Ko90,Ko91}. Relativistic transport
coefficients are discussed in \cite{Ha93b,Ay94,Mo94}.

The connection of the scalar and vector mean
fields to the underlying relativistic two-body theory
is explored in \cite{El92,Fu92,We92}.  Effective cross
sections in the medium are studied in \cite{Li93,Ma94b}.
Additional momentum-dependent scalar and vector potentials, which
provide a more accurate description of the optical potential,
are introduced in \cite{We92,We93a}.  Shock waves are discussed in
\cite{Mo95}. The role of the Dirac sea in such collisions is
discussed in \cite{Ju92}.
The production of kaons is examined in
\cite{Fa93,Fa93a,Fa94} and of antinucleons, in \cite{Te93,Te94,Li94}.

Interesting extensions include studies of a classical version of QHD
\cite{Bu93,Bu93a,Bu95} and of a transport theory for quarks and
mesons \cite{Zh92a}.  Experimental implications of a relativistic,
mean-field, two-fluid model are explored in \cite{Iv94,Ru94}.

\subsection{Extrapolation and Connections to QCD}

The original motivation for the development of QHD was to find a
theoretical framework that allows extrapolation of
the properties of observed nuclear matter to extreme
conditions of density and temperature, while retaining general
principles of physics, such as quantum mechanics, Lorentz
invariance, and microscopic causality.  The only consistent
theoretical framework we have for describing such a strongly
interacting quantum many-body system is relativistic quantum
field theory based on a local lagrangian density.  The first
model attempt using hadronic degrees of freedom, QHD--I,
consisted of a renormalizable theory with baryons and neutral
scalar and vector mesons.  When solved at the level of mean-field theory
(RMFT), one finds an equation of state with a minimal number of
parameters that can be fitted to the equilibrium point of
nuclear matter, which then provides a simple equation of
state at all densities $\rhoB$ and temperatures $T$.
These results have been discussed in Section~\ref{sec:qhd}.

As noted earlier, when viewed as an effective
hadronic field theory describing the underlying QCD, all
possible couplings consistent with the relevant symmetries should
be included in the lagrangian.  
Without some organizing principle, one soon loses predictive power.  
The organizing principle presented here is that while the mean
fields are large on the scale of the nuclear binding
energy, the dimensionless ratios $\Phi/M$ and $W/M$ are still
relatively small, and one can sensibly expand in these.  
This observation allows one to understand the success of
the RMFT treatment of nuclear structure. 
While providing a systematic basis for extrapolating away
from the properties of observed nuclear matter, this approach clearly limits
the extent over which one can reliably carry out this extrapolation. 

A major goal of the extrapolation to high baryon density
$\rhoB$ is to describe the properties of neutron stars. 
Extrapolations within the framework of RMFT with effective
couplings are studied in \cite{Ro92,Gl92,Ba94a,Mo94a,Pr94,Su95,Mu96}.  
In \cite{Mo94a}, cold nuclear matter is
investigated in nonlinear, mean-field, scalar--vector models
that include density-dependent meson parameters.  
This dependence can be both explicit and implicit
through the nucleon effective mass.
Interactions between the scalar and vector fields are included,
and the properties of neutron stars are investigated using the
resulting equation of state.  The implications of the model of
Zimanyi and Moszkowski \cite{Zi90} with derivative scalar couplings are
examined in \cite{Gl92a,Sa94a,De95}.  The connection between the RMFT
results and Dirac--Brueckner theory is studied in \cite{Li92}. 

A study of the uncertainties in extrapolation is contained in \cite{Mu96}, 
where the properties of high-density nuclear and neutron matter are
computed using a relativistic mean-field approximation to the
energy functional.  
Various types of nonlinearities involving
scalar-isoscalar $(\sigma)$, vector-isoscalar $(\omega)$, and
vector-isovector $(\vecrho)$ fields are introduced to parametrize 
the density dependence of the energy functional.  
The model parameters are calibrated at equilibrium nuclear matter 
density, and it is possible to build different models that reproduce 
{\em exactly\/} the same nuclear matter properties, but which yield
maximum neutron star masses that differ by as much as one solar mass,
even when the nonlinear parameters are restricted to be of natural size.
Moreover, with enough nonlinear couplings, one can reduce the predicted
maximum neutron star mass to $M_{\rm max}/M_{\odot} \approx 1.6$, 
which is only 10\%
larger than that of the most massive observed neutron stars.
Implications for the existence of kaon condensates or quark cores
in neutron stars are discussed. 

Properties of neutron stars are examined within a similar RMFT framework
in \cite{Ma92c} and within DBHF theory in \cite{Hu94}. 
The interesting possibility of the transition of nuclear matter
into a ``Peierls' type'' of periodic structure is investigated
in the RMFT in \cite{Lo92a,Lo94a}.   

Another goal of QHD is to provide a reliable description of the 
hadronic phase of matter as one approaches the transition to the
quark-gluon plasma (QGP) \cite{Se92,Wa95}, assuming that such a transition
exists.
The role of the QGP in the neutron matter equation of state is examined in
\cite{Ro92} and more generally in \cite{Ma93b,Dr95,Ka95,Ri95}. 

The less spectacular, but more easily accessed liquid-gas phase
transition is studied within the framework of RMFT
in \cite{So93,Ha94a,So94,Mu95}.  
In \cite{Mu95}, a RMFT model of nuclear matter with arbitrary proton
fraction is studied at finite temperature.  An analysis is
performed of the liquid-gas phase transition in a system with
two conserved charges (baryon and isospin) using the stability
conditions on the free energy, the conservation laws, and Gibbs'
criteria for phase equilibrium.  For a binary system with two
phases, the coexistence surface (binodal) is two-dimensional, and thus
the liquid-gas phase transition is continuous (second order by Ehrenfest's
definition) rather than discontinuous (first order), as in
familiar one-component systems.  Using a mean-field equation of
state calibrated to the properties of nuclear matter and finite
nuclei, various phase-separation scenarios are considered, and the
model is applied to the liquid-gas phase transition that
may occur in the warm, dilute matter produced in energetic
heavy-ion collisions.  In asymmetric matter, instabilities that
produce a liquid-gas phase separation arise from fluctuations in
the proton concentration, rather than from fluctuations in the
baryon density. 

The description of nuclear matter at higher temperature is also
of great interest; for example, very hot nuclear matter existed 
in the early universe.
A similar substance will be produced at
the Relativistic Heavy-Ion Collider (RHIC), and supernovae also
involve a large range of $\rhoB$ and $T$.
An analysis of hot nuclear matter within the framework of QHD is
developed in \cite{Fu90}.  
General principles of covariant thermodynamics and thermodynamic
consistency are
discussed, and these principles are illustrated by computing
nuclear matter properties in an arbitrary reference frame, using
the RMFT of QHD--I.  The results are shown to
be Lorentz covariant, and thermodynamic consistency is
demonstrated by proving the equality of the ``thermodynamic''
and ``hydrostatic'' pressures.  The mean-field results are used
in a simple hydrodynamic picture to discuss the phenomenology of
heavy-ion collisions and astrophysical systems, with an emphasis
on new features that arise in a covariant approach. 
Covariant Feynman rules for going beyond the RMFT are discussed in
\cite{Fu91,Fu91b}.

The nucleon mean free path is considered in \cite{Ha92,Ha94b}, 
thermal fluctuations and quantum corrections are
examined in \cite{Ku91,Su92a,Qi93,Ag94,Su95a}, and screening is
discussed in \cite{Ga94c}.
Supernovae and neutron stars are studied
together in \cite{Su92}.  Statistical properties are examined
in \cite{Vo92,Ra93}, and collective modes at finite $T$ in
\cite{Ni93}. Various aspects of relativistic heavy-ion physics
are discussed in \cite{Du95}. 

A novel $1/N$ expansion of the theory is pursued in
\cite{Ta92a,He93,Ta93,He95}, where, for isospin, one has $SU(N)$ with $N=2$.
The thermodynamic potential with exchange and RPA corrections to the RMFT is
calculated within this framework.   

A key element of the RMFT of nuclear matter in QHD--I is that the baryon 
acquires a density-dependent effective mass $\Mstar$ due to the classical
scalar field, which in turn arises from the baryons themselves.  
$\Mstar$ must be determined self-consistently at all
densities and temperatures, and in this model, $\Mstar /M$
goes to zero at high $\rhoB$ and high $T$.  
The successful description of the location and shape of the quasielastic 
peak in $(e,e^{\prime})$ at the RMFT level gives additional evidence for
the model values of $\Mstar /M$ in the nuclear medium \cite{Se86,Wa95}.  
There is no shift in the meson masses at this level of approximation,
although meson masses are modified when one includes
polarization insertions \cite{Se86,Wa95}.  
The modification of meson masses is crucial,
for example, for deciding whether pion or kaon condensation
takes place in the nuclear medium, and at what density.  The
modification of the masses of vector mesons can be studied
experimentally by looking at in-medium lepton-pair formation;
there are plans in place to do so at CEBAF. 

A great deal of work in the past few years has gone into the
investigation of the modification of hadron masses in the
nuclear medium  \cite{Ga94b,We94,Ga95a,Ko95c,Li95c,Sa95}.  The
effect on the nuclear force is examined in \cite{Ga95a} and on
relativistic transport in \cite{Ko95c,Li95c}.  In \cite{We94},
aspects of chiral symmetry and their implications for changes in
hadron structure in the nuclear medium are summarized and
discussed.  This includes issues such as the density dependence
of the chiral (quark) condensate, the related appearance of
strong scalar mean fields in nuclei, the stability of the pion
mass against compression in dense matter, and recent explorations
of s-wave kaon--nuclear interactions.  Kaon condensation is also
discussed in \cite{Ma94d,Sc94a,El95}.
Fortunately, there is an extensive survey article available on
this subject in \cite{Ad93}.  Here 
various scenarios for chiral symmetry restoration and deconfinement at 
finite temperature or density are studied assuming universal scaling 
relations for some hadron masses.

An interesting extension of QHD focuses on the
incorporation of the broken scale invariance of QCD. Reference
\cite{Ja93a} examines various chiral lagrangians in which the
QCD scale anomaly is implemented by introducing a dilaton field
representing the gluon condensate.  The lagrangians are used to
study the chiral phase transition and the propagators of both
scalar and vector mesons in dense baryonic matter.  A hybrid
model is proposed that allows for an unambiguous definition of
meson masses and coupling constants in dense matter.

At the level of QCD, there is a quark condensate, that is, a vacuum
expectation value $\langle 0 |\bar{q} q | 0 \rangle$, whose presence signals
the spontaneous breakdown of chiral symmetry and leads to a baryon mass.
QCD sum rules combine hadronic amplitudes, quark and gluon condensates, and
asymptotic freedom to obtain constraints on the properties of hadrons.
This quark condensate will be modified in the presence of other baryons;
using sum rules, one can relate the change in baryon (or more generally,
hadron) masses in nuclear matter to the underlying QCD.
Important papers in this regard
include \cite{Co91a,El91,Ce92a,Co92a,Fu92a,Ce93,Ji93a}.   

In \cite{El91}, chiral and scale symmetries of QCD are used to
describe the interaction between these condensates and hadrons. 
The resulting equations are solved self-consistently in the RMFT
approximation.  For these QCD condensates to be driven
towards zero at high density, their coupling to scalar and vector
mesons must be such that the masses of these mesons do not
decrease with density.  In this case, a physically sensible phase
transition to quark matter ensues. 

In \cite{Co91a,Co92a,Fu92a}, the self-energies for quasinucleon states in
nuclear matter are studied using QCD sum rules.
A correlator of nucleon interpolating fields, evaluated in the
finite-density ground state, is calculated using both an
operator product expansion and a dispersion relation with a
spectral {\em Ansatz}.  This approach relates the nucleon spectral
properties (such as the quasinucleon self-energies) to matrix
elements of QCD composite operators (condensates).  With
increasing nucleon density, large changes in Lorentz scalar and
vector self-energies arise naturally; the self-energies are
found to be comparable to those arising in RMFT phenomenology. 
The most important phenomenological inputs are the baryon density and the 
value of the nucleon ``sigma term'' ($\sigma_{\rm N}$)  divided by the
average current mass of the light quarks.  
The successful comparison to RMFT phenomenology is, however, sensitive to 
assumptions about the density dependence of certain four-quark 
condensates \cite{Ji93a,Ji94a}.

One extension of QHD involves developing model field theories
where the mesons couple directly to quarks.
Three important papers in this regard are \cite{Ja92,Sa94,Sa94b}.
In \cite{Ja92}, chiral symmetry restoration at finite baryon density is
studied in a quark model involving both scalar and vector interactions.
The presence of vector interactions makes chiral symmetry restoration more
difficult.
On-shell masses and coupling constants are calculated for the $\vecpi, 
\omega, \vecrho,$ and ${\bf a}_1$ mesons.
An attempt is made to relate the quark--meson interactions to the observed
nucleon--meson coupling constants. 

In {\em summary}, relativistic quantum field theories of the 
many-hadron system, with local lagrangian densities, appear
to provide the appropriate framework for extrapolating the properties 
of ordinary hadronic nuclear matter to all temperatures and
densities up to the transition to the quark-gluon phase.
We can understand this based on our earlier discussion.
Since the truncation of an effective lagrangian involves an expansion
in powers of $\kfermi /M$, then even at 10 times equilibrium density,
this ratio is only 0.6.
Moreover, the relevant temperatures are $T\lesssim 200\MeV$, which are
small compared to the baryon and heavy meson masses.
Nevertheless, interaction terms that are small at normal densities and
temperatures, and thus difficult to calibrate, can become important at
densities and temperatures relevant to the phase transition, thus producing
significant uncertainties in the extrapolation \cite{Mu96,Mu96a}.
Finding appropriate ways to calibrate these interactions is a major
challenge for future investigations.



\section{Loops in QHD}
\label{sec:loops}


We have seen that the mean-field approximation to
QHD gives a concise and highly successful nuclear phenomenology.
We have also noted, based on ideas from density functional theory, that
a mean-field energy functional fitted to nuclear properties implicitly
includes some effects of higher-order, many-body corrections.
Nevertheless, mean fields are insufficient for a detailed understanding
of nuclear structure and its relation to the underlying NN, NNN, 
\dots\ interactions.
Moreover, various experimental observables probe aspects of nuclear dynamics 
that go beyond a single-particle description, which can tell us about the
modification of the NN interaction inside a nucleus.
We must therefore develop reliable, relativistic techniques to extend the 
RMFT calculations discussed above.
Useful tools in this endeavor are Feynman diagrams, dispersion relations,
and path integrals, as discussed in \cite{Se86}, where a historical
development is presented, and in \cite{Se92}, which updates the 1986
volume.
References to the original literature are given in these review articles.
Here we briefly summarize some of those results and discuss some of the 
work that has been done since that time.  
A more detailed discussion of the theoretical background is contained in the 
recent text \cite{Wa95}.

Corrections to the RMFT will generically involve ``loop diagrams'', which
are important for including several different aspects of the quantum nature 
of the system.
For example, loop diagrams must be included to ensure the unitarity of
scattering amplitudes.
Baryon loops are necessary for incorporating familiar many-body effects,
like the exchange of identical nucleons or the summation of ladder and ring
diagrams \cite{Fe71}.
Moreover, loops introduce effects arising from the modification of the
quantum vacuum in the presence of valence nucleons,
and meson loops in particular generate contributions to the extended 
structure of the nucleon.
The relevant question is how to treat these loops most efficiently, 
consistent with the notion that QHD is intended to be a large-distance,
hadronic model of the underlying QCD.
We begin the discussion by considering pion loops and then turn to the
more difficult question of heavy-meson and baryon loops.

\subsection{Pion Loops}
\label{sec:piloops}

Because the pion has a small mass, pion loops make significant contributions
to both scattering and nuclear-structure observables.
Moreover, since the energies and momenta of interest are typically of
the order of the pion mass (or even larger), pion-loop contributions
generally involve nonanalytic functions of the external four-momenta.
Thus these contributions cannot be absorbed in a local effective lagrangian,
which, by definition, contains only finite powers of derivatives
\cite{Ge93,Ba94b}.
They must be computed explicitly, whether one is using a renormalizable 
model or an effective, nonrenormalizable lagrangian.

A highly developed framework for systematically including pion loops in
scattering processes is chiral perturbation theory (ChPT) \cite{Ga84,Me93a}.
Chiral symmetry implies that pion interactions in the lagrangian
can be grouped 
order-by-order in the number of derivatives and powers of $\mpisq$
(as in the nonlinear models
discussed earlier), so that there is a systematic expansion for scattering
amplitudes at low energies.
Moreover, an expansion in the number of loops also proceeds in
powers of momenta \cite{We67,We90a}, so one can systematically include loop 
corrections.
ChPT has been successful in describing scattering in both the $B=0$ and $B=1$
sectors of low-energy QCD \cite{Ga84,Me88,Me93a}.
Studies of two- and many-nucleon systems have been initiated and are
currently under active investigation \cite{Or92,Va93,Or96}.

Properties of the baryon also arise through pion-loop integrals in QHD.  
For example, the vertex diagram consisting of
the emission of a pion, its interaction with the virtual
electromagnetic field, and its reabsorption by the baryon
contributes to the baryon's anomalous magnetic moment.
The two-pion contribution gives the low-mass, or long-distance, part
of the spectral weight function $\rho_2 (\sigma^2 )$ for the anomalous 
magnetic form factor $F_2(q^2)$:
\beq
  F_2(q^2)  =  \frac{1}{\pi} \int_{ 4 \mpisq}^{\infty}
           \frac{\rho_2 (\sigma^2)}{\sigma^2 - q^2} \,{\rm d} \sigma^2
                                   \ .   \label{eq:8ac}
\eeq
Assume that the two-pion contribution arising from this vertex diagram 
dominates the spectral weight function everywhere.  
This yields a qualitative description
of the isovector anomalous magnetic moment and its mean-square
radius \cite{Se86,Wa95}.\footnote{The {\em isoscalar\/}
anomalous moment vanishes in this approximation; experimentally,
it is indeed very small.}

Pion loops also give hadronic contributions to vacuum polarization.
For example, a spectral analysis of the strong-interaction contribution to
electromagnetic vacuum polarization shows that the spectral
weight function starts at $ 4\mpisq$:
\begin{eqnarray}
  \Pi_{\mu \nu}^{\rm str} (q) & = & ( q_{\mu} q_{\nu} - q^2 g_{\mu \nu})
                         \Pi ( q^2) \ ,
                                               \\[4pt]
  \Pi ( q^2) & = & \frac{1}{\pi} \int_{ 4 m_{\pi}^2}^{\infty}
    \frac{\rho (\sigma^2)}{ \sigma^2 - q^2}\ {\rm d} \sigma^2 \ .
                                                \label{eq:8ab}
\end{eqnarray}
In the complex $q^2$ plane, $\Pi(q^2)$ is an analytic function
with a branch cut running along the real axis from $ 4 \mpisq$ to infinity.  
The discontinuity across that cut for
$4\mpisq \leq q^2 \leq 9 \mpisq $ comes from the electroproduction of
two real pions.
Thus the low-mass singularities of propagator and vertex functions are most
efficiently expressed in terms of hadronic variables.
The higher-mass singularities are more complicated, and hadrons are less 
efficient.
Nevertheless, if one is interested in studying the low-momentum behavior
of the vacuum polarization, one can emphasize the low-mass part of the
spectral integral by making several subtractions from the integral and 
by determining the unknown coefficients empirically \cite{Do96}.

The exchange of two correlated pions between nucleons also involves multiple
pion loops, which can be treated with dispersion 
relations \cite{Ja75,Du77,Du80,Li89,Li90,Ki94}.
The result is a strong NN attraction that can be simulated by introducing
an effective scalar-isoscalar field coupled directly to the nucleon, as
discussed in Sections~\ref{sec:nlsigma} and \ref{sec:VMD}.
More detailed studies of the scalar-isoscalar NN interaction can be 
performed by returning to the description with explicit pion 
loops \cite{Du93,Ao95}.
Similar observations can be made in exchange channels with vector meson
quantum numbers.
In particular, a tree-level effective lagrangian with vector mesons is
found to be essentially equivalent to ChPT in the pion sector 
at one-pion-loop order \cite{Me88,Do89,Ec89,Ec89a}.
Thus these non-Goldstone (heavy) bosons can be introduced into an
effective lagrangian containing baryons to efficiently account for the
intermediate-range NN interactions and to conveniently describe
nonvanishing expectation values of nuclear bilinears (\eg ${\overline N}
N$ and ${\overline N}\gammamul N$), as in Section~\ref{sec:VMD}.

Moreover, as noted in Section~\ref{sec:isobar}, the sum of $\pi$N ladder
diagrams with nucleon exchange can be investigated with partial-wave
dispersion relations, leading to a resonance in the $\Delta$ channel.
Thus the $\Delta(1232)$ can be included in an effective hadronic lagrangian
to incorporate this important physics \cite{De92a,We93,Ta96}.
Alternatively, the dynamical model of the resonance provides a means to
investigate many interesting questions concerning the behavior
of the $\Delta$ in the nuclear many-body system, such as
its binding energy in nuclear matter, its optical potential, and the 
modification of its electroweak properties.

Virtual pion effects have been studied extensively.  Pion loops and pion 
dressing are examined in \cite{Pa92} using a coherent-state approach. 
Reference \cite{Fr93} studies in depth the contribution of the virtual 
pion cloud to the electromagnetic properties of the nucleon.

\subsection{Loops in Renormalizable QHD Theories}
\label{sec:Nloops}

By starting with the Feynman rules for the Green's functions in a
renormalizable QHD theory, one can {\em in principle\/} go beyond the
RMFT, compute observables in terms of a finite number of couplings and
masses, and then compare with experiment.
{\em In practice\/}, this program is extremely difficult, since QHD is 
a strong-coupling relativistic quantum field theory.
Moreover, renormalizable theories explicitly include contributions from 
all length scales, and the QHD couplings get stronger at short distances, 
since these theories are not asymptotically free.

Nevertheless, just as in nonrelativistic many-body theory, one can use 
intuition to sum selected infinite sets of diagrams, determine the 
renormalized coupling constants by refitting nuclear matter properties, and
then see whether the RMFT results are {\em stable\/} under the
inclusion of these additional contributions, while investigating
new physical phenomena.  
All of the extensions we discuss involve loop corrections
to the RMFT of one sort or another. 
These corrections include familiar many-body effects, where the loop
momenta are typically of the size of $\kfermi$ (or at most, several times
$\kfermi$), and also corrections from the dynamical quantum vacuum,
which involve shorter distance scales.

In the RMFT, the baryon Green's function can be written as \cite{Se86}
\begin{eqnarray}
 G(k) & = & (\gammamul k^{* \mu} + \Mstar) \Big\{ \frac{1}{ k^{*2} -
            \Mstarsq + i \epsilon}
      + \frac{i \pi}{\Estark}\, \delta [ k_0^* - \Estark ]
            \theta( \kfermi - |{\bf k}|) \Big\}
                                             \nonumber \\
    & \equiv & G_F(k) + G_D(k) \ .
                                             \label{eq:8aa}
\end{eqnarray}
in the rest frame of the nuclear matter.
Here $k^{* \mu} \!\equiv \! (k_0 - \gv \Vzero , {\bf k})$ is the kinetic
four-momentum\footnote{Note that in closed-loop integrals, such
as those involved in computing the ground-state energy, a simple
shift of integration variables allows one to eliminate the dependence 
on $ \gv\Vzero$ \protect\cite{Se86,Fu89}.} 
and $\Estark\! \equiv \! \sqrt{{\bf k}^2 + \Mstarsq}$,
with $\Mstar \!\equiv\! M - \gs\phizero$.
The first term $G_F(k)$ is the Feynman propagator for a baryon of mass 
$\Mstar$, and the second term is the contribution arising from baryons 
already present at finite density; this latter contribution reproduces the 
RMFT results.  
In discussing the following extensions, we shall frequently distinguish
results obtained with the full baryon propagator $G(k)$ from those obtained
with just the second, or ``density-dependent'', contribution $G_D(k)$.
Since the three- and four-momenta are
constrained in $G_D(k)$, loop integrals over this
second term give well-defined, finite results
that are direct analogues of the terms arising in
nonrelativistic many-body theory.

For example, relativistic Hartree theory is
obtained by self-consistently summing the tadpole graphs
in the baryon self-energy \cite{Se86}.  
Retention of $G_D$ in the tadpoles
gives rise to the MFT, while the full $G$ (together with 
appropriate counterterms $\delta\lagrang$)
produces the RHA, as discussed previously.\footnote{%
One can also solve 1+1 dimensional QHD exactly for {\em finite\/} systems in
the one-loop approximation; this is carried out in \protect\cite{Fe93}.}
A characteristic result of QHD--I is that the
Lorentz scalar and vector self-energies are large on
the scale of nuclear energies; these
contributions {\em cancel\/} in the binding energy but {\em
add\/} in the spin-orbit interaction.

Hartree--Fock (HF) theory
is obtained by including the meson emission and reabsorption
(``exchange'') graphs in the baryon proper self-energy \cite{Se86,Se92}.
Retention of $G_D$ only in this HF theory leads to
exchange terms that are the direct relativistic
generalization of those arising when Slater determinants are used
to determine the best single-particle wave functions in nonrelativistic
many-body theory.
The inclusion of the exchange terms does not qualitatively alter the size
of the large scalar and vector self-energies found in the RMFT, and thus
the RMFT is {\em stable\/} with respect to these corrections.
In fact, after renormalization to equilibrium nuclear matter properties,
the binding energy curves in the RMFT and HF approximations are 
almost indistinguishable \cite{Se86}.
Moreover, the equation of state approaches that of the RMFT at high baryon 
density.
Further discussion is contained in \cite{Se92}.

Relativistic HF calculations that include charged
mesons and that examine the relation of the scalar and vector self-energies 
to DBHF calculations in nuclear matter are carried out in 
\cite{Bo87,Ce92,Be93,Fr93a,Bo94a,Fr94a,Ma94c,Zh94a,Be96}.
The existence of large Lorentz scalar and vector self-energies appears to
be a firm conclusion.
(Recall our discussion of DBHF results in Section~\ref{sec:DBHF}.) 
The connection to semiclassical approximations is given in \cite{Vo94}.
The general structure of the self-energy in matter is analyzed in
\cite{He91,Ru95a}.

The fully self-consistent HF theory retaining the complete
$G(k)$ and meson retardation is complicated \cite{Se92}; it has
not yet been successfully solved.
The summation of exchange diagrams only (\ie HF theory at
zero baryon density) is discussed in \cite{Bi83,Bi84,Kr93}.

The sum of fermion ring diagrams involving $G(k)$ is equivalent to
the random-phase approximation.
We shall refer to the calculation of the rings that keeps only terms with 
at least one factor of $G_D$ as the RPA. 
This includes loops that are at least linear in the density (particle-hole 
parts and admixtures between filled valence states and those in the Dirac 
sea) and is the direct relativistic extension of the
RPA in nonrelativistic many-body theory.  
The calculation that also includes the modification of the strong 
vacuum polarization in the nuclear medium due to the shift 
$M \rightarrow \Mstar$ will be called RRPA.  

The RPA as applied to finite nuclei was discussed in Section~\ref{sec:RPA}.
We emphasized the need to include contributions from negative-energy states
in order to maintain the conservation laws in the theory.
Some important results from RPA studies of nuclear matter are as follows.
The scalar and vector propagators mix in nuclear matter, and
at high density, vector meson exchange dominates in QHD--I.
The excitation spectrum of nuclear matter in the RPA is that of zero sound, 
where the sound velocity $c_0$ approaches the speed of light from below 
as the baryon density gets large ($c_0 \rightarrow 1^-$
as $\kfermi \rightarrow \infty$).  
This implies that signals in the medium cannot propagate faster than the
speed of light, in accord with special relativity.  
There are other branches in the excitation spectrum corresponding to 
meson propagation.

We turn now to calculations involving strong vacuum polarization (RRPA).
This polarization does not have any explicit density dependence, but 
depends on it
implicitly through the baryon mass $\Mstar$.
These vacuum loops correspond to the one-baryon loops 
contained in the RHA (Section~\ref{sec:RHA}), and just as we found 
unnaturally large contributions in the RHA, analogous large effects appear
in the RRPA.
In consequence, in nuclear matter, Landau ``ghost'' poles appear in the 
meson propagators at zero frequency $q_0 = 0$ and finite wave number 
${\bf |q|} \neq 0 $; the value of this wave number is a few times the 
nucleon mass in QHD--I \cite{Co87,Pe87,Fu88,We90,Li90a}.
Such poles imply an instability of the system against density
fluctuations of the corresponding wavelength and are a manifestation
of the lack of asymptotic freedom in the theory.
Apparently, a description of the quantum vacuum by summing simple baryon
loops is inadequate.

Similar problems occur when one extends the RHA to include two-loop
contributions \cite{Fu89}, which
consist of a closed baryon loop with an internal meson line.  
When the full propagator $G$ is used in this calculation, one incorporates
the many-body modifications to both the strong vacuum polarization and the
baryon self-energy that arise from the Pauli exclusion principle and from
the shifted baryon mass $\Mstar$.
One again finds unnaturally large contributions to the nuclear matter energy.

The problem in all of these calculations of vacuum effects is that an 
expansion in powers of loops is basically an expansion in the dimensionless 
coupling constants, which are large in QHD.  
The quantum corrections are correspondingly large, the series is not
converging, and the RMFT {\em is not stable against this
perturbative loop expansion}.  
Clearly, an alternative procedure must be found to systematically and 
reliably calculate the vacuum-fluctuation corrections to the RMFT in 
renormalizable QHD, if that is indeed possible.

One idea is to include corrections that sum diagrams to {\em all 
orders\/} in loops, such as the ``ladder'' summation discussed in 
Section~\ref{sec:DBHF}.
Alternatively, one observes that in a theory with a vector boson coupled to
a fermion, the vertex form factor is highly damped at large spacelike
momentum transfers $q^2 < 0$; this is analogous to the result first
derived by Sudakov in quantum electrodynamics \cite{Mi91}.
Such a form factor would decrease the sensitivity to
high-momentum (or short-distance) contributions to loop integrals
and would provide a favorable situation for QHD.  
Moreover, it is essential to include vertex corrections, as these 
reflect the internal hadron structure present in renormalizable QHD.
Remember that baryons are complicated objects in the full field theory;
they are surrounded by a cloud of virtual mesons and
baryon-antibaryon  pairs.  
It was an initial hope that the theory might contain vertex functions that
would damp the contributions from loop integrals before one reaches
distance scales where the lack of asymptotic freedom becomes 
manifest.\footnote{Note that in a theory with a vector field coupled to
a conserved fermion current, Ward's identity implies that the structure
of the vertex function is unrelated to the running of the coupling constant.}

In \cite{Al92}, vacuum polarization is studied in QHD--I
(without the scalar meson).  
The lowest-order (one-baryon-loop) polarization produces a ``ghost''
pole when summed to all orders in the RRPA, as discussed above.  
It is first verified that the infrared structure of
the meson--baryon vertex in this model produces an on-shell
proper vertex function that is strongly damped at large
spacelike momentum transfer.
When the model vertex function is approximated by its on-shell form and
combined with the lowest-order polarization, the vacuum
contributions are significantly reduced.  
The resulting RRPA meson propagator has no ``ghost'' poles and is finite at 
large spacelike momenta.  

In \cite{Se95}, it is shown that a similarly damped form factor arises
for the scalar--nucleon vertex in QHD--I due to the vector-meson dressing.
The on-shell approximations to the vertex functions are used to investigate
the two-loop contributions to the properties of nuclear matter, and it
is found that they are greatly reduced by the inclusion of 
vertex functions.
Similar results are found using {\em ad hoc\/} form factors in
\cite{Pr92,Fr92}.

RRPA calculations are performed with {\em ad hoc\/} vertex cutoffs in 
\cite{Pr92a} and with the on-shell model described above in \cite{Ta93a}.  
A two-loop calculation that also consistently treats the electromagnetic 
interaction to the same level of approximation is developed in \cite{Be92}.  
Extensions of RPA and RRPA to include the self-consistent sum of ring 
diagrams with additional $\Delta$-hole propagation are examined in
\cite{Be93a}.  The relation of the density dependence of
zero sound to the renormalization prescription is discussed in
\cite{Ca95}.

The fully off-shell vertex function is complicated in any field theory.
An attempt to include the off-shell vertex in the RRPA calculation for
QHD--I has been initiated in \cite{Al95}; no concrete results exist yet
for this very difficult calculation.
A simplified approach to the baryon self-energy, using an {\em ad hoc\/}
off-shell form factor and dispersion relations, is contained in \cite{Kr93}.

Nuclear Schwinger--Dyson equations, which provide a basis
for an analysis of the relativistic field theory content of the
nuclear many-body problem in terms of propagators and vertices,
are developed in \cite{Na91,Ko93,Na94,Na94a}.  Truncated applications
usually use some sort of spectral representation or dispersion
relation \cite{Ta91} to eliminate the Landau ``ghost'' poles in the meson
propagators discussed earlier.

Pion propagation in the nuclear medium, which involves 
baryon-antibaryon loops, is studied in \cite{He92a,Ka92}; 
again, the ``ghost'' poles are generally removed through one of 
the mechanisms described above. 
Vector meson propagation is examined in \cite{Je94} and vector meson 
mixing through baryon loops is examined in \cite{Pi93,Pi93a}.
(Note that here the strength of the mixing is {\em finite\/} without
any renormalization.) 
Meson modes in nuclear matter are also examined in \cite{Ja93,Ja93a}.

A chirally symmetric, renormalizable, anomaly-free theory that contains
baryons and ($\vecpi, \sigma, \omega, \vecrho, {\bf a}_1$)
mesons is developed in
\cite{Se92b}; we refer to this theory as QHD--III.  This model provides
a hadronic description of strongly interacting matter that includes
isovector, pseudoscalar and vector fields in addition to the
isoscalar, scalar and vector fields of QHD--I.  Although this
theory provides a consistent, self-contained description of
nuclear physics, including loop processes, its phenomenology
remains to be investigated \cite{Wa95}.

To summarize, the systematic calculation of loop corrections to the RMFT is 
an important goal in renormalizable QHD.
At present, vacuum contributions evaluated at various orders in loops are
all unnaturally large, signaling the inadequacy of these approximations
for representing the vacuum dynamics.
Although these large effects can be reduced by introducing vertex form
factors {\em external\/} to the renormalizable theory, the question of
whether such form factors can be generated {\em internally}, and indeed,
whether calculations with off-shell vertices are even feasible, is still open.
At best, even if renormalizable QHD provides a realistic description of
vacuum dynamics, the results will involve sensitive cancellations between
unnaturally large contributions.
We are thus motivated to consider other options.

\subsection{Loops in Effective QHD Theories}
\label{sec:loopseft}

The results discussed above present significant evidence that the requirement
of renormalizability is too restrictive, and that it is more appropriate to
consider QHD as an effective, nonrenormalizable hadronic field theory.
In this approach, short-distance effects and vacuum modifications are 
parametrized through nonrenormalizable interaction terms in the lagrangian,
with couplings that are determined by fitting to data.\footnote{These
couplings also implicitly include the effects of heavier degrees of freedom
not included in the model lagrangian.}
Nucleon and non-Goldstone boson fields are still needed in the lagrangian
to account for the valence nucleons and to treat the large mean fields
conveniently.
Loop diagrams involving these degrees of freedom must also be included to
ensure the unitarity of scattering amplitudes and to incorporate many-body,
density-dependent effects.
The relevant question is how to separate the short-range contributions, which
are to be absorbed in the model parameters, from the long-range, many-body
effects.

The baryon propagator in Eq.~\eqref{eq:8aa} separates into an explicitly
density-dependent part $G_D$ and a part $G_F$ that has no explicit density
dependence.
The familiar many-body contributions arise from loop integrals involving $G_D$
alone (as in the RMFT or HF approximations) or combinations
of $G_D$ and $G_F$ (as in the DBHF approximation or the RPA); these are to
be computed explicitly.
One must also retain various density-dependent contributions that contain
negative-energy states, but that are finite, such as those needed to preserve
the conservation laws in the RPA.
The remaining contributions involve
divergent loop integrals that contain vacuum 
dynamics and short-distance behavior; these are to be absorbed in the
parameters of the lagrangian that are fitted to empirical data.
Some of these contributions will have explicit density dependence (like the
strong Lamb shift), while some will not (vacuum polarization).
We emphasize that this is a more difficult problem than the inclusion of
loops in ChPT, since the loop corrections do not
produce an expansion in powers of external momenta, and thus corrections
at a given order in loops renormalize the parameters at {\em all lower 
orders\/} in loops.

In a renormalizable theory, the divergences can be removed by cancelling them
against a finite number of counterterms.
In a nonrenormalizable theory, however, the number of divergences is, in
principle, infinite.
Nevertheless, these divergences can always be formally cancelled by the
appropriate counterterms, {\em since all possible interaction terms consistent
with the symmetries are already included in the effective lagrangian}.
For example, in infinite nuclear matter, any vacuum contribution to the
energy arising from loops containing only factors of $G_F$ must appear in
the form of a polynomial in the scalar field $\phizero$.\footnote{Although
nonanalytic functions of $\phizero$ may arise in principle, we assume that
any function involving heavy mesons can be expanded in a Taylor series in
the relevant density regime.}
As another example, contributions from the nucleon's Lamb shift can
be (formally) cancelled by counterterms of the form $\bar{N}N\phi^n$ or
$\bar{N}(i\gammamul\dmu - M)^m N\phi^n$.
The important issue thus becomes how to limit the required counterterms
to a finite and manageable number.
Moreover, due to the freedom one has to redefine the field variables, the
most efficient way to write the counterterms is still an unsolved problem.

One way to constrain the parameters is to rely on the broken scale invariance
of QCD, which restricts the interactions in the scalar-isoscalar channel.
In \cite{Fu95}, an effective relativistic hadronic
model for nuclear matter that incorporates nonlinear chiral
symmetry and broken scale invariance is applied at the one-baryon-loop level.
The model contains an effective light scalar field that is
responsible for the mid-range NN attraction and that has anomalous 
scaling behavior.  
One-loop vacuum contributions in this background scalar field at finite 
density are constrained by low-energy theorems that reflect the
broken scale invariance of QCD, so that the scalar effective potential
contains only three free parameters.
The resulting mean-field energy functional for nuclear matter and nuclei 
has only a finite number of parameters and yields good fits to the bulk
and single-particle properties of nuclei.

These results are consistent with the discussions in Sections~\ref{sec:eft}
and \ref{sec:rmft}, where it was observed that only the first few terms
in an expansion in powers of the fields (and their derivatives) are
relevant up to moderate densities; in particular, only three parameters
are needed in the scalar potential  [$\mssq$, $\kappa_3$, and $\kappa_4$ in
Eq.~\eqref{eq:NLag}].
This result obtains because the empirically fitted parameters are of natural
size.
Thus one may expect that the assumption of naturalness is sufficient to
limit the unknown parameters in the lagrangian to a manageable number.
This expectation remains to be confirmed by calculations beyond one-baryon-loop
order, which are currently in progress.

To {\em summarize} this section on loop processes in QHD: although
there are many qualitative and even semi-quantitative insights,
and applications of the ideas of nonrenormalizable effective field theory
look promising,
{\em there is still no consistent, reliable, practical approach
to the relativistic nuclear many-body problem that includes all loop terms}.



\section{Summary}
\label{sec:summ}

Our goals in this paper are to describe consistent microscopic treatments of 
the relativistic nuclear many-body problem based on hadrons (quantum
hadrodynamics) and to summarize work in this field from early 1992 through
1995.
Although QCD is the underlying theory of the strong interaction, the QCD 
couplings are strong at large distances, and hadrons are more efficient
degrees of freedom.
But since it is still impossible to derive the low-energy 
hadronic lagrangian directly from QCD, we must rely on more indirect
information, such as the symmetries of the QCD lagrangian, and on
well-known nuclear phenomenology to guide us in the construction of
the low-energy theory.
The framework is based on local, Lorentz-invariant lagrangian densities,
as this is the most general way to parametrize observables consistent with
the desired constraints of quantum mechanics, special relativity, unitarity,
causality, cluster decomposition, and the intrinsic QCD symmetries.
Historically, the hadronic lagrangian was required to be renormalizable, so
that one could calibrate the theory and then extrapolate without the
appearance of new, unknown parameters.
There are now strong indications, however, that the constraint of
renormalizability is too restrictive, so it is important to generalize
our viewpoint to include nonrenormalizable, effective field theories, which 
can still provide a consistent treatment of the relativistic nuclear 
many-body problem.

The simplest model that we study (QHD--I) contains protons, neutrons, and
neutral scalar and vector mesons, and is renormalizable.
At large enough densities, the meson field operators can
be replaced by their expectation values, and the result is a relativistic
mean-field theory (RMFT), which must be solved self-consistently.
In the original version of the model, which omits scalar self-couplings,
one obtains a simple, two-parameter description of the equilibrium 
properties of nuclear matter that can be extrapolated to arbitrary density,
temperature, and proton fraction.
The classical meson fields are large (several hundred MeV),
and nuclear saturation occurs because the nucleon effective mass $\Mstar$
decreases as the density increases,
so that the attractive forces saturate, and the binding curve develops a
minimum.
If this RMFT is applied to finite nuclei (with the addition of classical
fields for the neutral rho meson and Coulomb potential), then from this
minimal set of parameters fitted to the properties of nuclear matter, one
derives the nuclear shell model.
Moreover, when cubic and quartic scalar self-interactions are included,
one can tune the equilibrium nuclear matter properties to provide a
realistic description of nuclear charge densities, charge radii, binding
energies, single-particle spectra, and quadrupole deformations throughout
the periodic table.

To explicitly include pions, we need a lagrangian consistent with global
chiral $SU(2)_{\rm L} \times SU(2)_{\rm R}$ symmetry, which is a symmetry of
QCD in the limit of massless $u$ and $d$ quarks.
It is simplest to use a linear representation of the symmetry, in which
the fields enter as chiral multiplets.
If one also demands renormalizability, one is led to the linear sigma model,
with an additional neutral vector field.
The neutral scalar field plays a dual role as the chiral partner of the
pion and the mediator of the attractive NN force.
One finds, however, that the mean-field approximation to this model is
unable to provide a realistic description of nuclei, primarily due to the
strong constraints on the scalar self-interactions arising from the
scalar potential used to induce spontaneous symmetry breaking.
This is evidence, even at the level of the RMFT, that the simultaneous
constraints of linear chiral symmetry and renormalizability are 
too restrictive.

We observe, however, that it is possible to make a chiral, point 
transformation of the fields and to recast the lagrangian so that the 
symmetry is realized {\em nonlinearly}.
The new scalar field is a chiral singlet that can be decoupled from the
theory by taking its mass to be very large, thus removing the unwanted
nonlinear interactions.
(In the limit of infinite scalar mass, the model becomes nonrenormalizable.)
Although the mid-range NN attraction is apparently destroyed by this
procedure, it can be restored by including correlated two-pion exchange in
the scalar-isoscalar channel.
Moreover, this two-pion exchange can be efficiently and adequately simulated
by introducing a {\em new\/} scalar field into the theory with a mass of
roughly $500\MeV$.\footnote{We observe with some amusement that the
possibility of introducing such an additional scalar into a chiral theory
was already noted in the early work of Coleman, Wess, and Zumino on
nonlinear realizations of chiral symmetry.
See footnote 9 in \protect\cite{Co69}.}
The new field is also a chiral scalar, so the lagrangian remains chirally
invariant, and the RMFT of this chiral model is identical to that of QHD--I.
Thus we draw the important conclusion that the scalar field in
QHD--I is to be interpreted as an {\em effective\/} field that incorporates
the pion-exchange contributions that are the most important for describing
the bulk properties of nuclear matter.

To get a deeper understanding of the large mean fields and to connect
the RMFT description to the underlying NN interaction, we consider 
corrections from nucleon exchange and short-range correlations in nuclear
matter.
This can be done in the Dirac--Brueckner--Hartree--Fock (DBHF) framework.
Here an NN quasipotential fitted to two-nucleon data is used to determine
a self-consistent NN interaction in the medium, which includes the
effects of the Pauli exclusion principle and the modifications to the 
single-nucleon Dirac wave functions at finite density.
Three important conclusions from this work are: the nucleon scalar and
vector self-energies (which are analogous to the scalar and vector fields
in the RMFT) are essentially the same size as the scalar and vector mean
fields studied earlier, and the state dependence of the self-energies is
small;
the depletion of the Fermi sea due to correlations is considerably smaller
in the DBHF theory than in nonrelativistic Brueckner--Goldstone theory;
the DBHF effective NN interaction contains density dependence that goes
beyond what is included in nonrelativistic
Brueckner--Goldstone theory, and it is therefore
possible to simultaneously fit both the NN phase shifts and the nuclear
matter equilibrium point at the two-hole-line level.
This last result can be understood by considering an expansion of the RMFT
energy/particle in powers of the Fermi momentum; one finds that even
two-body interactions involving a Lorentz scalar field lead to terms that
would be interpreted as many-body forces in a nonrelativistic framework.

Motivated by the appearance of an effective scalar field in our discussion
of chiral symmetry, we generalize the QHD framework to embrace the ideas
of nonrenormalizable, effective field theory.
We still retain a local, Lorentz-invariant lagrangian, but now the lagrangian
must contain all possible (non-redundant) interaction terms consistent with
the symmetries of the underlying QCD.
The coefficients of these terms parametrize the short-distance dynamics that
will be modelled by the effective hadronic lagrangian.
Since we cannot calculate these coefficients directly from QCD, they must be
determined by fitting to data, and we have the important freedom to choose
our generalized coordinates (fields) to make this fitting as efficient as
possible.
Moreover, since the effective lagrangian has, in principle, an infinite
number of terms, we must have an organizing principle to retain some
predictive power, and this is also influenced by the choice of dynamical
variables.

Based on the successes of the hadronic description of both the NN interaction
and the nuclear-structure observables, we choose as degrees of freedom the
nucleon, pion, and low-mass scalar and vector fields.
By retaining the ``heavy'' non-Goldstone bosons, we can describe the NN
interaction without the explicit calculation of multi-pion loops, and
we can efficiently describe the expectation values of nucleon bilinears
using mean fields.
We then construct a lagrangian consistent with the underlying symmetries of
QCD, in which the chiral symmetry is realized nonlinearly and 
electromagnetic interactions are introduced both through minimal-coupling
terms and through derivative couplings of the nucleon to the photon, which 
allows us to incorporate the nucleon electromagnetic structure directly 
in the theory.

To organize the lagrangian, we rely on naive dimensional analysis, which
allows us to extract the dimensional scales of any term, on the assumption
of naturalness, which says that the remaining dimensionless coefficient
for each term should be of order unity, and on the observation that the
ratios of mean meson fields $\gs \phizero$ and $\gv \Vzero$
(and their derivatives) to the nucleon mass $M$
(which generically represents the ``heavy'' mass scale) are good expansion
parameters.
Since the meson fields are roughly proportional to the nuclear density,
and since the spatial variations in nuclei are determined by the momentum
distributions of the valence-nucleon wave functions, this organizational
scheme is essentially an expansion in $\kfermi /M$, where $\kfermi$ is a
Fermi wavenumber corresponding to ordinary nuclear densities.
If naive dimensional analysis and the naturalness assumption are valid,
we can expand the lagrangian in powers of the fields and their derivatives,
truncate at some finite order, and thus have some predictive power for
the properties of nuclei.
The model parameters fitted to bulk and single-particle nuclear properties
show that this is indeed the case.

We also observe, based on concepts from density functional theory, that
RMFT parameters fitted to nuclear properties implicitly include not only
the short-distance effects parametrized in the lagrangian, but also
long-range, many-body effects arising from corrections to the RMFT.
For evidence that mean fields should provide a reasonably accurate way to
parametrize these many-body effects, we rely on the DBHF results, which
show that exchange and short-range-correlation corrections do not 
significantly modify the mean fields obtained at the RMFT level.
Thus we conclude that densities ``large enough'' to justify the use of
mean-field theory are present in the interiors of medium and heavy nuclei,
since the Hartree contributions dominate the nucleon self-energies
already at these densities.
The numerous successful RMFT calculations of nuclei throughout the periodic
table should be interpreted within this density functional context.
By explicitly calculating higher-order corrections, one can improve the
description of the density functional, and by using the properties of nuclei
to determine the parameters, one will ultimately obtain the values that 
should appear in the effective lagrangian.

This observation leads to another important conclusion.
No matter how well one can compute the many-body contributions, there
will {\em always\/} be additional terms in the energy functional that depend
(for example) on powers of the scalar field (or scalar density).
The coefficients of these terms, which parametrize the short-distance
behavior of QCD, are unknown and are presently impossible to calculate.
Moreover, these contributions will be significant on the scale of the 
nuclear binding energy when their coefficients are of natural size.
Unless one has a concrete argument for why these coefficients should be
{\em anomalously small}, they will always be present to adjust the energy
functional in a way that allows one to fit nuclear matter saturation or
nuclear properties.
The goal of predicting the nuclear matter equilibrium point from an NN
interaction fitted to two-body data therefore loses some of its importance.
In the conventional nonrelativistic language, these additional terms 
correspond to the short-range parts of the three-body (and many-body)
forces that cannot be calculated and that must be fitted to the desired 
observables.\footnote{The situation has been succinctly described by
Jackson \cite{Ja92a}: ``The problem is that there is too little dirt and too
much rug under which to sweep it.''}

The problems of extrapolation to extreme conditions of density and
temperature, or to large energy-momentum transfers,
and the systematic computation of loop corrections still
pose challenges for QHD.
Although one can argue that the effective lagrangian can be
truncated at a few terms for studies of neutron stars or the 
transition to the quark-gluon plasma, interaction terms that are small at
normal densities and temperatures, and thus difficult to calibrate, can
become important under extreme conditions.
This incomplete calibration leads to significant uncertainties in the
extrapolation, even for parameters of natural size.
As for loop calculations, there is now significant evidence that it is
insufficient to represent vacuum modifications by simple loop integrals in 
renormalizable QHD theories.
Although it may be possible to reduce the size of these corrections by
incorporating vertex modifications, calculations involving off-shell
vertex functions computed entirely within the context of the theory remain
to be done.

Nonrenormalizable effective theories appear more promising, since short-range
and vacuum effects are absorbed into the parameters of the lagrangian,
and long-range, many-body effects are (typically) straightforward
relativistic generalizations of their nonrelativistic 
counterparts.\footnote{%
There are, however, some nontrivial aspects to these generalizations from
a nonrelativistic to a relativistic description.
For example, the dynamical nature of the mesons introduces retardation
effects that have not yet been adequately studied.}
Moreover, calculations of pion loops based on chiral perturbation theory
or dispersion relations can be performed systematically.
Nevertheless, a fully consistent, practical, relativistic 
many-body description of nuclei based on effective hadronic field theory 
remains to be formulated.

In {\em summary}, nuclear physics is the study of the structure of strongly
interacting baryonic matter ($B \geq 1$), 
and the only consistent theoretical framework
we have for describing such a relativistic, interacting, quantum-mechanical,
many-body system is relativistic quantum field theory based on a local
lagrangian density.  
Although QCD of quarks and gluons provides the basic underlying theory,
lagrangians based on hadronic degrees of freedom (QHD), which are the
particles observed in the
laboratory, provide the most efficient description of the physics in the
strong-coupling, nuclear domain.
While it is not surprising that initial QHD attempts to model the system
within the subset of renormalizable lagrangians appears to be too
restrictive, some results of the initial simple models remain robust.  
In particular, the notion of strong isoscalar, Lorentz scalar and vector
mean fields remains valid in a density-functional framework where
correlations and other higher-order effects are incorporated in the
functional.  
In the modern effective field theory approach to QCD, one incorporates only
the underlying symmetries of QCD in the hadronic lagrangian. 
With the crucial observation that while the scalar and vector fields in
nuclei are comparable to the rest mass of the nucleon, the ratios
$\gs \phizero (x) /M$ and $\gv \Vzero (x) /M$ still provide expansion
parameters less than unity,
one is able to understand the multitude of successful applications of the
relativistic mean-field treatment of nuclei (RMFT). 
Applications of QHD to nuclear structure, electroweak interactions with
nuclei, the hadronic region of the nuclear phase diagram, relativistic
heavy-ion reactions, and to many other phenomena, now abound.  
The challenge for the future is to further understand the successful
applications, the failures where they occur, the full role of hadronic
loops, and the deeper connection to QCD.

\vspace{0.5in}
\section*{Acknowledgements}

We thank our colleagues
R. J. Furnstahl,
B. R. Holstein,
H. M\"uller,
L. N. Savushkin,
and H.-B. Tang
for useful discussions and for comments on a draft of the manuscript.
This work was supported in part by the Department of Energy under
Contract No.~DE--FG02--87ER40365.




\end{document}